\documentclass[final]{jpp}
\usepackage{graphicx}
\usepackage{xcolor}
\usepackage{hyperref}
\hypersetup{colorlinks=true,linkcolor=black,citecolor=black,filecolor=black,urlcolor=black}
\newcommand{\V}[1]{\mathbf{#1}} 
\newcommand{\T}[1]{\texttt{#1}}
\newcommand\Alfven{Alfv\'en }
\newcommand\Alfvenic{Alfv\'enic }
\newcommand{\figref}[1]{Fig.~\ref{#1}}
\newcommand{\secref}[1]{\S\ref{#1}}
\newcommand{\eqref}[1]{Equation~(\ref{#1})}

\newcommand{\zhat}{\mbox{$\hat{\mathbf{z}}$}} 

\title[Field-Particle Correlations in Fourier Space]{Collisionless energy transfer in kinetic turbulence: field-particle correlations in Fourier space}

\author[ T.~C.~Li, G.~G.~Howes, K.~G.~Klein, Y.-H.~Liu, and J.~M.~TenBarge]%
{Tak~Chu~Li$^1$\thanks{Email address for correspondence: tak.chu.li@dartmouth.edu}, Gregory~G. Howes$^2$, Kristopher~G.~Klein$^3$,\\
 Yi-Hsin Liu$^1$ and Jason M. TenBarge$^4$}

\affiliation{$^1$Department of Physics and Astronomy, Dartmouth College, Hanover, New Hampshire 03755, USA\\[\affilskip]
$^2$Department of Physics and Astronomy, University of Iowa,
Iowa City, IA 52242, USA\\[\affilskip]
$^3$Lunar and Planetary Laboratory, University of Arizona, Tucson, AZ 85719, USA\\[\affilskip]
$^4$Department of Astrophysical Sciences, Princeton University, Princeton, NJ 08544, USA
}

\begin{document}
\maketitle

\date{\today}
\begin{abstract}
Turbulence is ubiquitously observed in nearly collisionless heliospheric
plasmas, including the solar wind and corona and the Earth's 
magnetosphere. Understanding the collisionless mechanisms
responsible for the energy transfer from the turbulent fluctuations
to the particles is a frontier in kinetic turbulence research. 
Collisionless energy transfer from the turbulence to the particles
can take place reversibly, resulting in non-thermal energy in 
the particle velocity distribution functions (VDFs) before eventual collisional
thermalization is realized. Exploiting the information contained in 
the fluctuations in the VDFs is valuable. 
Here we apply a recently developed method based on VDFs, the
\emph{field-particle correlation technique}, to a $\beta=1$, solar-wind-like, low-frequency \Alfvenic turbulence 
simulation with well resolved phase space 
to identify the field-particle energy transfer in velocity
space. The field-particle correlations
reveal that the energy transfer, mediated by the parallel electric field, results in significant structuring of the ion and electron VDFs in the direction parallel to the magnetic field. 
Fourier modes representing the length scales 
between the ion and electron gyroradii show that energy transfer 
is resonant in nature, localized in velocity space to the Landau resonances for each Fourier mode. The energy transfer closely follows the Landau resonant velocities with varying perpendicular wavenumber $k_\perp$ and plasma $\beta$. This resonant signature, consistent with Landau damping, is observed in all diagnosed Fourier modes that cover the dissipation range of the simulation.


\end{abstract}

\section{Introduction}\label{sec:intro}

Plasma turbulence is ubiquitous in the universe. In near-Earth
collisionless plasmas---such as the solar corona, the solar wind, and
the Earth's magnetosphere---turbulence plays a fundamental role in the
transport of energy from large fluid scales to small kinetic
scales. This cascade of turbulence energy across scales is usually
characterized by a power law in the energy spectrum of fluctuating
quantities. A spectral break (change of the power law index) in the
energy spectrum near the proton scale defines the transition into the
dissipation range where the turbulence energy is converted into plasma
heat and/or particle energization. How this process occurs remains a
key subject of current kinetic plasma research in space observations
\citep{He:2012,Alexandrova:2013b,Chen:2013b,Narita:2016,Perrone:2016,Roberts:2017Nov,Vech:2017,Wang:2018},
numerical simulations
\citep{Howes:2008a,Parashar:2009,TenBarge:2013b,Karimabadi:2013,Chang:2014,Vásconez:2014,Franci:2015b,Told:2015,Li:2016,Navarro:2016,Wan:2016,Parashar:2016,Hughes:2017,Grošelj:2017,Yang:2017,Cerri:2018,Grošelj:2018,Kawazura:2018b,Arzamasskiy:2019}
and theoretical efforts
\citep{Howes:2008b,Schekochihin:2009,Boldyrev:2013,Passot:2015,Howes:2015b,Schekochihin:2016,Adkins:2018,Schekochihin:2019}.

Recent advances in temporal and phase-space resolution in spacecraft
observations \citep{Servidio:2017,Chen:2019a} and numerical simulations
\citep{Li:2016,KleinHT:2017,Cerri:2018} have opened up new pathways
for characterizing turbulence: using the velocity distribution
functions (VDFs) to investigate non-thermal structures in the velocity
space arising from the dissipation mechanisms. These velocity-space
structures contain important information about the transfer of energy
to particles from turbulent fields, ultimately leading to the thermodynamic
heating of the plasma. Indeed, 
information in the VDFs underlying the nature of dissipation
has been explored and 
identified in recent three-dimensional (3D) kinetic turbulence
simulations \citep{Li:2016,KleinHT:2017,Cerri:2018}. Current
space missions such as the Multiscale Magnetosphere Mission (MMS)
\citep{Burch:2016} have the capibility to sample well resolved VDFs at
high time cadence. Utilizing and understanding the crucial information
contained in the phase-space represents a new avenue in kinetic turbulence
research \citep{Howes:2017b}.

Recently, an innovative technique tracking the net energy
transfer between turbulent fields and plasma particles in velocity
space was developed, the \emph{field-particle correlation technique},
\citep{Klein:2016,Howes:2017}, and first illustrated in a nonlinear
one-dimensional-one-velocity (1D-1V) Vlasov-Poisson plasma. Clear
velocity-space signatures of direct energy exchange between the
electric field and the particles was demonstrated. The 3D
implementation of this technique and application to single-point data
sets was performed in gyrokinetic turbulence simulations, which showed
velocity-space structures associated with ion Landau resonance
\citep{KleinHT:2017}.  It was also applied to diagnose the particle
energization in current sheets arising self-consistently from \Alfven
wave collisions \citep{Howes:2016b}, showing spatially
localized energization by Landau damping \citep{Howes:2018a}. 
It has recently been applied to well resolved data from the MMS, 
which has provided evidence for electron
Landau damping in the turbulent magnetosheath \citep{Chen:2019a}.

Both of these previous numerical studies sampled the 3D electromagnetic
turbulence simulation at single spatial points. Here we
apply this technique in Fourier space for the first
time. Sampling in Fourier space is a complementary approach to 
sampling in spatial space in numerical simulations, but it requires
full spatial information of the system and therefore cannot be applied to 
spacecraft data that often make in-situ measurements from only one or 
a few spatial locations at a given time. Nevertheless, application of the 
field-particle correlation technique in Fourier space can be
advantageous. 
It can be particularly useful in revealing information on the 
length scale dependence of the energy transfer mechansims.
In \Alfvenic turbulence, the
nonlinear turbulent cascade of \Alfven waves may undergo significant
collisionless damping when it reaches the kinetic regime 
$k_\perp\rho_i \gtrsim 1$ (where $k_\perp$ is the perpendicular wavenumber and 
$\rho_i$ the ion gyroradius), where the kinetic \Alfven wave becomes dispersive.
In this dispersive regime, the resonant velocity $v_{p \parallel}$ associated with the collisionless damping of kinetic \Alfven waves depends on $k_\perp$
linearly. Sampling at a single point will measure
contributions from a broad range of $k_\perp$ modes, each with a
different resonant velocity, potentially smearing out any resonant
signatures of the energy transfer in velocity space.  
Analyzing in Fourier space, however, each turbulent wave mode specified by
$\V{k}=(k_x,k_y,k_z)$ has $v_{p \parallel}$ that directly depends on $k_\perp$,
 producing a relatively clean resonant signature at that $v_{p \parallel}$ even in a turbulent plasma containing a broadband spectrum of modes.  


In this work, we study the properties of energy transfer in a
$\beta$=1, gyrokinetic turbulence simulation. Resonant signatures
associated with Landau resonances are observed in particle velocity
space in all diagnosed Fourier modes spanning the dissipation range of
the simulation. The energy transfer signatures also show a clear
$k_\perp$ dependece of the resonant velocity (moving to higher
resonant velocities with increasing $k_\perp$), reflecting the
characteristic of phase velocity of kinetic \Alfven waves. Remarkable
agreement with the $\beta$ dependence of the resonant velocity is
observed in the energy transfer signatures in identical simulations
with lower $\beta$'s. Our results demonstrate the resonant nature of
field-particle energy transfer by collisionless wave-particle
interactions with \Alfven waves.

This paper is organized as follows: the theory of nonlinear
field-particle energy transfer in collisionless plasmas in 3D and in
Fourier space is outlined in \secref{sec:et-theory}; field-particle
correlations are defined and constructed in the gyrokinetic framework
in \secref{sec:fpc}; the form of field-particle correlations in Fourier space is derived in \secref{sec:fpc_fourier}; the sampled spectrum in Fourier space is
illustrated in \secref{sec:diagnose}; the simulation setup and
parameters are described in \secref{sec:setup}; the results on energy
evolution of the systems are presented in \secref{sec:energy};
field-particle correlations in gyrotropic velocity
space $(v_\parallel,v_\perp)$ are examined in \secref{sec:gyro}; 
primary results of this work
including (a) examples of $v_\perp$-integrated, reduced correlations
of individual Fourier modes for electrons and ions, (b) correlations
summing over a range of $k_z$ modes for all diagnosed $(k_x,k_y)$
Fourier modes, (c) $k_\perp$ dependence in summed-$k_z$ modes and (d)
$\beta_i$ dependence of energy transfer signatures are presented and
discussed in \secref{sec:individual}; we further discuss and summarize
our results in \secref{sec:diss_conc}.

\section{Techniques and Simulation setup}\label{sec:tech}
Here we describe the field-particle correlation technique in terms of its formulation in physical and Fourier space, its implementation within the gyrokinetic framework, and provide the setup of the simulation.

\subsection{Rate of Energy Transfer between Fields and Particles in a Collisionless Plasma}\label{sec:et-theory}

Dissipation in weakly collisional plasmas consists of a two-step
process \citep{Howes:2017c}. First, energy is transferred from the
electromagnetic fields to plasma particles, appearing as nonthermal
energy in the particle velocity distributions. This collisionless
process is reversible. Then, the nonthermal energy is cascaded in
phase space, generating fine structures in the velocity space 
that are eventually smoothed out as thermal energy in the particle
distributions via the action of collisions, and hence realizing
irreversible heating of the plasma \citep{Howes:2008c,Schekochihin:2008,Schekochihin:2009}. The
energy transfer directly measured by the field-particle correlation
technique is the net energy transferred between the electromagnetic
fields and plasma particles in the first step.

The theory describing the energy transfer between fields and plasma
particles, which is directly measured in field-particle correlations,
was first derived for a 1D-1V Vlasov-Poisson plasma
\citep{Klein:2016,Howes:2017}, and subsequently for a 3D-3V
electromagnetic plasma \citep{KleinHT:2017}. Here we briefly summarize
the key points of the theory in 3D-3V, using insights from the 1D-1V
model. Consider the Boltzmann's equation for species $s$($=i,e$) in a
proton-electron plasma:
\begin{eqnarray}
\frac{\partial f_s}{\partial t} + \V{v}\cdot\nabla f_s +\frac{q_s}{m_s}\left( \V{E}+\frac{\V{v}\times\V{B}}{c} \right) \cdot \frac{\partial f_s}{\partial \V{v}} = \left( \frac{\partial f_s}{\partial t} \right)_c.
\label{eq:boltz}
\end{eqnarray}
Focusing on the net collsionless energy transfer in the first step of
dissipation, we neglect the collision term, $(\partial f_s/\partial
t)_c$, on the right. We then multiply the whole equation by $m_sv^2/2$
to obtain an equation for the rate of change of \emph{phase-space
  energy density}, $w_s \equiv (m_sv^2/2) f_s$:
\begin{eqnarray}
\frac{\partial w_s}{\partial t} = - \V{v}\cdot\nabla w_s - q_s\frac{v^2}{2}\V{E}\cdot\frac{\partial f_s}{\partial \V{v}} - \frac{q_s}{c}\frac{v^2}{2}(\V{v}\times\V{B})\cdot \frac{\partial f_s}{\partial \V{v}}.
\label{eq:mv2}
\end{eqnarray}
 When integrating over all space and velocity, $w_s$ gives the microscopic kinetic energy $W_s$:
\begin{eqnarray}
W_s \equiv \int\!d^3\V{x}\int\!d^3\V{v}\frac{1}{2}m_sv^2f_s.
\label{eq:W_s}
\end{eqnarray}
When integrating \eqref{eq:mv2} over all space, the first term on the
right vanishes for periodic or infinite spatial boundaries. When
integrating over velocity space, the magnetic force term involves
$\V{v}\cdot (\V{v}\times\V{B})$=0 and therefore does not contribute to
the change of the microscopic kinetic energy $W_s$, as expected. The
change of the microscopic kinetic energy, or equivalently, the energy
transfer to species $s$, comes from the electric field term. We can
further separate the $E_\parallel$ and $\V{E}_\perp$ contributions to the
rate of change of the microscopic kinetic energy as:
\begin{eqnarray}
\frac{\partial W_s}{\partial t} = -\frac{q_s}{2}\int\!d^3\V{x}\int\!d^3\V{v} \, v^2 \left( E_\parallel \frac{\partial f_s}{\partial v_\parallel} + \V{E_\perp}\cdot\frac{\partial f_s}{\partial \V{v_\perp}} \right),
\label{eq:dW_dt}
\end{eqnarray}
where "$\parallel$" is with respect to the local magnetic field. The first and second terms on the right represent energy
transfer mediated by $E_\parallel$ and $\V{E}_\perp$, respectively. Note
that this rate of energy density transfer, when integrated over
velocity, simply yields the rate the work done by the electric field,
$\V{J}_s \cdot \V{E} = J_{s\parallel} \cdot E_\parallel +
\V{J}_{s\perp} \cdot \V{E}_{\perp}$
\citep{Klein:2016,Howes:2017,KleinHT:2017}.

\subsection{Field-Particle Correlations and Gyrokinetics}\label{sec:fpc}
We can now construct the field-particle correlations based on
\eqref{eq:dW_dt}. Without integrating over space or velocity, the
transfer rate of phase-space energy density can be directly computed
by correlating $E_\parallel$ or $\V{E}_\perp$ with the corresponding
velocity derivative of $f_s$ at a point $\V{x}$ and time $t$ over a
correlation interval $\tau$ as:
\begin{eqnarray}
C_{E_\parallel}(\V{x},\V{v},t,\tau) = C\left(-q_s \frac{v_\parallel^2}{2}\frac{\partial f_s(\V{x},\V{v},t)}{\partial v_\parallel}, E_\parallel(\V{x},t)\right)
\label{eq:corr_par}  
\end{eqnarray}
\begin{equation}
C_{E_\perp}(\V{x},\V{v},t,\tau) = C\left(-q_s \frac{v_{\perp1}^2}{2}\frac{\partial f_s(\V{x},\V{v},t)}{\partial v_{\perp1}}, E_{\perp1}(\V{x},t)\right) \\
+ C\left(-q_s \frac{v_{\perp2}^2}{2}\frac{\partial f_s(\V{x},\V{v},t)}{\partial v_{\perp2}}, E_{\perp2}(\V{x},t)\right),\\ 
\label{eq:corr_perp}  
\end{equation}
where the unnormalized correlation on the right is defined as:
\begin{eqnarray}
C(A,B) = \frac{1}{N}\sum\limits^{i+N}_{j=i+1} A_jB_j
\label{eq:corr_nonnorm}  
\end{eqnarray}
for real quantities $A_j$ and $B_j$ measured at discrete times
$t_j=j\Delta t$ over a correlation interval $\tau=N\Delta t$ that starts from
an initial time $t_{i+1}$. The
factor $v^2=v_\parallel^2+v_\perp^2$ in \eqref{eq:dW_dt} is reduced to
$v_\parallel^2$ in \eqref{eq:corr_par} upon integration
$\int\!d\V{v_\perp}$ as it can be separated from
$\int\!dv_\parallel$. The correlation interval $\tau$ is an important
parameter in the unnormalized correlation. When $\tau$ is set to zero,
the unnormalized correlation measures the
instantaneous energy transfer at each time $t$, which often contains
a large oscillatory component due to undamped wave motions 
To measure the net secular or long-term energy transer, 
$\tau$ is chosen to be
sufficiently long such that the oscillatory component in the
instantaneous energy transfer, which is often large
but does not contribute to the net energy transfer, can be averaged
out. Normally, one wave period of the outer-scale mode is sufficient
\citep{Howes:2017}. The parallel correlation
$C_{E_\parallel}(\V{x},\V{v},t,\tau)$ measures the net energy transfer
rate mediated by the parallel electric field, and is therefore a
suitable measure for Landau damping \citep{Landau:1946}, or mechansims
like strong-guide-field magnetic reconnection that may be dominated by
$E_\parallel$ \citep{Dahlin:16}; it, however, does not capture transit time damping \citep{Barnes:1966,Quataert:1998} that arises from the change in the magnetic field strength. Similarly, the perpendicular correlation
$C_{E_\perp}(\V{x},\V{v},t,\tau)$ captures the net transfer rate
mediated by the perpendicular electric field, and is therefore
suitable for determining cyclotron damping \citep{Coleman:1968,Isenberg:1983} 
or stochastic ion heating
\citep{Chen:2001,Chandran:2010a}. In this study using gyrokinetic simulations,
we cannot explore perpendicular energization due to the conservation
of the magnetic moment in gyrokinetic theory, so we focus here on the
net energy transfer rate accomplished by $E_\parallel$.

In gyrokinetics, the system evolves in a three-spatial and
two-velocity dimension (3D-2V) phase-space, where the two velocity
coordinates are $v_\parallel$ and $v_\perp$.
The distribution function to the first order in gyrokinetics \citep{Howes:2006} is given by
\begin{eqnarray}
f_s(v_\parallel,v_\perp) =
\left(1-\frac{q_s\varphi}{T_{0s}}\right)F_{0s}(v)+
h_s(v_\parallel,v_\perp),
\label{eq:f_s}
\end{eqnarray} 
where $F_{0s} = (n_{0s}/\pi^{3/2}v_{ts}^3)\exp(-v^2/v_{ts}^2)$ is the
equilibrium Maxwellian distribution, $q_s\varphi/T_{0s}$ is the
Boltzmann term ($q_s$ the species charge and $\varphi$ the electric
potential), $h_s$ is the first-order gyroaveraged part of the
perturbed distribution. When substituting $f_s(v_\parallel,v_\perp)$
into the first term in \eqref{eq:dW_dt}, we see that 
$\partial F_{0s}(v)/\partial v_\parallel$ is odd in $v_\parallel$, and is
multiplied by an even power $v_\parallel^2$. Thus, this term vanishes
upon integration over all $v_\parallel$ (again, $\int\!d\V{v_\perp}$
can be separated from $\int\!dv_\parallel$ here). Hence, the
equilibrium Maxwellian distribution and the Boltzmann term in
\eqref{eq:f_s} do not contribute to the net energy transfer in the
field-particle correlation in \eqref{eq:corr_par}. The perturbed,
gyroaveraged contribution $h_s$ can be used in place of $f_s$ for the
purpose of calculating field-particle correlations.

Another convenient transformation used is the complementary
distribution function $g_s$ \citep{Schekochihin:2009}
\begin{eqnarray}
 g_s(v_\parallel,v_\perp) = h_s(v_\parallel,v_\perp) - \frac{q_s F_{0s}}{T_{0s}}\left\langle \varphi - \frac{\V{v}_\perp\cdot\V{A}_\perp}{c} \right\rangle_{\V{R}_s},
\label{eq:g_s}
\end{eqnarray} 
where $\langle ... \rangle$ represents gyroaveraging at constant
guiding center $\V{R}_s$, capturing the perturbations to the
Maxwellian distribution in the moving frame of \Alfven waves. It
retains the parallel perturbations of $\delta f_s (\equiv f_s -
F_{0s})$ in \eqref{eq:f_s}, and is therefore appropriate for
calculating the net energy transfer.  Note that the term $\langle
\varphi - \V{v}_\perp\cdot\V{A}_\perp/c \rangle_{\V{R}_s}$ is
independent of $v_\parallel$, and therefore yields zero upon
integration $\int\!dv_\parallel$ when substituted into the first term
in \eqref{eq:dW_dt}. Therefore, this term is the same,
 whether evaluated with $h_s$ or $g_s$. 
 As a result, the correlation $C_{E_\parallel}$ in
\eqref{eq:corr_par}, which is based on this term,
is quantitatively similar using $h_s$ or
$g_s$, but $g_s$ contains less additional terms and can more
clearly reveal structures near the resonant velocity in the 
distribution functions themselves.

Here we will use $g_s$ in the calculation of
field-particle correlations, so the specific parallel field-particle
correlation used to analyze the gyrokinetic turbulence simulations
here takes the form
\begin{eqnarray}
C_{E_\parallel}(\V{x},v_\parallel,v_\perp,t,\tau) = C\left(-q_s \frac{v_\parallel^2}{2}\frac{\partial g_s(\V{x},v_\parallel,v_\perp,t)}{\partial v_\parallel}, E_\parallel(\V{x},t)\right),
\label{eq:corr_par_g}  
\end{eqnarray}
where $E_\parallel$ is gyroaveraged.
For simplicity, we refer to this form of parallel correlation that
depends on gyrotropic velocity space $(v_\parallel,v_\perp)$ as
$C_{E_\parallel}(v_\parallel,v_\perp,t)$. The parallel reduced
correlation $C_{E_\parallel}(v_\parallel,t)$, defined by integrating
$C_{E_\parallel}(v_\parallel,v_\perp,t)$ over all $v_\perp$, is given
by
\begin{eqnarray}
 C_{E_\parallel}(v_\parallel,t,\tau)=\int\! v_\perp dv_\perp C_{E_\parallel}(v_\parallel,v_\perp,t,\tau).
\label{eq:corr_par_reduced}  
\end{eqnarray}

\subsection{Field-Particle Correlations in Fourier Space}\label{sec:fpc_fourier}
To derive the appropriate form of the field-particle correlation in
Fourier space, we first express the electric field at a single position
$\V{x}$ by a Fourier series
\begin{equation}
  \V{E}(\V{x}) = \sum_{\V{k}}  \V{E}_{\V{k}} e^{i \V{k}\cdot \V{x}},
\end{equation}
summed over all Fourier modes $\V{k}$ in a 3D domain of volume
$L^3$. The particle velocity distribution function $f_s(\V{x})$ is
expressed similarly in terms of its Fourier coefficients $f_{s\V{k}}$.
The product of two real quantities $A(\V{x})$ and $B(\V{x})$
integrated over a volume $L^3$ can be converted to a sum over the
product of the Fourier coefficients
\begin{eqnarray}
\int\!d^3\V{x} \, A(\V{x}) B(\V{x}) = L^3\sum_{\V{k}} A^{*}_{\V{k}}
B_{\V{k}} = L^3\sum_{\V{k}} A_{\V{k}} B^*_{\V{k}},
\label{eq:dx_dk}
\end{eqnarray}
where the complex Fourier coefficients $A_{\V{k}}$ and $B_{\V{k}}$
both satisfy the reality condition $A_{\V{k}}= A^*_{-\V{k}}$.
Substituting these Fourier series into \eqref{eq:dW_dt} and using this
relation, we obtain the rate of change of microscopic kinetic energy
in terms of the Fourier coefficients of the electric field and
particle distribution function,
\begin{eqnarray}
  \frac{\partial W_s}{\partial t} =  \sum_{\V{k}} \left(-\frac{q_s}{2}\right) L^3\int\!d^3\V{v} \, v^2
  \left( E_{\parallel \V{k}} \frac{\partial f^{*}_{s \V{k}}}{\partial v_\parallel}
  + \V{E}_{\perp\V{k}}\cdot\frac{\partial f^{*}_{s \V{k}}}{\partial \V{v_\perp}} \right),
\label{eq:dW_dt_fourier}
\end{eqnarray}

Without summing over all Fourier modes, the contribution to the change
of particle energy from each Fourier mode $\V{k}$ is given by the
summand, leading to the following forms of the field-particle
correlations in Fourier space corresponding to \eqref{eq:corr_par} and
\eqref{eq:corr_perp}:
\begin{eqnarray}
C_{E_\parallel}(\V{k},\V{v},t,\tau) = C\left(-q_s \frac{v_\parallel^2}{2}\frac{\partial f^{*}_{s \V{k}}(\V{v},t)}{\partial v_\parallel}, E_{\parallel\V{k}}(t)\right)
\label{eq:corr_par_k}  
\end{eqnarray}
\begin{equation}
C_{E_\perp}(\V{k},\V{v},t,\tau) = C\left(-q_s \frac{v_{\perp1}^2}{2}\frac{\partial f^{*}_{s \V{k}}(\V{v},t)}{\partial v_{\perp1}}, E_{\perp1\V{k}}(t)\right)\\
+ C\left(-q_s \frac{v_{\perp2}^2}{2}\frac{\partial f^{*}_{s \V{k}}(\V{v},t)}{\partial v_{\perp2}}, E_{\perp2\V{k}}(t)\right).\\
\label{eq:corr_perp_k}  
\end{equation}

The discussion on using $g_s$ in gyrokinetics in place of $f_s$, which
is based on velocity integrals, applies also to the correlations in
Fourier space (with the gyroaveraging operation $\langle ... \rangle$
in \eqref{eq:g_s} reduced to multiplications by Bessel functions
\citep{Howes:2006,Numata:2010}). The parallel correlation in
\eqref{eq:corr_par_g} then becomes
\begin{eqnarray}
C_{E_\parallel}(\V{k},v_\parallel,v_\perp,t,\tau) = C\left(-q_s \frac{v_\parallel^2}{2}\frac{\partial g^{*}_{s \V{k}}(v_\parallel,v_\perp,t)}{\partial v_\parallel}, E_{\parallel\V{k}}(t)\right).
\label{eq:corr_par_g_k}  
\end{eqnarray}
This represents the net energy transfer rate in Fourier and
gyrotropic velocity space. Together with the parallel reduced
correlation in \eqref{eq:corr_par_reduced}, it is the form of
field-particle correlations we use in this work.

\subsection{Diagnosing Energy Transfer in Fourier Space}\label{sec:diagnose}

\begin{figure} \centering
\hbox{{(a)\hspace{-1.em}} \hfill \resizebox{2.65in}{!}{\includegraphics[scale=1,trim=0cm 0cm 0cm 0cm, clip=true]{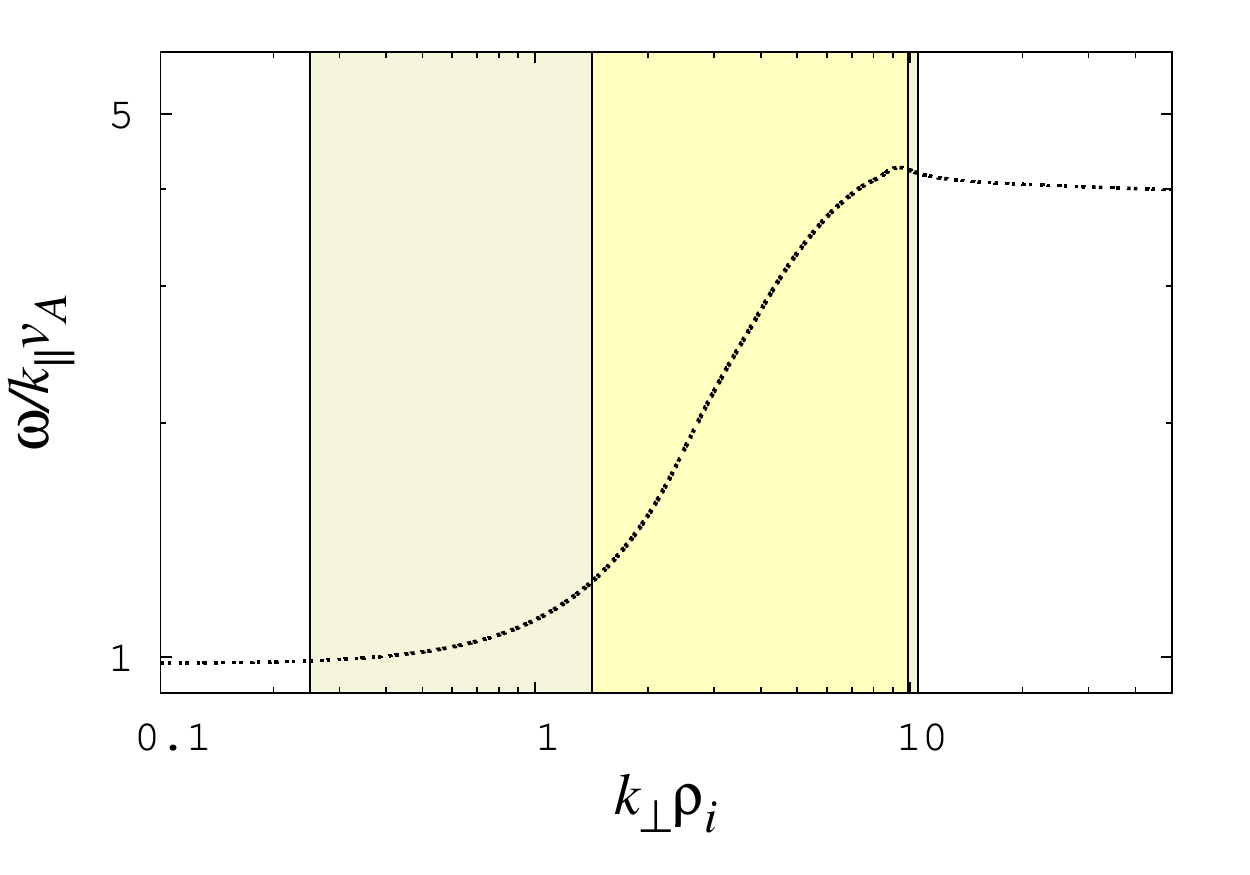}} \hfill  
{(b)\hspace{-1.em}} \hfill \resizebox{2.65in}{!}{\includegraphics[scale=1.]{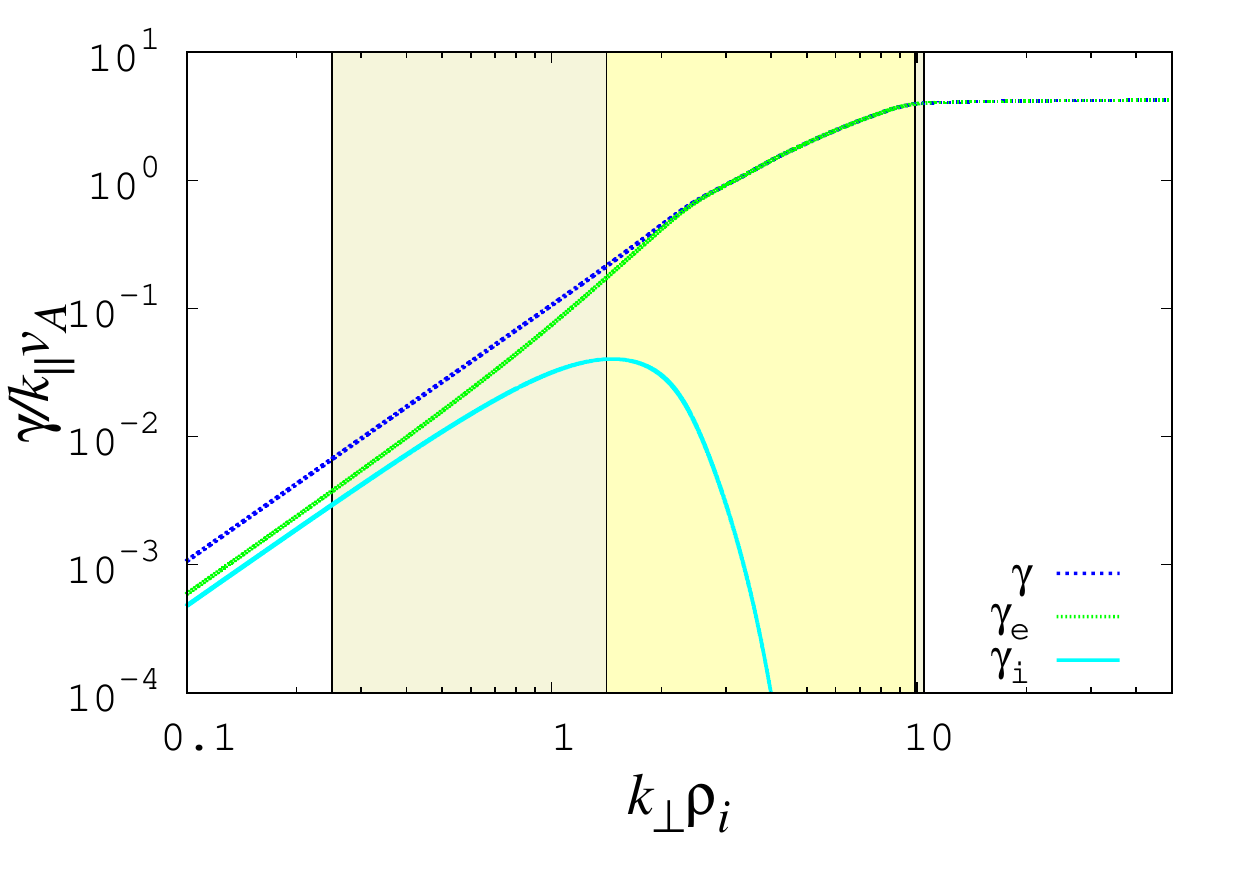}} \hfill  
  } 
\vspace{.5em}
\caption{ \label{fig:1om_g} The linear gyrokinetic dispersion relation
  for a $\beta_i=1$ and $T_i/T_e=1$ plasma, showing (a) the normalized
  real frequency or phase velocity $\omega/k_\parallel v_A$ (dashed
  black) and (b) damping rates $\gamma/k_\parallel v_A$ for the
  electrons (green), ions (cyan) and the total damping rate
  (blue). The total shaded region represents the resolved dynamic
  range $0.25 \le k_\perp\rho_i \le 10.5$. The inner yellow shaded
  region indicates the range of the sampled $k_\perp$ spectrum given
  in Table \ref{tab:kspace}, $1.4 \le k_\perp\rho_i \le 9.9$. }
\end{figure}

Previous studies have used the field-particle
correlation technique at a single point in physical space to evaluate the energy transfer between fields and particles \citep{Klein:2016,Howes:2017,Klein:2017,Howes:2017b,KleinHT:2017,Howes:2018a,Chen:2019a}.
Here we diagnose the particle energization in Fourier space, where
each Fourier mode $\V{k}$ is specified by a normalized wavevector 
$(k_x\rho_i,k_y\rho_i,k_zL_z/2 \pi)$, where $L_z$ is the system size
in $z$. There are a number of advantages when applying
field-particle correlations in Fourier space compared to physical
space.  First, by utilizing full spatial information of the whole domain
as opposed to a single point, we can determine how the length scale of
fluctuations, within the broadband turbulent spectrum, influences the
collisionless field-particle energy transfer.
Second, in the gyrokinetic limit $k_\parallel \ll k_\perp$, well
justified for the anisotropic turbulence observed in the solar wind 
at kinetic scales \citep{Sahraoui:2010b,
  Narita:2011,Roberts:2013,Roberts:2015b}, the parallel phase velocity
$v_{p \parallel}= \omega/k_\parallel$ of the linear wave modes depends
only on the perpendicular wavenumber and the plasma parameters,
$\overline{\omega} \equiv \omega/(k_\parallel v_A) = v_{p
  \parallel}/v_A= \overline{\omega} (k_\perp \rho_i,\beta_i, T_i/T_e)$
\citep{Howes:2006}. Therefore, the parallel phase velocity that
governs resonant collisionless interactions is well defined for each
mode $\V{k}$ in Fourier space. In contrast, at a single point in physical
space, the dispersive nature of kinetic \Alfven waves (see
\figref{fig:1om_g}) would lead to a range of resonant velocities,
broadening the energy transfer signal and potentially smearing out
any resonant energy transfer signatures in velocity
space. Third, the broadband turbulent fluctuations have decreasing
amplitudes with increasing perpendicular wavenumber $k_\perp \rho_i$,
but in Fourier space one can isolate the fluctuations at higher
$k_\perp$ from the rest of the fluctuations containing much larger amplitude fluctuations at lower $k_\perp$.

In \figref{fig:1om_g}, we plot (a) the normalized frequency
$\omega/k_\parallel v_A$ and (b) the normalized damping rate
$\gamma/k_\parallel v_A$ as a function of $k_\perp\rho_i$ from the
linear gyrokinetic dispersion relation \citep{Howes:2006} for a plasma
with parameters $\beta_i \equiv 8 \pi n_i T_i/B^2 =
v_{ti}^2/v_{A}^2=1$, $T_i/T_e=1$, and $m_i/m_e=25$.  The total shaded
region represents the resolved dynamic range of $k_\perp$ in the
simulation, $0.25 \le k_\perp\rho_i \le 10.5$. Table~\ref{tab:kspace}
lists the perpendicular wavevectors $(k_x\rho_i,k_y\rho_i)$ of Fourier
modes that are sampled in the simulation, along with the corresponding
normalized perpendicular wavenumber $k_\perp \rho_i$, where
$k_\perp\equiv \sqrt{k_x^2+k_y^2}$.  The range of these sampled
perpendicular wavenumbers $k_\perp \rho_i$ is indicated in
\figref{fig:1om_g} by the inner yellow shaded region, covering $1.4 \le
k_\perp\rho_i \le 9.9$ or 0.3 $\leq k_\perp\rho_e \le$ 2.0.  For each
$(k_x\rho_i,k_y\rho_i)$ mode, seven $k_zL_z/2 \pi \in$ [-3,3] modes
are selected, for a total sampling of 42 modes in
$(k_x\rho_i,k_y\rho_i,k_zL_z/2 \pi)$ Fourier space.  Also tabulated in
Table~\ref{tab:kspace} are the parallel phase velocities normalized to
the ion thermal velocity, $v_{p \parallel}/v_{ti} = \omega/k_\parallel v_A$
 (valid for a $\beta_i=1$ plasma) and $v_{p \parallel}/v_{te}$. 
The last
column gives the linear wave period $\tau_k$ of a kinetic \Alfven wave
specified by $k_\perp \rho_i$.

\begin{table}
\begin{center}
\def~{\hphantom{0}}
\begin{tabular}{ccccc}
 $(k_x\rho_i,k_y\rho_i)$  & \hspace*{0.15in} $k_\perp\rho_i$  \hspace{0.15in}&  \hspace*{0.25in}$\omega/k_\parallel v_A = v_{p \parallel}/v_{ti}$ \hspace{0.25in} &  \hspace*{0.1in}$v_{p \parallel}/v_{te}$  \hspace*{0.1in}& $\tau_k/\tau_0$ \\
\hline
       (1,1)   &   1.4   &  1.3   &   0.25 & 0.79 \\  
       (2,2)   &   2.8   &  2.1   &   0.42 & 0.47 \\ 
       (3,4)   &   5.0   &  3.3   &   0.66 & 0.30 \\
       (5,5)   &   7.1   &  4.0   &   0.79 & 0.25 \\
       (6,6)   &   8.5   &  4.2   &   0.84 & 0.24 \\
       (7,7)   &   9.9   &  4.2   &   0.85 & 0.23 \\ 
\end{tabular}
\caption{List of diagnosed $(k_x,k_y)$ modes in Fourier space.}\label{tab:kspace}
\end{center}
\end{table}


The dispersion relation plot in \figref{fig:1om_g}(a) shows shows the
characteristic dispersion of kinetic \Alfven waves, where the parallel
phase velocity begins increasing at the transition of
$k_\perp\rho_i\sim$ 1. Given the energy transfer from fields to
particles due to resonant collisionless wave damping is governed by
the parallel phase velocity, one would expect to observe that the
region of velocity space in which the particles exchange energy with
the fields, known as the {\it velocity-space signature} of the particle 
energization, will shift to higher parallel velocities,
tracking the increased parallel phase velocity as the perpendicular
wavenumber $k_\perp \rho_i$ increases. Our analysis in
Sections~\secref{sec:gyro} and \secref{sec:individual} confirms this
expectation.

Finally, the collisionless damping rates for ions (cyan) and electrons
(green) and the total damping rate (blue) are plotted in
\figref{fig:1om_g}(b). The total damping of the waves is primarily
attributed to electron damping for $k_\perp\rho_i >$ 2 while both ion
and electron damping contribute for $k_\perp\rho_i <$ 2. Particularly,
ion collisionless damping becomes significant and peaks around
$k_\perp\rho_i\sim$1, but drops off very rapidly for $k_\perp\rho_i
>2$.

\subsection{Simulation Setup}\label{sec:setup}
The simulation was performed with the gyrokinetic code \T{AstroGK}
\citep{Numata:2010}. It has been extensively used to investigate
turbulence in weakly collisional space plasmas
\citep{Howes:2008a,Tatsuno:2009,Howes:2011a,TenBarge:2012a,Nielson:2013a,TenBarge:2013a,Howes:2016b,Li:2016,Howes:2018a}
as well as collisionless magnetic reconnection in the
strong-guide-field limit
\citep{Numata:2011,TenBarge:2014b,Kobayashi:2014,Numata:2015}.
\T{AstroGK} is an Eulerian continuum code with triply periodic
boundary conditions. It has a slab geometry elongated along the
straight, uniform background magnetic field, $\V{B}_0=B_0 \zhat$. The
code evolves the perturbed gyroaveraged Vlasov-Maxwell equations in
five-dimensional phase space (3D-2V)
\citep{Frieman:1982,Howes:2006}. The evolved quantities are the
electromagnetic gyroaveraged complementary distribution function
$g_s(x,y,z,\lambda,\varepsilon)$ for each species $s$, the scalar
potential $\varphi$, parallel vector potential $A_\parallel$ and the
parallel magnetic field perturbation $\delta B_\parallel$, where
$\parallel$ is along the total local magnetic field $\V{B}=B_0\zhat+
\delta \V{B}$. Note that the total and background magnetic fields are the same to first-order accuracy, which is retained for perturbed fields in gyrokinetics. The 2V velocity grid is specified by pitch angle
$\lambda=v_\perp^2/v^2$ and energy $\varepsilon=v^2/2$. The
background distribution functions for both species are stationary
uniform Maxwellians. Collisions are incorporated using a fully
conservative, linearized gyro-averaged Landau collision operator
consisting of energy diffusion and pitch-angle scattering between
like-particles, electrons and ions, and inter-species scattering of
electrons off ions \citep{Abel:2008,Barnes:2009}.

The same simulation setup as \citet{Li:2016} is used, a 3D
generalization of the classic 2D Orszag-Tang Vortex (OTV) problem
\citep{Orszag:1979}. The 2D problem was widely used in fluid and 
magnetohydrodynamic turbulence simulations; various 3D generalizations 
have also been used for studying turbulence
\citep{Politano:1989,Dahlburg:1989,Picone:1991,Politano:1995b,Grauer:2000,Mininni:2006,Parashar:2009,Parashar:2014b}. This
3D OTV setup consists of counterpropagating \Alfven waves along
$\V{B}=B_0\zhat$ such that on the $z=0$ plane, its initial condition
reduces to that of the 2D OTV problem. An initial amplitude
$\mbox{z}_0$ of Els\"asser variables in the OTV setup \citep{Li:2016}
is chosen to yield a nonlinearity parameter $\chi =
k_\perp\mbox{z}_0/(k_\parallel v_A) = 1$ (where $v_A=B_0/\sqrt{4\pi
  m_in_0}$ is a characteristic \Alfven speed), corresponding to
\emph{critical balance} \citep{Goldreich:1995}, or a state of strong
turbulence. Note that previous studies using \T{AstroGK}
have shown consistency with the prediction of a critically balanced cascade 
in the dissiaption range \citep{TenBarge:2012a,TenBarge:2013b}.

To resolve the turbulent cascade from the inertial range to the
dissipation range for both ions and electrons, a reduced mass ratio of
$m_i/m_e=25$ is used. The simulation domain is $L_{\perp}=8\pi \rho_i$
and dimensions are
$(n_x,n_y,n_z,n_\lambda,n_\varepsilon,n_s)=(128,128,32,64,32,2)$,
which resolves a dynamic range of $0.25 \le k_\perp\rho_i \le 10.5$,
or $0.05 \le k_\perp\rho_e \le 2.1$, covering both the inertial and
dissipation ranges of the system. Typical conditions of solar wind
turbulence are considered with ion plasma beta $\beta_i = 8 \pi n_i
T_{0i}/B_0^2$ = 1, where $T_{0i}$ is the constant background ion
temperature, and equal temperatures are used for ions and electrons
($T_{0i}/T_{0e}$ = 1).

To realize collisional dissipation of turbulent energy in particle
velocity space, weak finite collision frequencies are required
\citep{Howes:2008c,Schekochihin:2008,Schekochihin:2009}.  Low collision frequencies
$\nu_s/\omega_{A0} \ll 1$ are chosen to avoid altering the weakly
collisional dynamics, where collision frequencies of $\nu_i$ =
3$\times$10$^{-3} \omega_{A0}$ and $\nu_e$ = 0.06 $\omega_{A0}$ (where
$\omega_{A0} \equiv k_{\parallel}v_A$ is a characteristic \Alfven wave
frequency) are used.  We also employ a constant ion hypercollision frequency of
$\nu_{Hi}$ = 6$\times$10$^{-3} \omega_{A0}$ and adaptive electron
hypercollision frequency of $\nu_{He}$ = 0.12 $\omega_{A0}$ to ensure
fluctuations in velocity space remain well resolved
\citep{Howes:2008c,Howes:2011a}. 
The electron hypercollisional coefficient is adjusted based on nonlinear estimations of the total collisional damping rate (including hypercollisions) $\gamma_{nl}$ and of the energy transfer frequency $\omega_{nl}$ (that is given by the cascade model \citep{Howes:2008b} and dependent on the magnitude of the magnetic field fluctuations) such that a value of $\gamma_{nl}/\omega_{nl}\simeq$1/2$\pi$ is achieved within some tolerance.  
The hypercollisionality chosen has the
form of a pitch-angle-scattering operator with a wave-number-dependent
collision rate $\nu_{Hs}(k_\perp/k_{\perp max})^{p_s}$, where
$k_{\perp max}$ is a grid-scale wave number with $p_i$=4 and
$p_e$=8. It produces positive-definite heating close to the grid scale
and is sufficient to terminate the turbulent cascade at the smallest
resolved scales.

\section{Energy Evolution and Current Density}\label{sec:energy}
Here we present an overview of the evolution of the 3D OTV simulation
in terms of the evolution of the energy and the self-consistent
development of current sheets in the simulation.

\begin{figure}\centering
  \resizebox{4.25in}{!}{\includegraphics[scale=0.6,trim=1cm 5cm 1cm 5cm, clip=true]{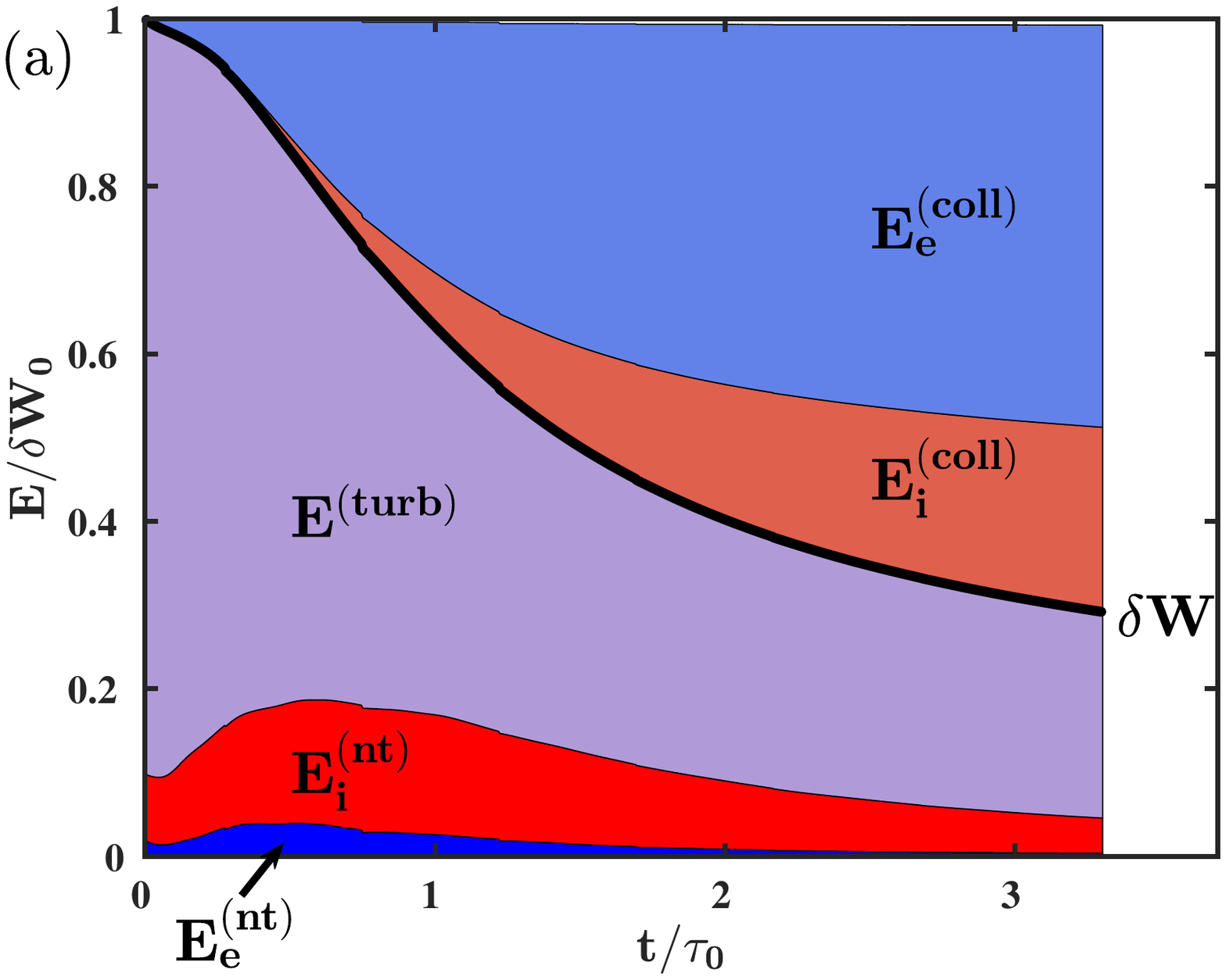}}
  \vskip-0.5in
  \resizebox{4.25in}{!}{\includegraphics[scale=0.6,trim=1cm 5cm 1cm 5cm, clip=true]{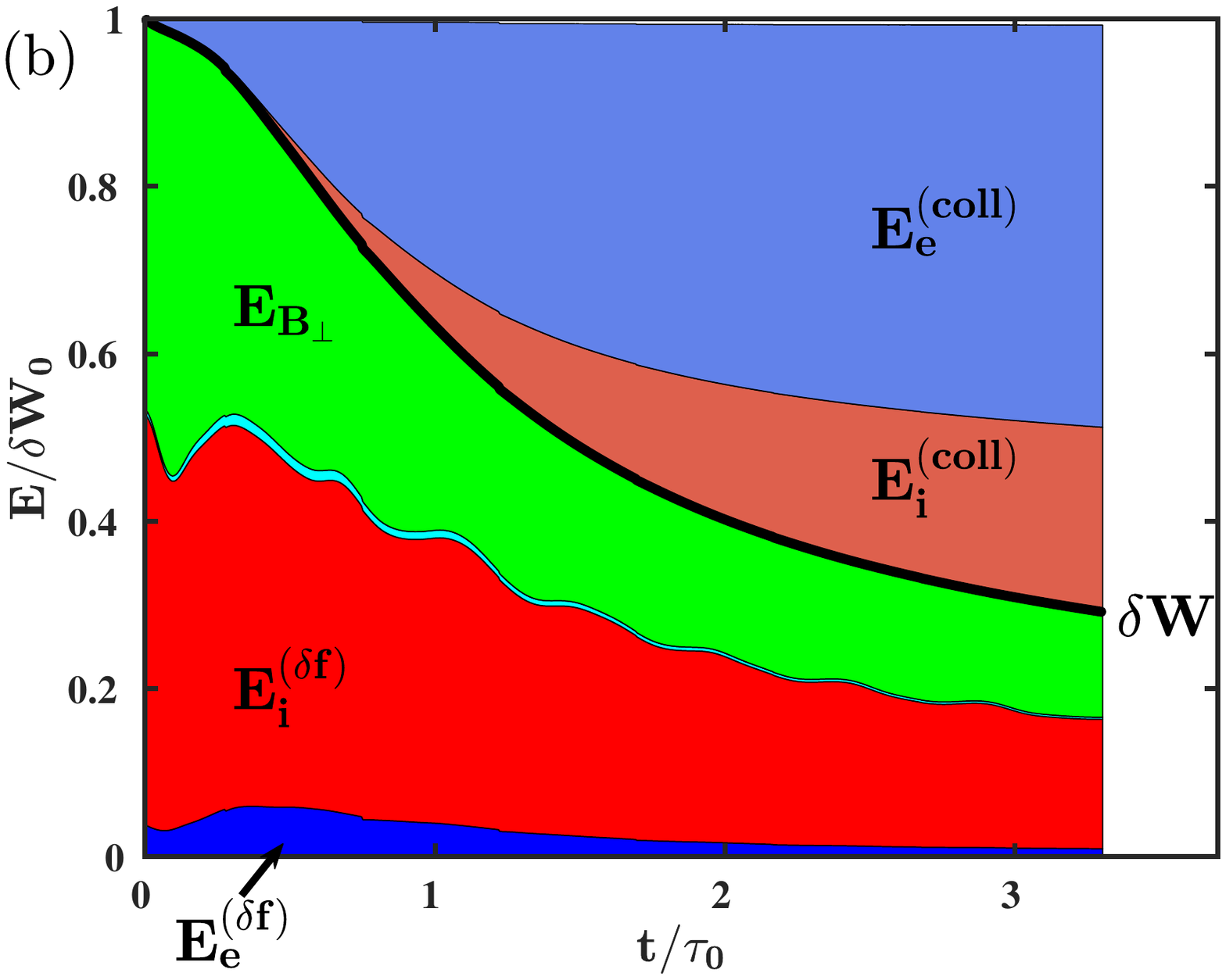}}
  \vskip-0.25in
 \caption{(a) The energy budget of the simulation vs.~time, showing
    the turbulent energy $E^{(turb)}$, non-thermal ion energy
    $E^{(nt)}_i$, non-thermal electron energy $E^{(nt)}_e$, ion heat
    $E^{(coll)}_i$ and electron heat $E^{(coll)}_e$. (b) The same
    energy budget decomposed according to \eqref{eq:deltaW_GK},
    showing the perpendicular magnetic field energy $E_{B_\perp}$,
    parallel magnetic field energy $E_{B_\parallel}$ (cyan, not
    labeled, appearing between $E_{B_\perp}$ and $E^{(\delta f)}_i$),
    total fluctuating ion kinetic energy $E^{(\delta f)}_i$, total
    fluctuating electron kinetic energy $E^{(\delta f)}_e$, ion heat
    $E^{(coll)}_i$ and electron heat $E^{(coll)}_e$. The total
    fluctuating energy $\delta W$ is shown in both panels (thick black
    line). }
 \label{fig:ft3_heat}
\end{figure}

\begin{figure}\centering
  \resizebox{5.0in}{!}{\includegraphics[scale=1,trim=0cm 6cm 0cm 0cm, clip=true]{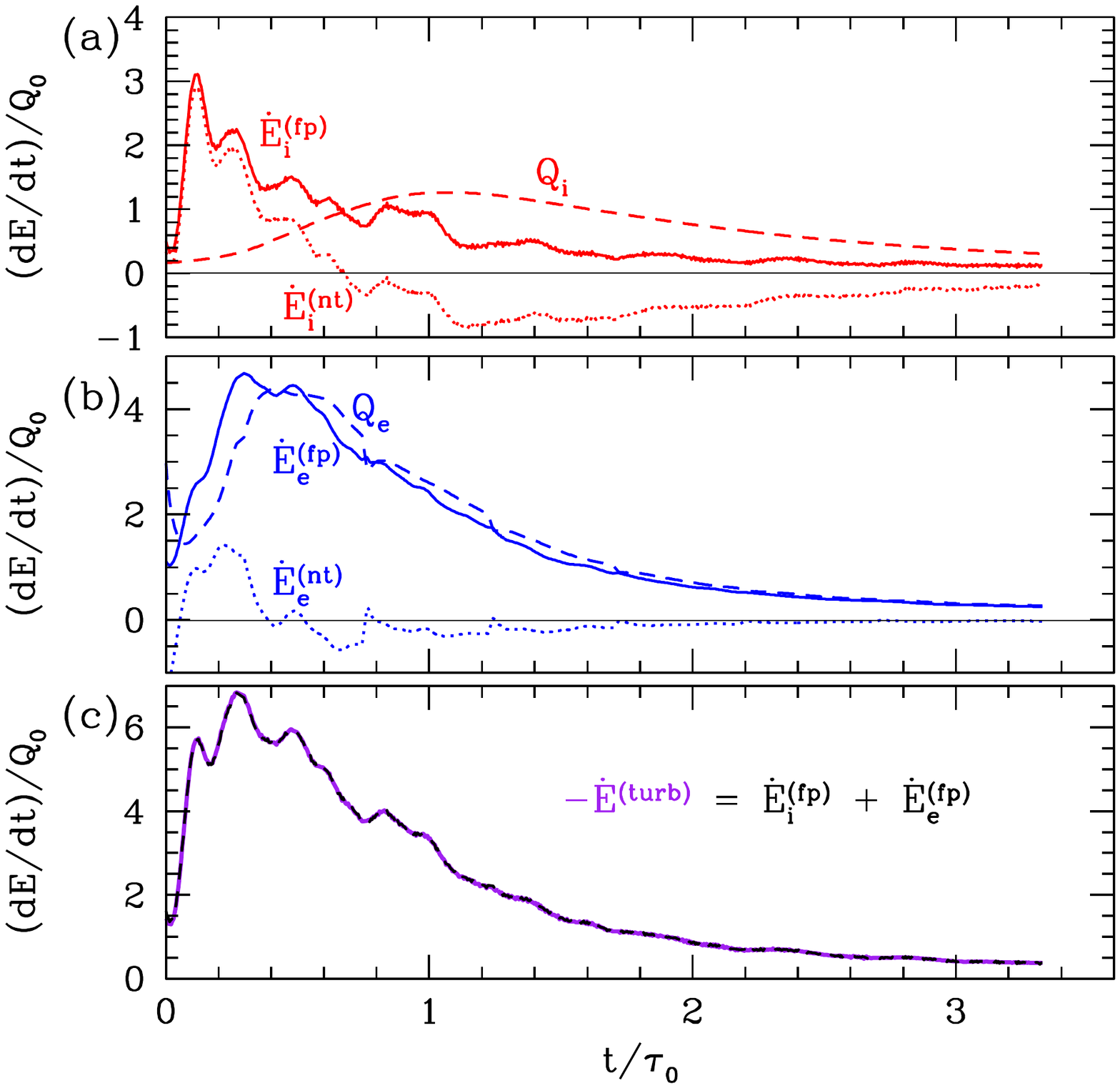}}
 \caption{The rate of energy transfer by field-particle interactions
    $\dot{E}^{(fp)}_s$ (solid), the rate of change of non-thermal
    energy $\dot{E}^{(nt)}_s$ (dotted), and the collisional heating
    rate $Q_s$ (dashed) for (a) ions (red) and (b) electrons (blue).  (c) The
    energy balance between the loss of turbulent energy $
    -\dot{E}^{(turb)}$ (purple solid) and the summed transfer of energy to both ions
    and electrons, $\dot{E}^{(fp)}_i + \dot{E}^{(fp)}_e$ (black dashed). }
 \label{fig:ft3_dedt}
\end{figure}

Under weakly collisional conditions, the removal of energy from
turbulent fluctuations and the eventual conversion of that energy into
plasma heat, unlike in the more familiar fluid limit, is a two-step
process \citep{Howes:2017c}: first, the turbulent fluctuations are
damped through reversible, collisionless interactions between the
electromagnetic fields and the plasma particles, leading to
energization of the particles; and second, this non-thermal
energization of the particle velocity distributions is subsequently
thermalized by arbitrarily weak collisions, thereby accomplishing the
ultimate conversion of the turbulent energy into particle heat.

In a gyrokinetic system, the \emph{total fluctuating energy} $\delta W$
\citep{Howes:2006,Brizard:2007,Schekochihin:2009} is given
by\footnote{Note that in the gyrokinetic approximation, the electric
  field energy is relativistically small relative to the magnetic
  field energy \citep{Howes:2006}.}
\begin{equation}
\delta W =
\int\!d^3\V{r} \left[ \frac{|\delta \V{B}|^2+|\delta \V{E}|^2}{8\pi}+\sum\limits_s 
\int d^3 \V{v}\ \frac{T_{0s}\delta f_s^2}{2F_{0s}} \right], 
\label{eq:deltaW_GK}
\end{equation}
where the index $s$ indicates the plasma species and $T_{0s}$ is the
temperature of each species' Maxwellian equilibrium. The left term
represents the electromagnetic energy and the right term represents
that microscopic fluctuating kinetic energy of the particles of each
plasma species $s$. As explained in \citet{Howes:2018a}, in the
standard form of gyrokinetic theory the appropriate
conserved quadratic quantity in gyrokinetics is the Kruskal-Obermann
energy, $E^{(\delta f)}_s\equiv \int d^3 \V{r} \int d^3
\V{v}\ T_{0s}\delta f_s^2/2F_{0s}$ \citep{Kruskal:1958,Morrison:1994},
in contrast to the usual kinetic theory definition of microscopic
kinetic energy, $ \int d^3 \V{r} \int d^3 \V{v} \ (m_s v^2/2) f_s$.
Note also that $\delta W$ includes neither the equilibrium thermal
energy, $\int d^3 \V{r} \frac{3}{2} n_{0s} T_{0s} = \int d^3 \V{r}
\int d^3 \V{v} \frac{1}{2} m_sv^2 F_{0s}$, nor the equilibrium
magnetic field energy, $\int d^3 \V{r} \ B_0^2/8\pi$. Thus, the terms
of $\delta W$ in \eqref{eq:deltaW_GK} represent the perturbed
electromagnetic field energies and the microscopic kinetic energy of
the deviations from the Maxwellian velocity distribution for each
species.

A more intuitive form of the total fluctuating energy $\delta W$ can
be obtained by separating out the kinetic energy of the bulk motion of
the plasma species from the non-thermal energy in the distribution
function that is not associated with bulk flows \citep{Li:2016},
\begin{equation}
\delta W =
\int\!d^3\V{r} \left[\frac{|\delta \V{B}|^2+|\delta \V{E}|^2}{8\pi}+\sum\limits_s
  \left( \frac{1}{2}n_{0s}m_s|\delta
  \V{u_s}|^2 + \frac{3}{2} \delta P_s  \right) \right]
\label{eq:deltaW}
\end{equation}
where $n_{0s}$ is the equilibrium density, $m_s$ is mass, and
$\delta\V{u_s}$ is the fluctuating bulk flow velocity, which includes 
the parallel and perpendicular bulk flows. The
\emph{non-thermal energy} in the distribution function (not including
the bulk kinetic energy) is defined by \citep{TenBarge:2014b}
\begin{equation}
E^{(nt)}_s\equiv \int d^3
\V{r}\frac{3}{2} \delta P_s \equiv \int d^3 \V{r} \left[\int d^3
  \V{v} \left(\frac{T_{0s}\delta f_s^2}{2F_{0s}}\right)  - \frac{1}{2}n_{0s}m_s|\delta
\V{u_s}|^2\right].
\label{eq:e_nt}
\end{equation}
The \emph{turbulent energy} is defined as the sum of the
electromagnetic field and the bulk flow kinetic energies
\citep{Howes:2015b,Li:2016,Howes:2018a},
\begin{equation}
  E^{(turb)}\equiv \int d^3 \V{r} \left[\frac{|\delta \V{B}|^2+|\delta \V{E}|^2}{8\pi}     + \sum_s
    \frac{1}{2}n_{0s}m_s|\delta \V{u_s}|^2 \right].
 \label{eq:e_turb}
\end{equation}
Therefore the total fluctuating energy is simply the sum of the
turbulent energy and species non-thermal energies, $\delta
W=E^{(turb)} + E^{(nt)}_i + E^{(nt)}_e$.

In \figref{fig:ft3_heat}, we present area plots of the components of
the energy in the simulation as a function of normalized time
$t/\tau_0$, where $\tau_0$ is the period of \Alfven waves at the
domain scale.  Note that collisions in \T{AstroGK}, as well as in real
plasma systems, convert non-thermal to thermal energy, representing
irreversible plasma heating with an associated increase of
entropy. The energy lost from $\delta W$ by collisions is tracked by
\T{AstroGK} and represents thermal heating of the plasma species, but
this energy is not fed back into the code to evolve the equilibrium
thermal temperature, $T_{0s}$
\citep{Howes:2006,Numata:2010,Li:2016}.  We account for the energy
lost from $\delta W$ in the \T{AstroGK} simulation to collisional
plasma heating by accumulating the thermalized energy in each species
over time, $E^{(coll)}_s (t) = \int_0^t dt' Q_s(t')$, where
\T{AstroGK} computes the collisional heating rate per unit volume for
ions $Q_i$ and electrons $Q_e$.

In \figref{fig:ft3_heat}(a), we plot the evolution of the energy
budget over the course of the simulation, showing that turbulent
energy $E^{(turb)}$, which dominates at the beginning of the
simulation, is largely converted to electron heat $E^{(coll)}_e$ and
ion heat $E^{(coll)}_i$ and by the end of the simulation, with about
70\% of the initial total fluctuating energy $\delta W_0$ lost by
$t/\tau_0 =3$.  Approximately one third of the energy dissipated is
channeled to ions and two thirds to electrons. This is in close
agreement with a recent gyrokinetic turbulent simulation using a
realistic mass ratio of $m_i/m_e=1836$ in which about 70\% of the
total dissipated energy is channeled through electrons
\citep{Told:2015}.  Total energy $E$ is conserved over the course of the
simulation to less than 1\%.

Another view of the energy budget, based on \eqref{eq:deltaW_GK}, is
presented in \figref{fig:ft3_heat}(b), where we plot the perpendicular
magnetic field energy $E_{B_\perp}$ (green) and parallel magnetic
field energy $E_{B_\parallel}$ (cyan) along with the total fluctuating
kinetic energy of the ions and electrons $E^{(\delta f)}_s$. An
interesting feature of the evolution of the energy budget, seen in
both panels (a) and (b), is that although the electrons gain the bulk
of the thermal energy, the total fluctuating kinetic energy
$E^{(\delta f)}_e$ and total non-thermal energy $ E^{(nt)}_e$ in the
electrons remains small. Thus, electrons do not contain much of the
non-thermal energy at any given time, and are an effective conduit for
channeling the turbulent energy into electron thermal energy.

The field-particle correlation technique diagnoses the collisionless,
reversible transfer of energy between the electromagnetic fields and
the particles. This first step of the two-step process in converting
turbulent energy into particle thermal energy is not easily discerned
in the energy budget plots shown in \figref{fig:ft3_heat}. To more
directly evaluate this rate of energy transfer, we compute the net
rate of energy transfer due to field-particle interactions using 
\begin{equation}
  \dot{E}^{(fp)}_s  =  \dot{E}^{(nt)}_s + Q_s,
  \label{eq:edotnt}
\end{equation}
where $\dot{E}^{(nt)}_s$ is the rate of change of non-thermal species
energy and $Q_s$ is the collisional heating of the species. To check
the energy conservation in the simulation, we can plot
\begin{equation}
  -\dot{E}^{(turb)} = \dot{E}^{(fp)}_i + \dot{E}^{(fp)}_e,
  \label{eq:balance}
\end{equation}
because the rate of change of turbulent energy must be the sum of the
collisionless field-particle energy transfer for each species.

In \figref{fig:ft3_dedt}, we plot these energy transfer rates,
normalized to the characteristic heating rate per unit volume,
$Q_0=(n_{0i} T_{0i}
v_{ti}/L_\parallel)(\pi/8)(L_\perp/L_\parallel)^2$, for (a) ions (red)
and (b) electrons (blue), as well as (c) the total energy transfer
rate balance given by \eqref{eq:balance}. Several features of these
energy transfer rate curves are notable.  First, with the exception of
a brief large energy transfer to ions within the first $0.2 \tau_0$,
the field-particle energy transfer rate to electrons is about twice of
that to ions on average. Second, the time lag between field-particle energy
transfer to electrons and the collisional thermalization of that
energy is very short, a delay of abot $0.1\tau_0$.  Thus, energy
transferred to electron non-thermal energy is rapidly converted to
electron heat by collisions. In contrast, for the ions, the time lag
between the peaks of the field-particle energization of the ions and the
collisional thermalization of that energy is much longer, about $1.0
\tau_0$.  This longer time lag for ions is consistent with
 the need for the ion entropy cascade
\citep{Schekochihin:2009,Tatsuno:2009,Cerri:2018} to transfer non-thermal energy
to sufficient small scales in velocity space before collisions can
effectively thermalize that energy. Finally, (c) the
sum of the field-particle energy transfer to ions and electrons is
in quantitative agreement with the loss of turbulent energy.


\begin{figure} \centering

\hbox{{(a)\hspace{-0.4em}} \resizebox{2.8in}{!}{\includegraphics[scale=0.6]{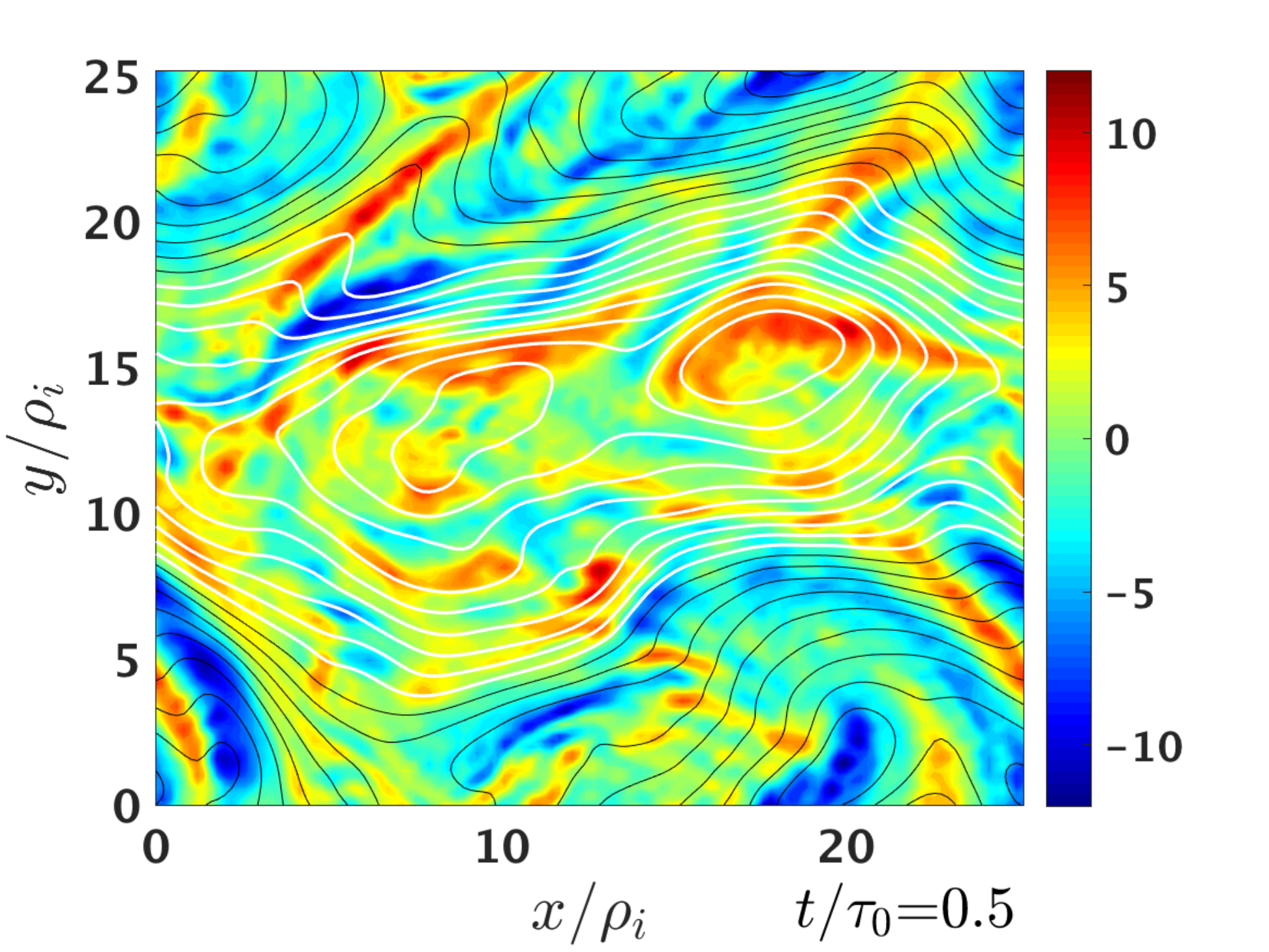}} \hfill 
{(b)\hspace{-0.4em}} \hfill \resizebox{2.8in}{!}{\includegraphics[scale=0.6]{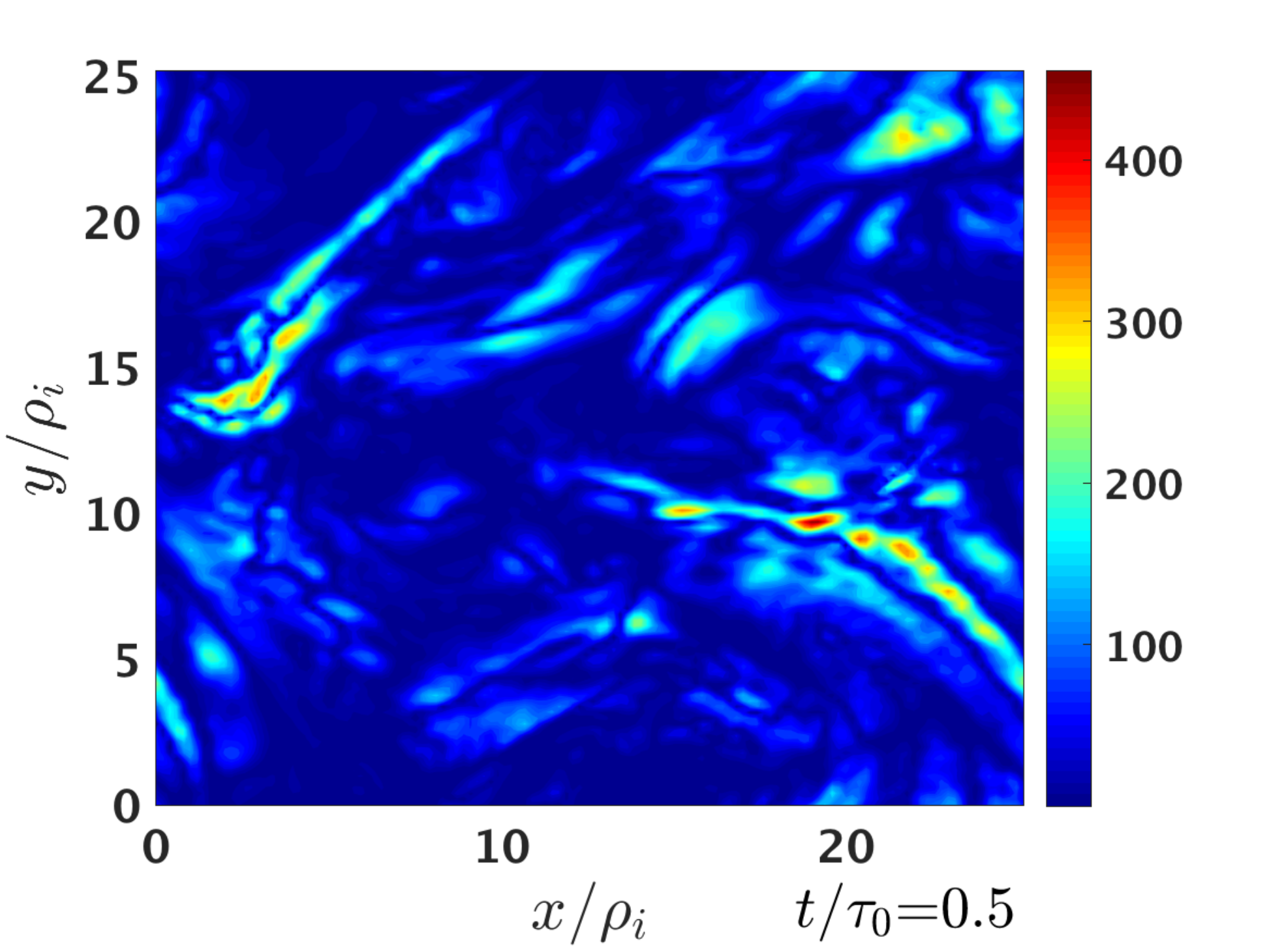}} \hfill
 }
{\hfill\hbox{{(c)\hspace{-0.4em}} \hfill \resizebox{2.8in}{!}{\includegraphics[scale=0.6]{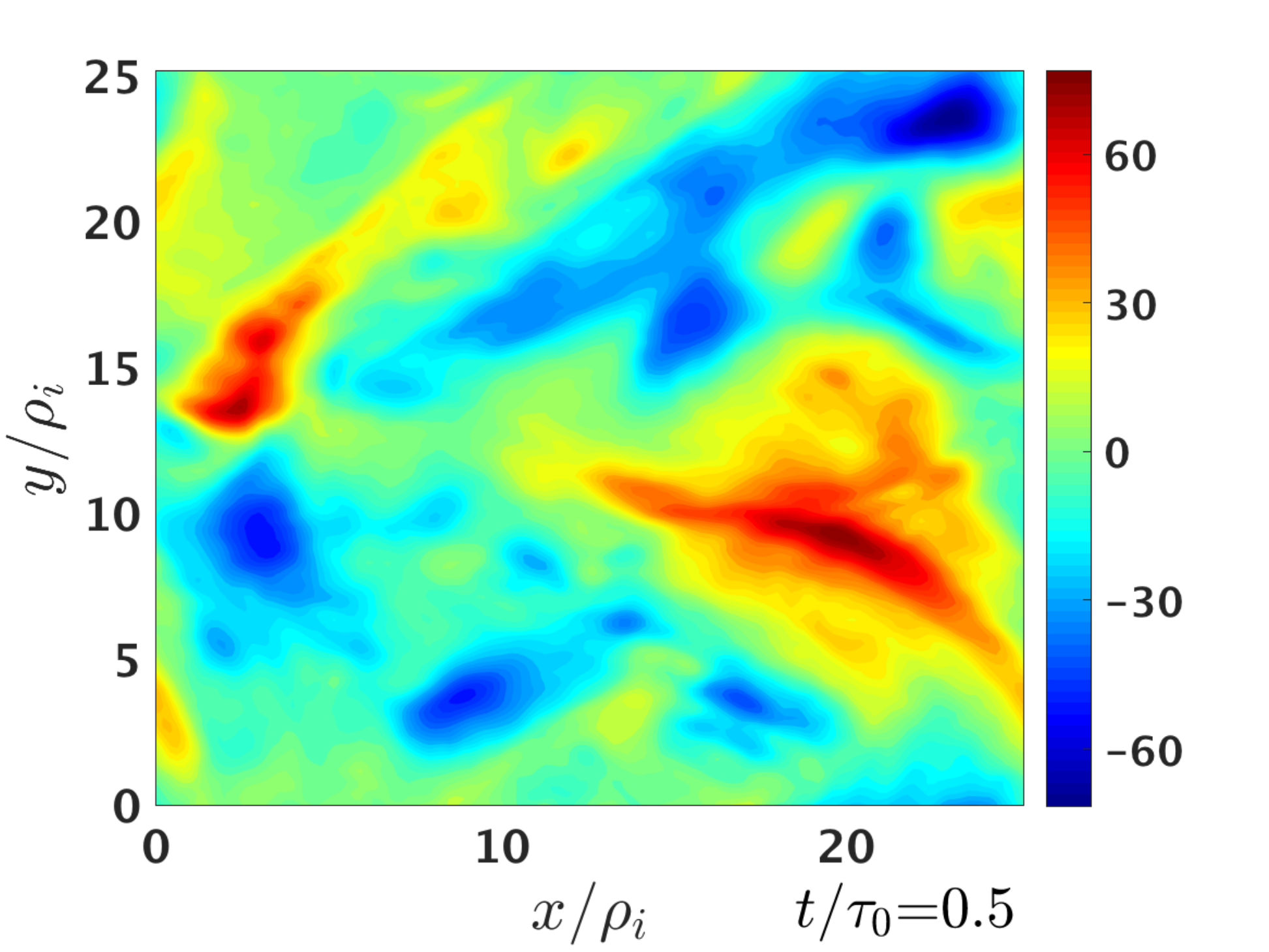}} 
 }\hfill}

 \caption{ \label{fig:ft3_1_otv} Spatial profile of (a) $J_\parallel$ (color) and
  $A_\parallel$ (contour), (b) $J_\parallel E_\parallel$ and (c) $E_\parallel$ on the $z=0$ plane at $t/\tau_0=0.5$. Contours represent positive (white) and negative (black) values of $A_\parallel$. }
\end{figure}


In \figref{fig:ft3_1_otv}, we plot the spatial profiles on the $z=0$
plane at $t/\tau_0=0.5$ of (a) the parallel current $J_\parallel$
(color) and vector potential $A_\parallel$ (contour), 
(b) $J_\parallel E_\parallel$ and (c) $E_\parallel$. The topology of the initial
double-vortex pattern, representative of the Orzag-Tang Vortex, is
traceable by the central white contours. Multiscale features
in $J_\parallel$, including self-consistently generated current
sheets, are seen over the entire $(x,y)$ plane, representing a
turbulent state containing a spectrum of nonlinearly generated
modes. $J_\parallel E_\parallel$ represents the net energy density
transfer rate mediated by the parallel electric field (the first term
on the right hand side of \eqref{eq:dW_dt} when integrated over
velocity). In (b), strong energy transfer density (red) occurs in
sub-ion or ion scale regions. Two particular strong $J_\parallel E_\parallel$ 
regions center at $(x,y)\sim$(3,14) and (20,10). They
co-locate with thin sheet-like currents and share similar shapes with
them, but not necessarily with the strongest $J_\parallel$
regions. They, however, are seen at the strongest $E_\parallel$ in (c), which
also occurs in localized ion scale regions. Otherwise, $E_\parallel$ has 
generally broad features. 

\section{Field-Particle Correlations}\label{sec:fpc_res}
Here we present results on identifying the collisionless energy transfer in the simulation using field-particle correlations.

\subsection{Field-Particle Correlations in Gyrotropic Velocity Space}\label{sec:gyro}
To identify the regions of gyrotropic velocity space
$(v_\parallel,v_\perp)$ where particles play a role in the
collisionless transfer of energy density from the parallel electric
field to the plasma ions and electrons, we employ the field-particle
correlation technique to compute
$C_{E_\parallel}(\V{k},v_\parallel,v_\perp,t,\tau)$, given by
\eqref{eq:corr_par_g_k}.




\begin{figure} \centering
\hbox{{(a)\hspace{-1.3em}} \hfill
  \resizebox{2.6in}{!}{\includegraphics*[scale=0.4,trim=1.5cm 0.5cm 0cm 0.5cm, clip=true]{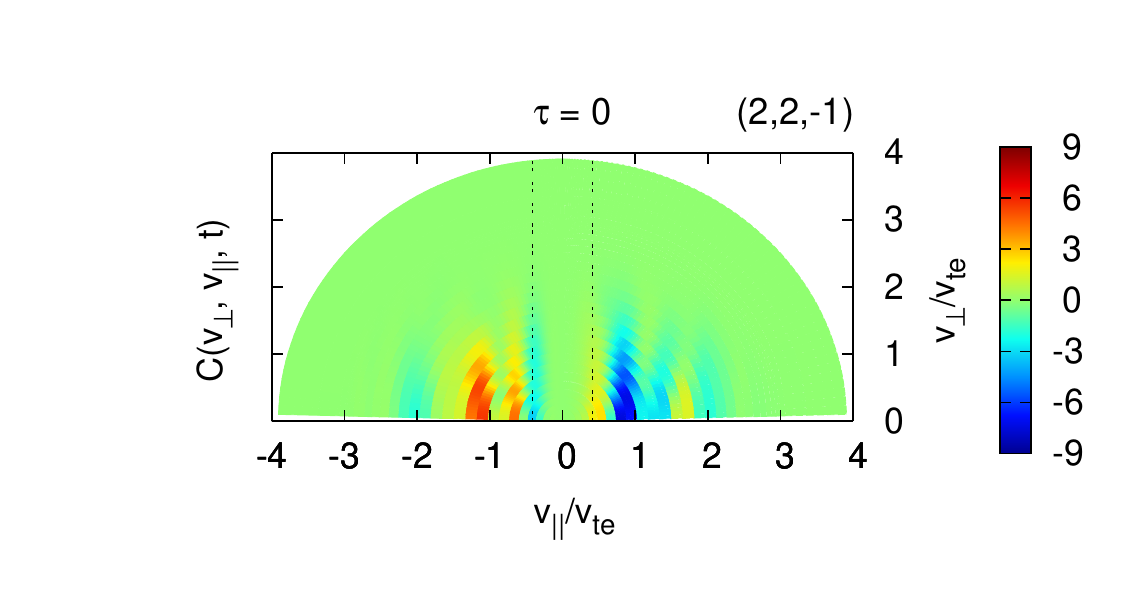}}
  \hfill \hspace{-1.em} {(b)\hspace{-1.3em}} \hfill
  \resizebox{2.6in}{!}{\includegraphics*[scale=0.4,trim=1.5cm 0.5cm 0cm 0.5cm, clip=true]{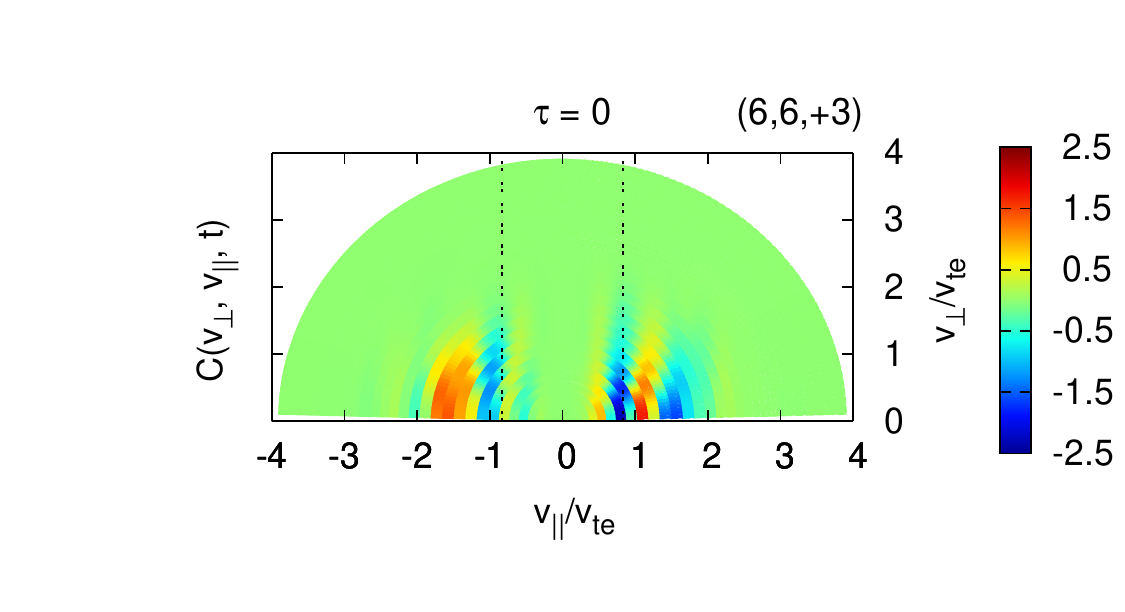}}
  \hfill }

\hbox{{(c)\hspace{-1.3em}} \hfill
  \resizebox{2.6in}{!}{\includegraphics*[scale=0.4,trim=1.5cm 0.5cm 0cm 0.5cm, clip=true]{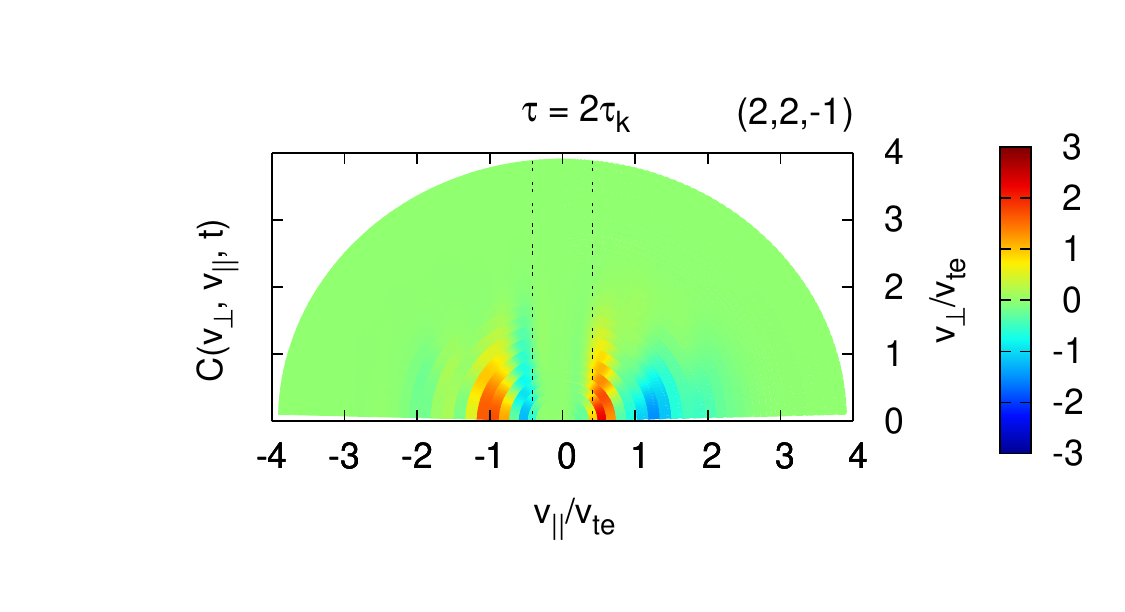}}
  \hfill \hspace{-1.em} {(d)\hspace{-1.3em}} \hfill
  \resizebox{2.6in}{!}{\includegraphics*[scale=0.4,trim=1.5cm 0.5cm 0cm 0.5cm, clip=true]{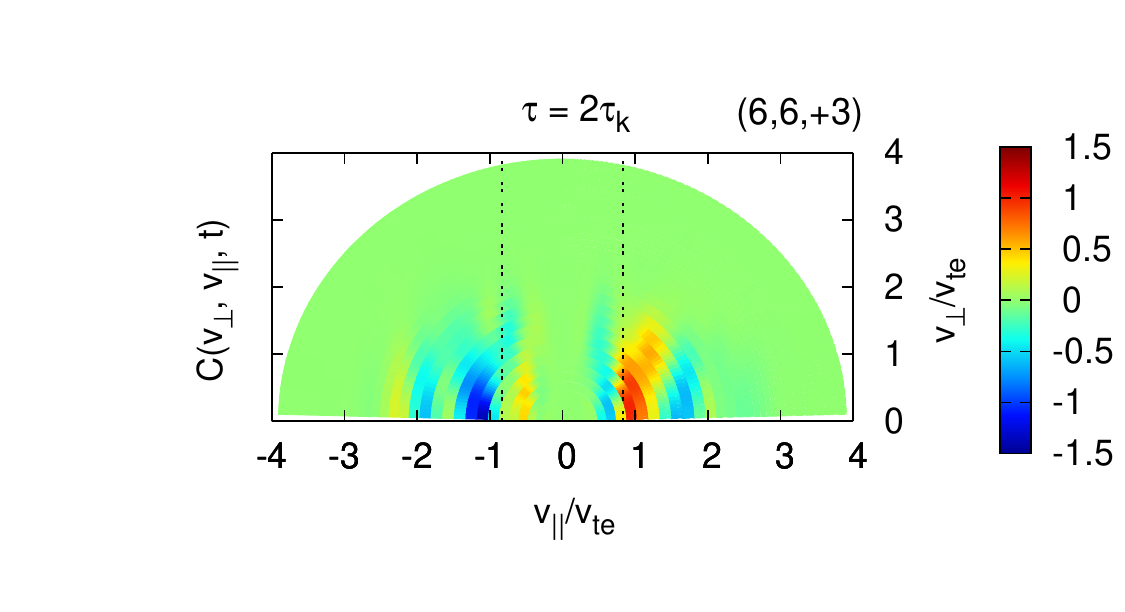}}
  \hfill }
 \caption{ \label{fig:1_22m1_gyro} Electron gyrotropic correlation
   $C_{E_\parallel}(v_\parallel,v_\perp,t)$ for the $(2,2,-1)$ Fourier mode using 
   (a) $\tau = 0$ and (c) $\tau = 2\tau_k$, plotted at 
   $t_c/\tau_0=0.86$, corresponding to $t/\tau_k=0.81$ 
   for $\tau = 2\tau_k$. Also plotted are
   $C_{E_\parallel}(v_\parallel,v_\perp,t)$ for the $(6,6,+3)$ Fourier mode 
   using (b) $\tau = 0$ and (d) $\tau = 2\tau_k$ 
   at $t_c/\tau_0=0.34$, corresponding to $t/\tau_k=0.42$ 
   for $\tau = 2\tau_k$. Dashed lines
   denote Landau resonant velocities: $v_{p \parallel}/v_{te}$ = $\pm$0.42
   for the $(2,2,-1)$ Fourier mode and $v_{p \parallel}/v_{te}$ = $\pm$0.84
   for the $(6,6,+3)$ Fourier mode. Arbitrary units are used in the colorbars 
   while the relative amplitudes between the $\tau = 0$ and $\tau = 2\tau_k$
   correlations are preserved for each Fourier mode. }
\end{figure}

\subsubsection{Electron Gyrotropic Correlations}

Here we explore the net energy transfer rate to the electrons by the
parallel electric field as a function of particle velocity in
gyrotropic velocity space $(v_\parallel,v_\perp)$ for particular Fourier modes,
each denoted by their normalized wavevector
$(k_x\rho_i,k_y\rho_i,k_zL_z/2 \pi)$. In the top row of
\figref{fig:1_22m1_gyro}, we plot the instantaneous correlation
$C_{E_\parallel}(\V{k},v_\parallel,v_\perp,t,\tau=0)$ for modes (a)
$(2,2,-1)$ and (b) $(6,6,+3)$. The instantaneous energy transfer is
spread over a wide range of $v_\parallel$ relative to $v_{p \parallel}$ 
of each Fourier mode of the kinetic \Alfven
wave, given by (a) $\omega/(k_\parallel v_{te})= \pm 0.42$ and (b)
$\omega/(k_\parallel v_{te})=\pm 0.84$, indicated by the vertical dashed
black lines. This shows that particles over a relatively broad range 
of $v_\parallel$ participate in energy exchange with the parallel electric 
field.

In order to eliminate the often large contribution to the instantaneous
energy transfer given by the oscillating energy transfer associated with
undamped wave motions (which is measured in a kinetic \Alfven wave \citep{Gershman:2017}),
 the field-particle correlation technique
performs a time-average of the correlation product 
(in the unnormalized correlation) over a suitably long time
period \citep{Klein:2016,Howes:2017}.  This effectively removes
the oscillating component of the energy transfer, isolating the often 
smaller net secular energy transfer from fields to particles. In both of
these cases, we choose a correlation interval $\tau= 2 \tau_k$ where
$\tau_k$ is the linear wave period of a kinetic \Alfven wave specified 
by the normalized wavevector, given by the final column in Table~\ref{tab:kspace}. The correlation interval is chosen such that the time evolution of
the parallel reduced correlation qualitatively converges, which 
corresponds to $\tau\geq\tau_k$ (see Appendix A).
 The resulted time-averaged field-particle correlation are plotted
in \figref{fig:1_22m1_gyro} for Fourier modes (c) $(2,2,-1)$ and (d)
$(6,6,+3)$.

For the time-averaged correlations, time $t$ is defined at the beginning of the correlation interval $\tau$. Another convenient way to specify time is at the center of the correlation interval, giving a centered time of $t_c=t+\tau/2$. Both $t$ and $t_c$ are given in \figref{fig:1_22m1_gyro} for reference.


Several key features of the net energy transfer over the correlation
interval $\tau$ are observed in \figref{fig:1_22m1_gyro}(c) and (d).
First, the region in velocity space of largest net transfer of energy
between $E_\parallel$ and the electrons is closely connected to $v_{p \parallel}$
. This contrasts with the 
broader spread over $v_\parallel$ of instantaneous energy transfer when
taking $\tau=0$ in panels (a) and (b). 
The resulting velocity-space characteristic of the energy transfer 
is consistent
with the \emph{velocity-space signature} of Landau damping of the turbulent fluctuations, as found in previous studies \citep{KleinHT:2017,Howes:2018a,Klein:2016,Howes:2017,Chen:2019a}. The contrast in the energy transfer signals
between the instantaneous and long-time-averaged correlations shows
that while electrons over a broad range of parallel velocities participate in 
instantaneous energy transfer, the net secular 
energy transfer is largely contributed
by "near-resonance" electrons that are closely connected to $v_{p \parallel}$.
This reflects the resonant nature of secular net energy transfer 
in Landau damping of the turbulent fluctuations. 
Second, the variation in the energy transfer is largely a function of
$v_\parallel$, with weak $v_\perp$ dependence other than the monotonic
drop off as $v_\perp$ increases. For the $(6,6,+3)$ mode in (d) which
has $k_\perp\rho_e\approx$ 1.7 $>$ 1, however, there is weak $v_\perp$
dependence near $v_\parallel/v_{te}\sim 1$, which lasts very briefly,
for $\sim$0.2$\tau_k$. This $v_\perp$ dependence in the energy transfer reveals structuring of the distribution in $v_\perp$ as $E_\parallel$
does not depend on velocity. This qualitative feature is consistent with
the expectation of the electron entropy cascade
\citep{Schekochihin:2009,Tatsuno:2009,Schoeffler:2014}, which predicts some
structuring of the distribution in $v_\perp$ due to nonlinear phase
mixing for $k_\perp\rho_e >$1.




\begin{figure} \centering
\hbox{{(a)\hspace{-1.3em}} \hfill
  \resizebox{2.6in}{!}{\includegraphics*[scale=0.4,trim=1.5cm 0.5cm 0cm 0.5cm, clip=true]{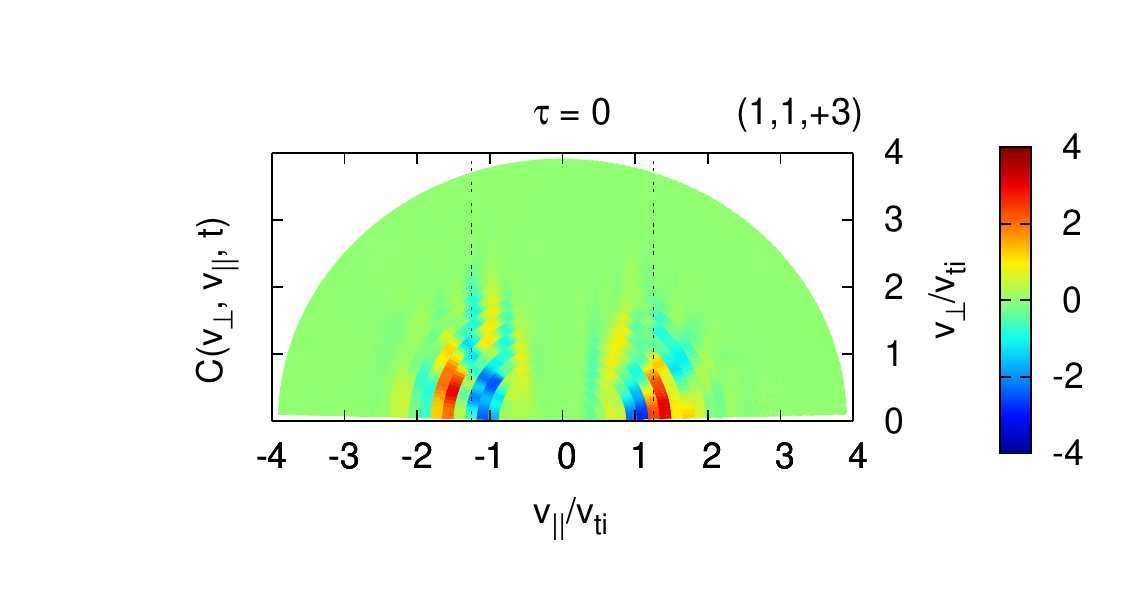}}
  \hfill \hspace{-1.em} {(b)\hspace{-1.3em}} \hfill
  \resizebox{2.6in}{!}{\includegraphics*[scale=0.4,trim=1.5cm 0.5cm 0cm 0.5cm, clip=true]{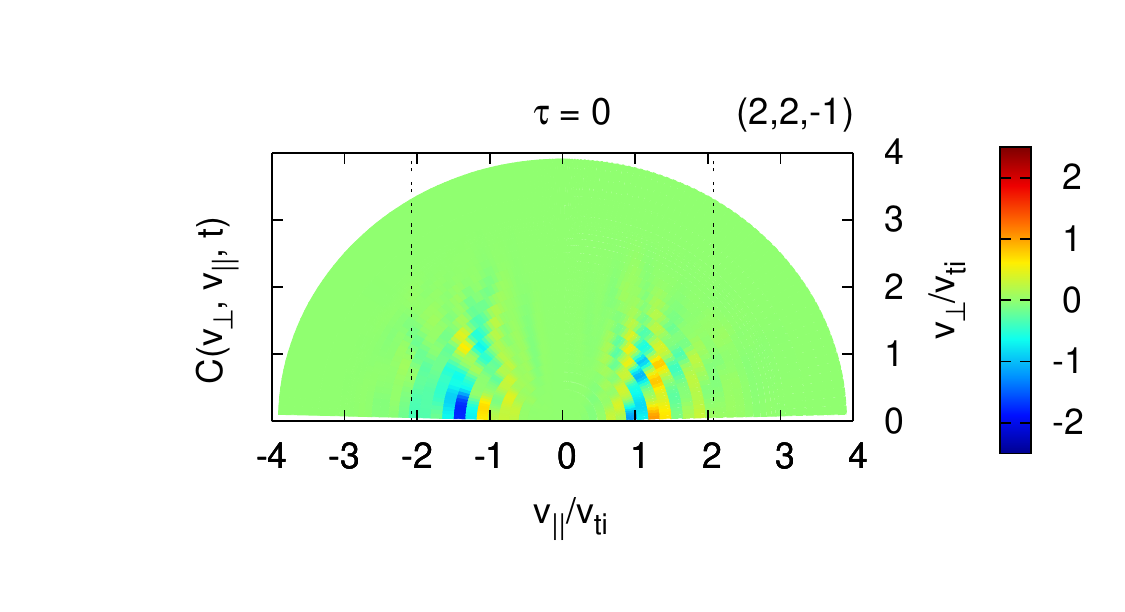}}
  \hfill }

\hbox{{(c)\hspace{-1.3em}} \hfill
  \resizebox{2.6in}{!}{\includegraphics*[scale=0.4,trim=1.5cm 0.5cm 0cm 0.5cm, clip=true]{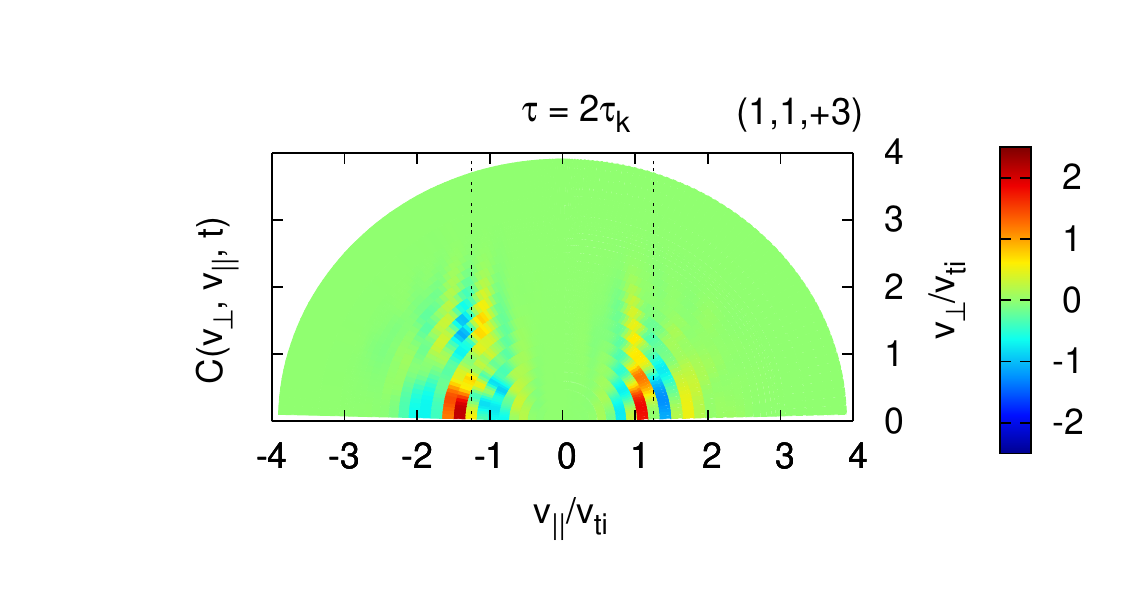}}
  \hfill \hspace{-1.em} {(d)\hspace{-1.3em}} \hfill
  \resizebox{2.6in}{!}{\includegraphics*[scale=0.4,trim=1.5cm 0.5cm 0cm 0.5cm, clip=true]{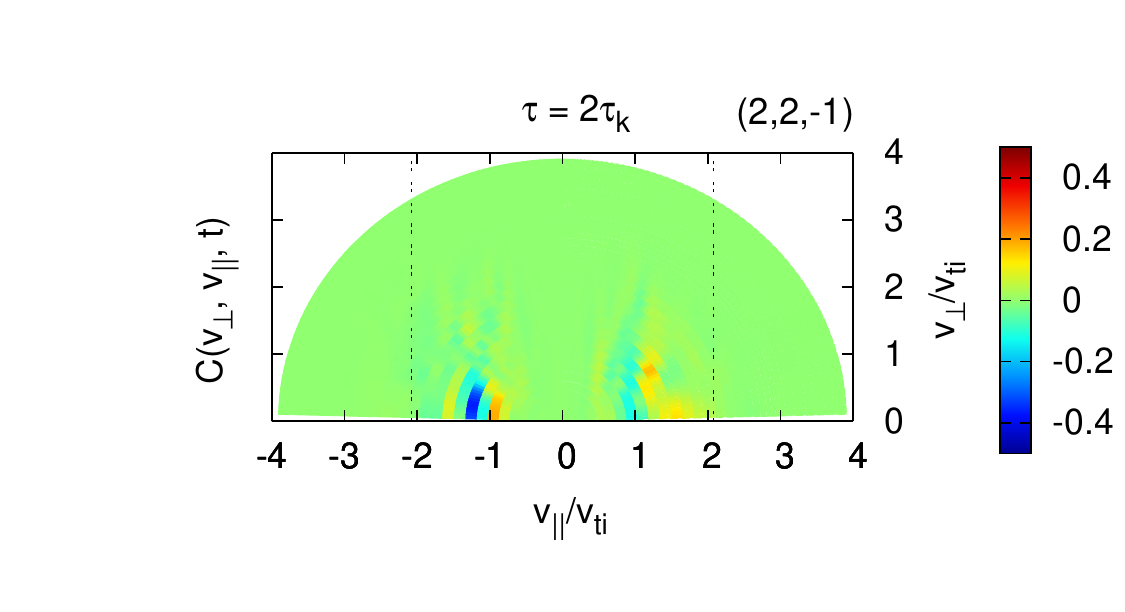}}
  \hfill }

 \caption{ \label{fig:1_11m1_gyro_i} Ion gyrotropic correlation
   $C_{E_\parallel}(v_\parallel,v_\perp,t)$ for the $(1,1,+3)$ Fourier mode using
   (a) $\tau = 0$ and (c) $\tau = 2\tau_k$, plotted at $t_c/\tau_0=1.2$,
   corresponding to $t/\tau_k=0.48$ for $\tau = 2\tau_k$. 
   (c) shows observable weak $v_\perp$ variations at $v_\parallel<0$ in addition
   to more prominent $v_\parallel$ dependence.
   Also plotted are $C_{E_\parallel}(v_\parallel,v_\perp,t)$ for the $(2,2,-1)$ Fourier mode 
   using (b) $\tau = 0$ and (d) $\tau = 2\tau_k$ at $t_c/\tau_0=0.90$,
   corresponding to $t/\tau_k=0.90$ for $\tau = 2\tau_k$.
   Dashed lines denote Landau resonant velocities: $v_{p \parallel}/v_{ti}$
   = $\pm$1.3 and $\pm$2.1 for the $(1,1,+3)$ and $(2,2,-1)$ Fourier modes,
   respectively. The same format is used as \figref{fig:1_22m1_gyro}.
 }  
\end{figure}

\subsubsection{Ion Gyrotropic Correlations}

In \figref{fig:1_11m1_gyro_i}, we plot the instantaneous correlation
$C_{E_\parallel}(\V{k},v_\parallel,v_\perp,t,\tau=0)$ for ions for
modes (a)$(1,1,+3)$ and (b) $(2,2,-1)$. As with the instantaneous
correlations for the electrons in \figref{fig:1_22m1_gyro}, we see
that the energy transfer is not tightly constrained in $v_\parallel$
to the resonant parallel phase velocities for these modes ( (a)
$\omega/(k_\parallel v_{ti})= \pm 1.2$ and (b) $\omega/(k_\parallel
v_{ti})=\pm 2.1$ (vertical dashed black lines) ). For the $(2,2,-1)$
mode, having resonant parallel phase velocities much higher than
the ion thermal speed, there is only very weak energy transfer.

Computing the unnormalized correlation (time average) over the
correlation interval $\tau=2 \tau_k$ for each mode, we obtain the
ion energization rate in gyrotropic velocity space  
shown in \figref{fig:1_11m1_gyro_i} for Fourier modes (c) $(1,1,+3)$ and (d)
$(2,2,-1)$.  The ion energy transfer rate results for (c) $(1,1,+3)$
have similar characterstics to the findings for the electrons: (i) the
secular energy transfer rate is closely connected to the parallel
resonant velocity; and (ii) the energy transfer rate varies strongly
as a function of $v_\parallel$, again consistent with the
velocity-space signature expected of Landau damping.  One notable
difference is that there is significantly more variation of the energy
transfer rate in the $v_\perp$ direction than for the electrons. This
is qualitatively consistent with the action of the ion entropy cascade
\citep{Schekochihin:2009,Tatsuno:2009} which leads to structuring of
the perturbed distribution function in the $v_\perp$ direction
\citep{Howes:2008c}.

Another notable difference from the electron case is that the energy
transfer rates are significantly weaker, with nearly no resonant
energy transfer signatures near $v_{p \parallel}/v_{ti}$,
 for the (d) $(2,2,-1)$ Fourier mode.
 In fact, the
$(1,1)$ Fourier modes are the only $k_\perp$ value in the sampled
spectrum (Table \ref{tab:kspace}) displaying resonant signatures in
the resolved range of $v_\parallel/v_{ti}$ plotted. The $(1,1)$ Fourier
modes also sit near the peak of the ion collisionless damping rate
($\gamma_i$ in \figref{fig:1om_g}). At $k_\perp\rho_i>$ 2, $\gamma_i$
drops off rapidly. This is because $v_{p \parallel}/v_{ti}$ moves to the tail
of the ion distribution (at greater than twice the ion thermal speed)
where there are very few particles to interact with the waves. For
instance, the $(2,2)$ and $(3,4)$ Fourier modes have $v_{p \parallel}/v_{ti}$
= $\pm$2.1 and $\pm$3.3, respectively, which are in
the tail of the distribution. Indeed, resonant energy transfer
signatures are not observed for the $(2,2)$ to $(7,7)$ Fourier modes.

\subsection{Parallel Reduced Correlations}\label{sec:individual}
Because the variation of the rate of energy density transfer, shown in
Figs.~\ref{fig:1_22m1_gyro} and \ref{fig:1_11m1_gyro_i}, is largely
a function of $v_\parallel$ with little variation in $v_\perp$, we may
integrate over $v_\perp$ to obtain reduced parallel correlations
$C_{E_\parallel}(v_\parallel,t,\tau)$, given by
\eqref{eq:corr_par_reduced}. These reduced correlations can
conveniently be plotted, for a given correlation interval
$\tau$, as timestack plots of $C_{E_\parallel}(v_\parallel,t)$ as a
function of $v_\parallel$ and time $t$ to illustrate the evolution of
the field-particle energy transfer due to a single Fourier mode over
the course of the entire simulation.

\subsubsection{Electron Parallel Reduced Correlations }\label{sec:1_elec_k22}

\begin{figure} \centering
\hbox{{(a)\hspace{-1.3em}} \hfill \resizebox{2.6in}{!}{\includegraphics*{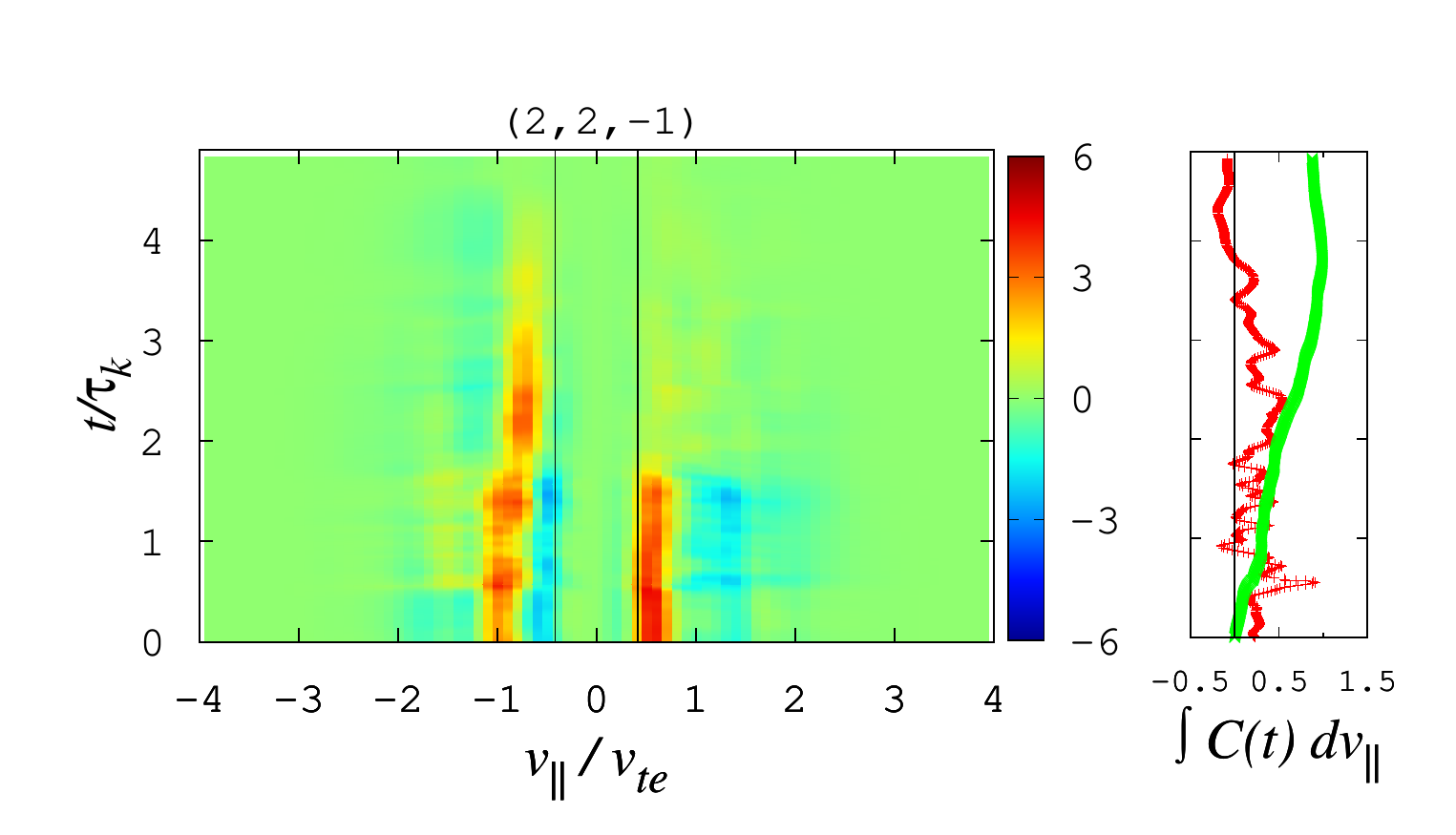}} \hfill  
{(b)\hspace{-1.3em}} \hfill \resizebox{2.6in}{!}{\includegraphics*{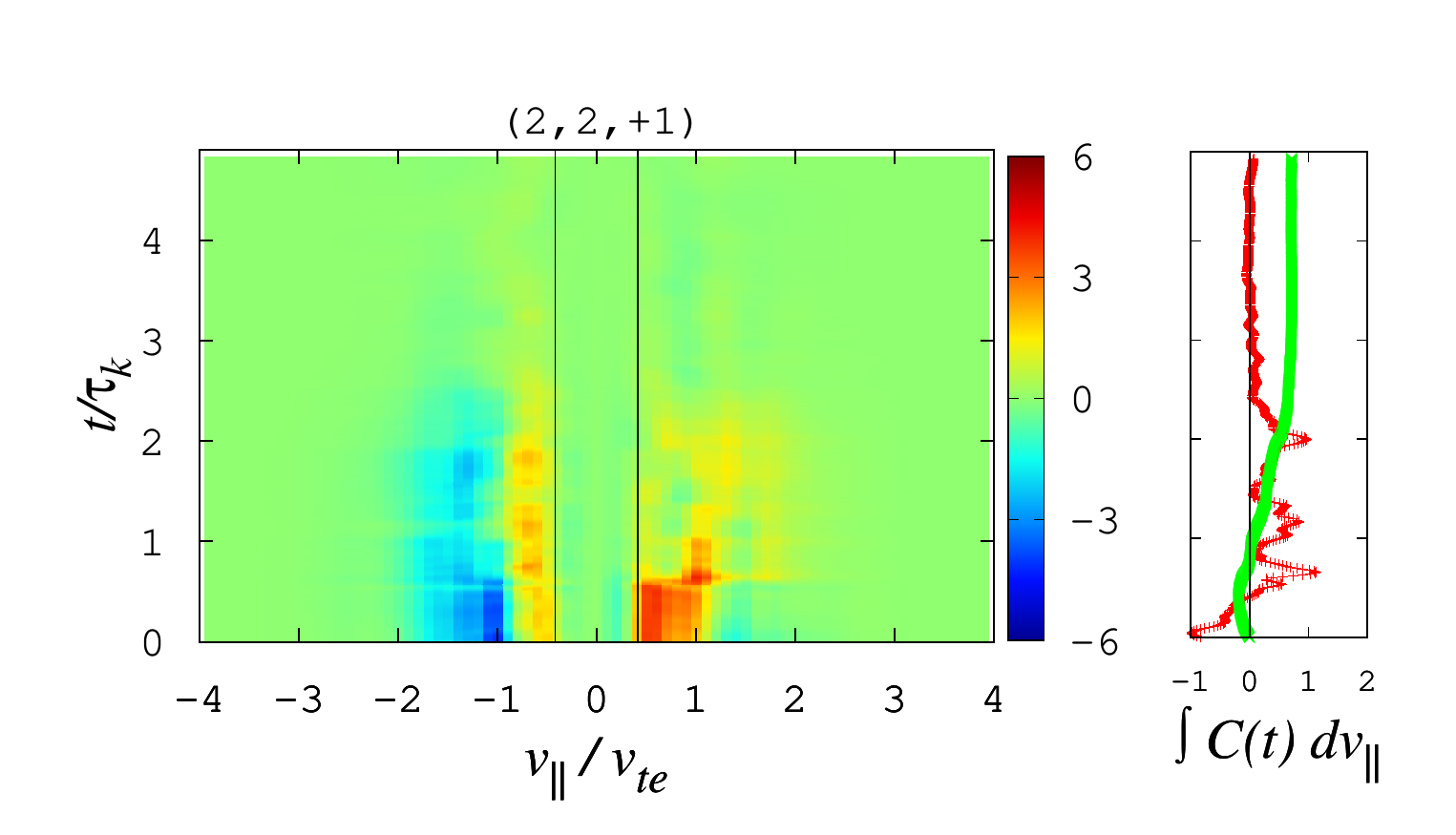}} \hfill  
  } 
\hbox{{(c)\hspace{-1.3em}} \hfill \resizebox{2.6in}{!}{\includegraphics*{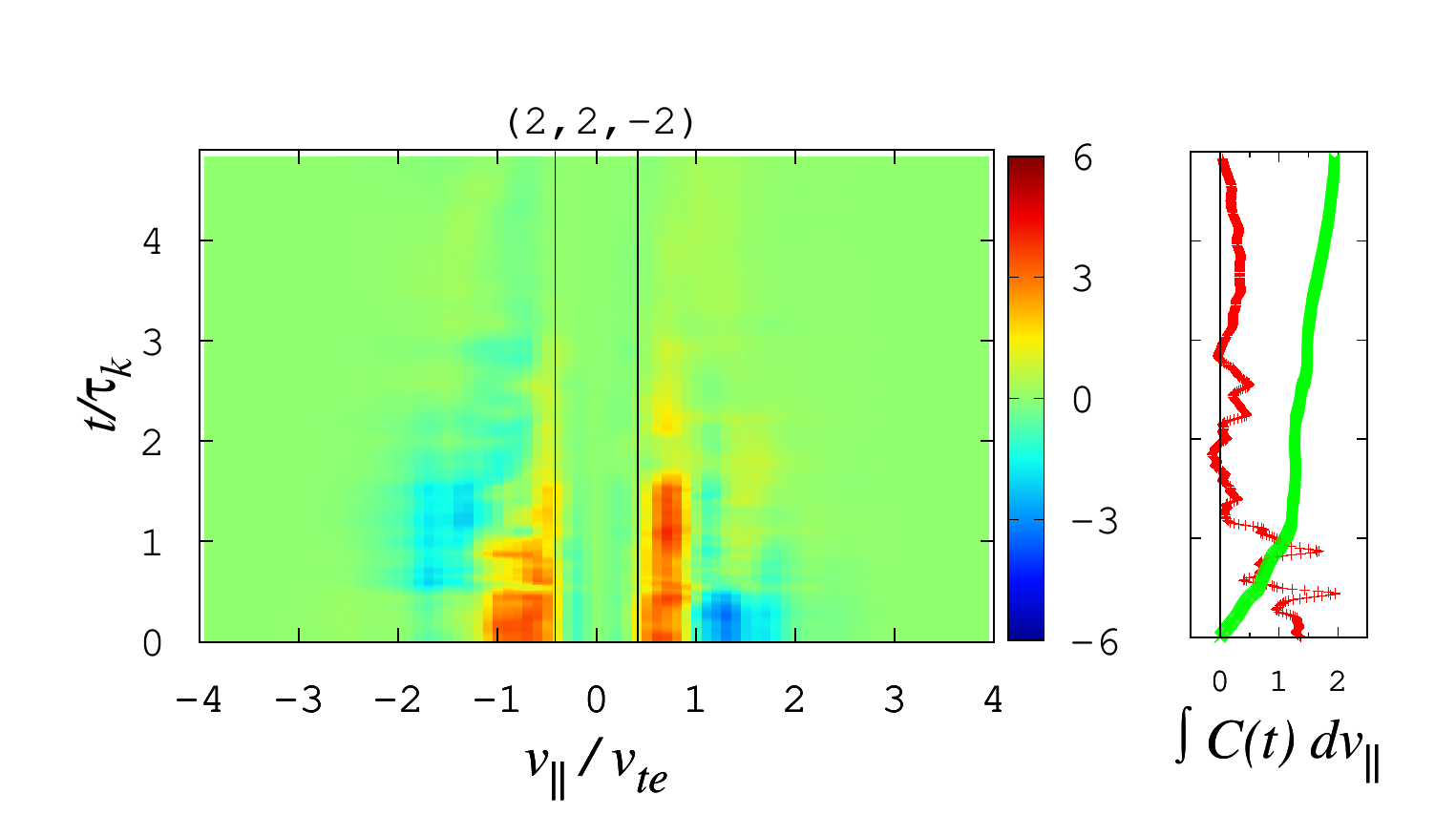}} \hfill  
{(d)\hspace{-1.3em}} \hfill \resizebox{2.6in}{!}{\includegraphics*{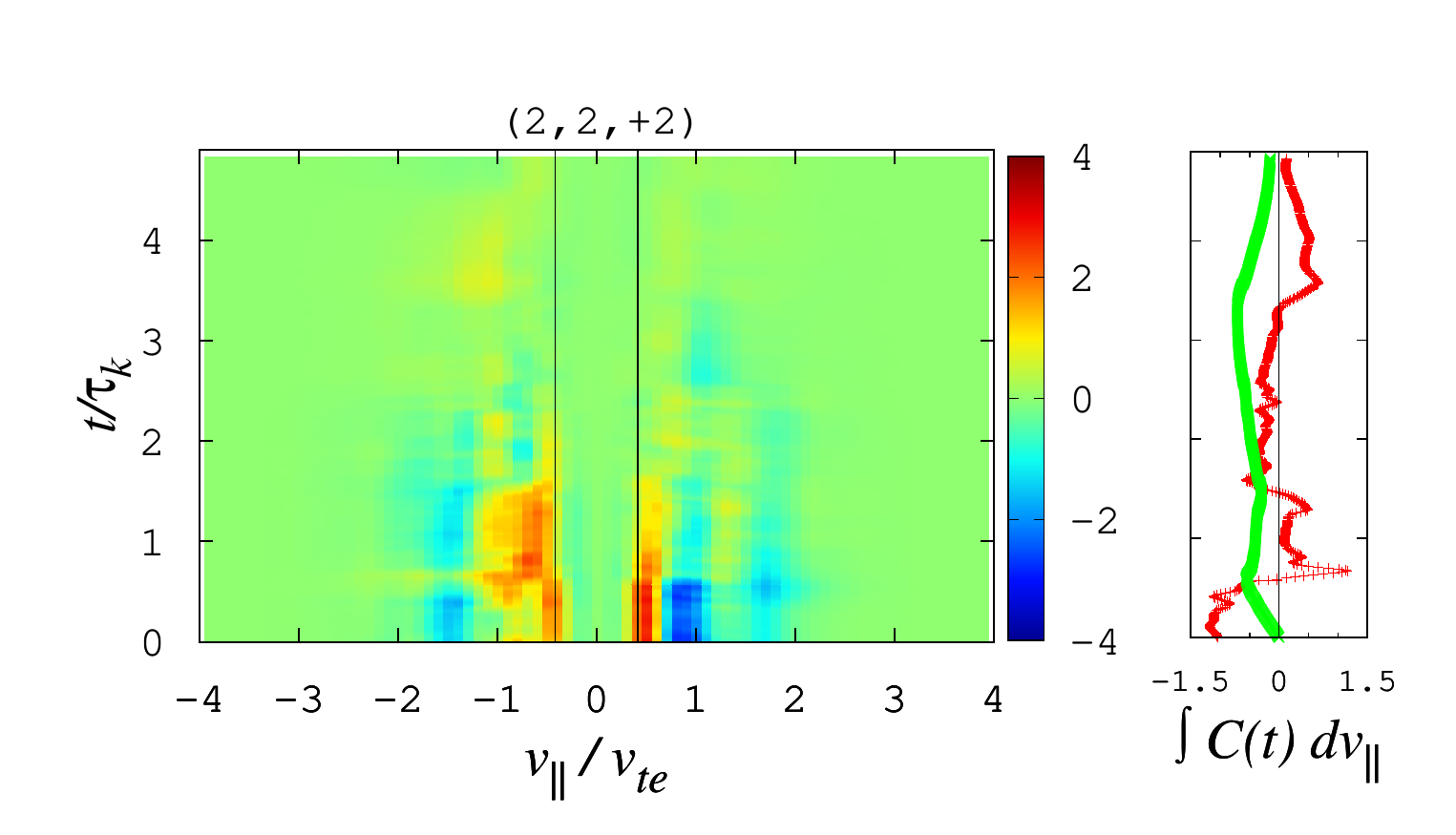}} \hfill  
  } 
\hbox{{(e)\hspace{-1.3em}} \hfill \resizebox{2.6in}{!}{\includegraphics*{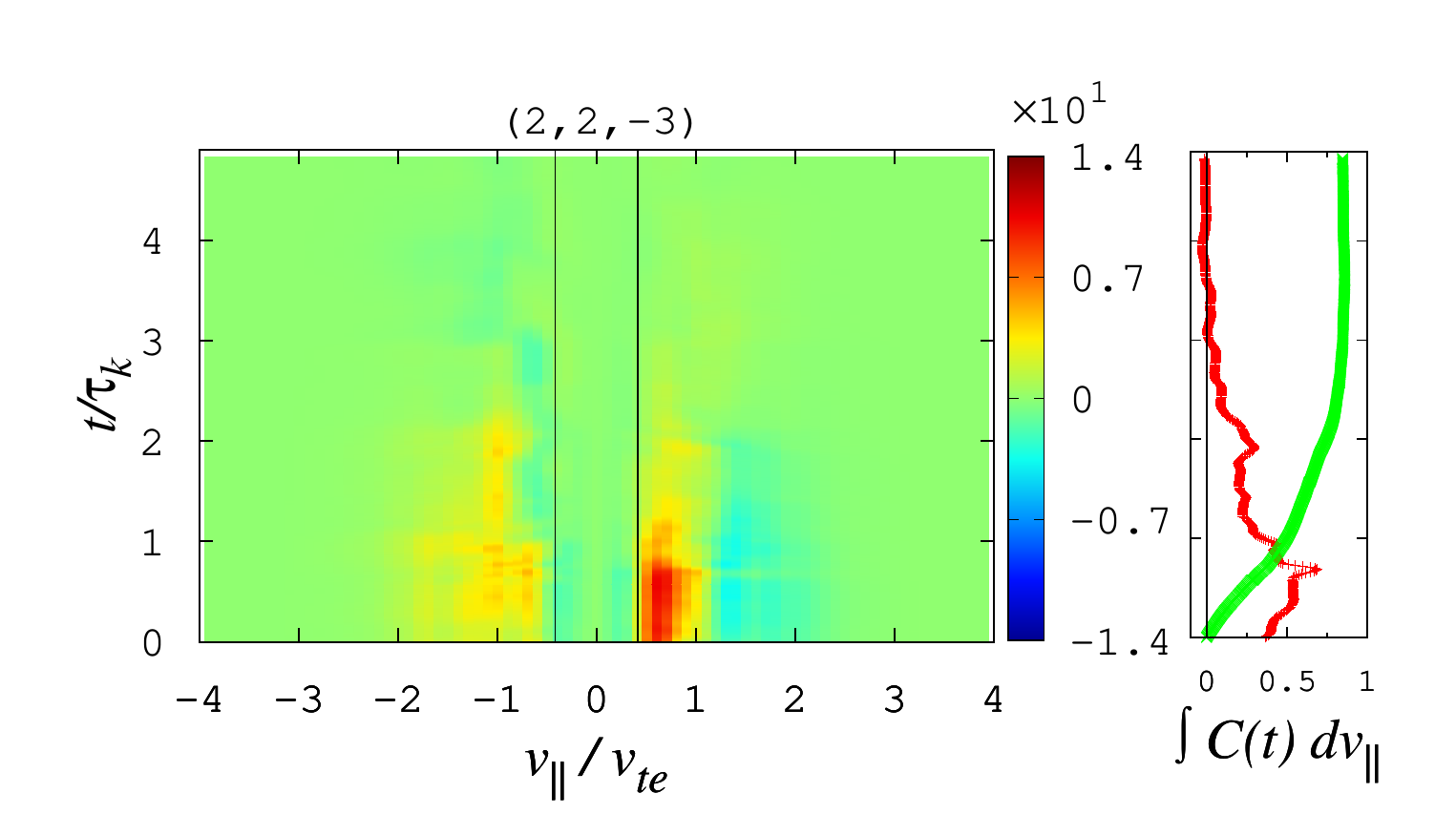}} \hfill  
{(f)\hspace{-1.3em}} \hfill \resizebox{2.6in}{!}{\includegraphics*{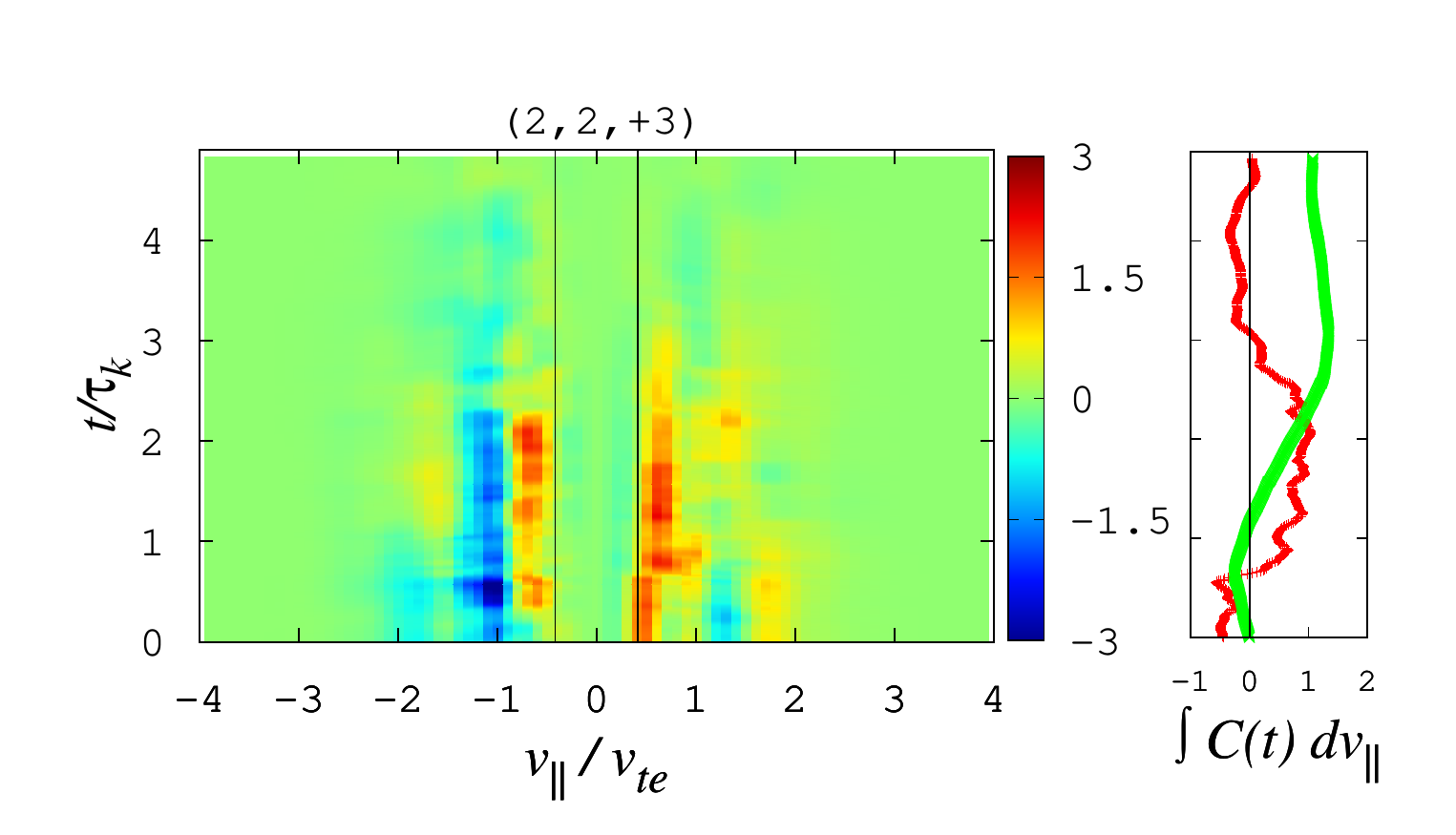}} \hfill  
  } 
\hbox{{(g)\hspace{-1.3em}} \hfill \resizebox{2.6in}{!}{\includegraphics*{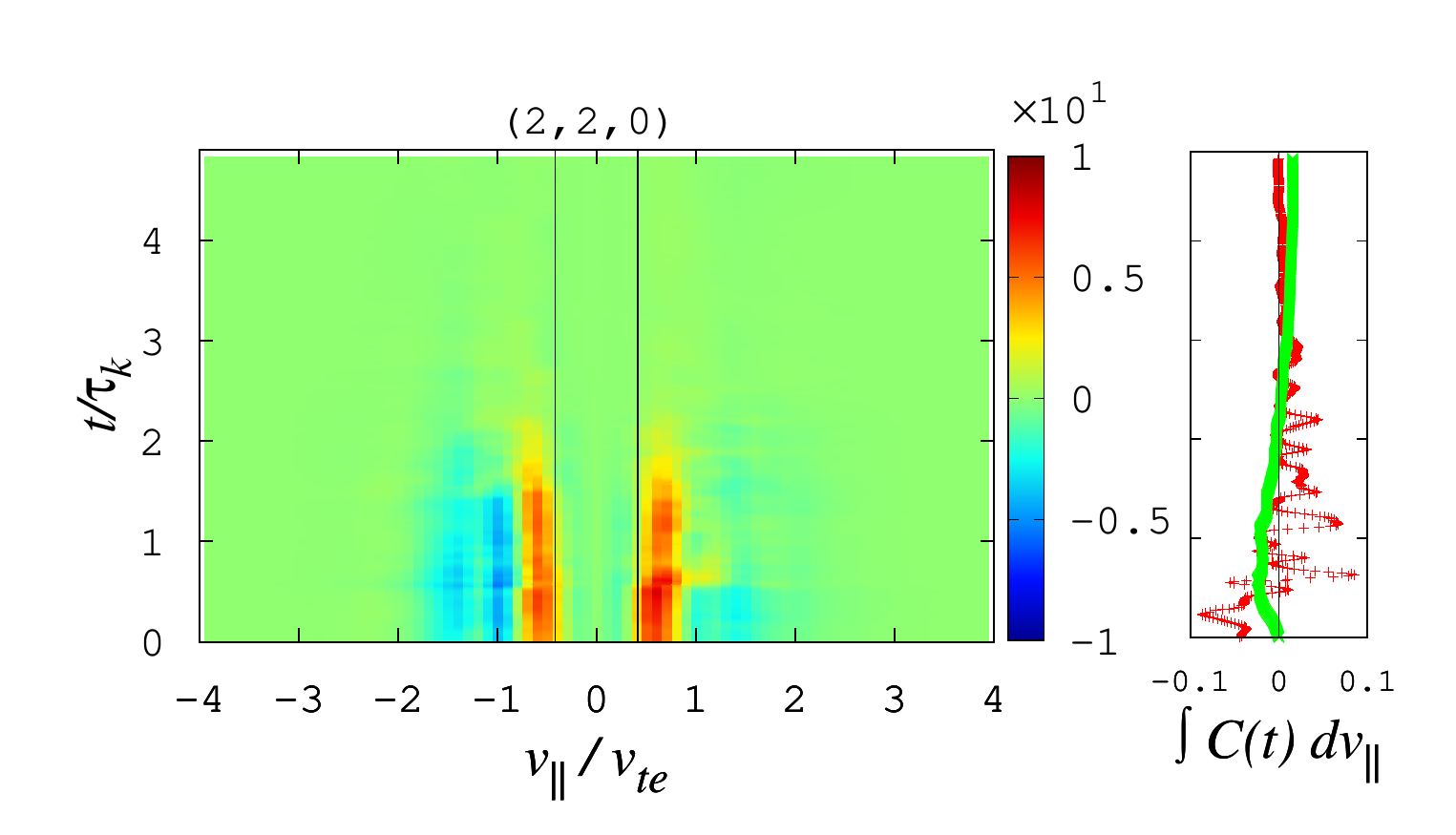}} \hfill  
{(h)\hspace{-1.3em}} \hfill \resizebox{2.8in}{!}{\includegraphics*[scale=1.,trim=0cm 0cm 0cm 0cm, clip=true]{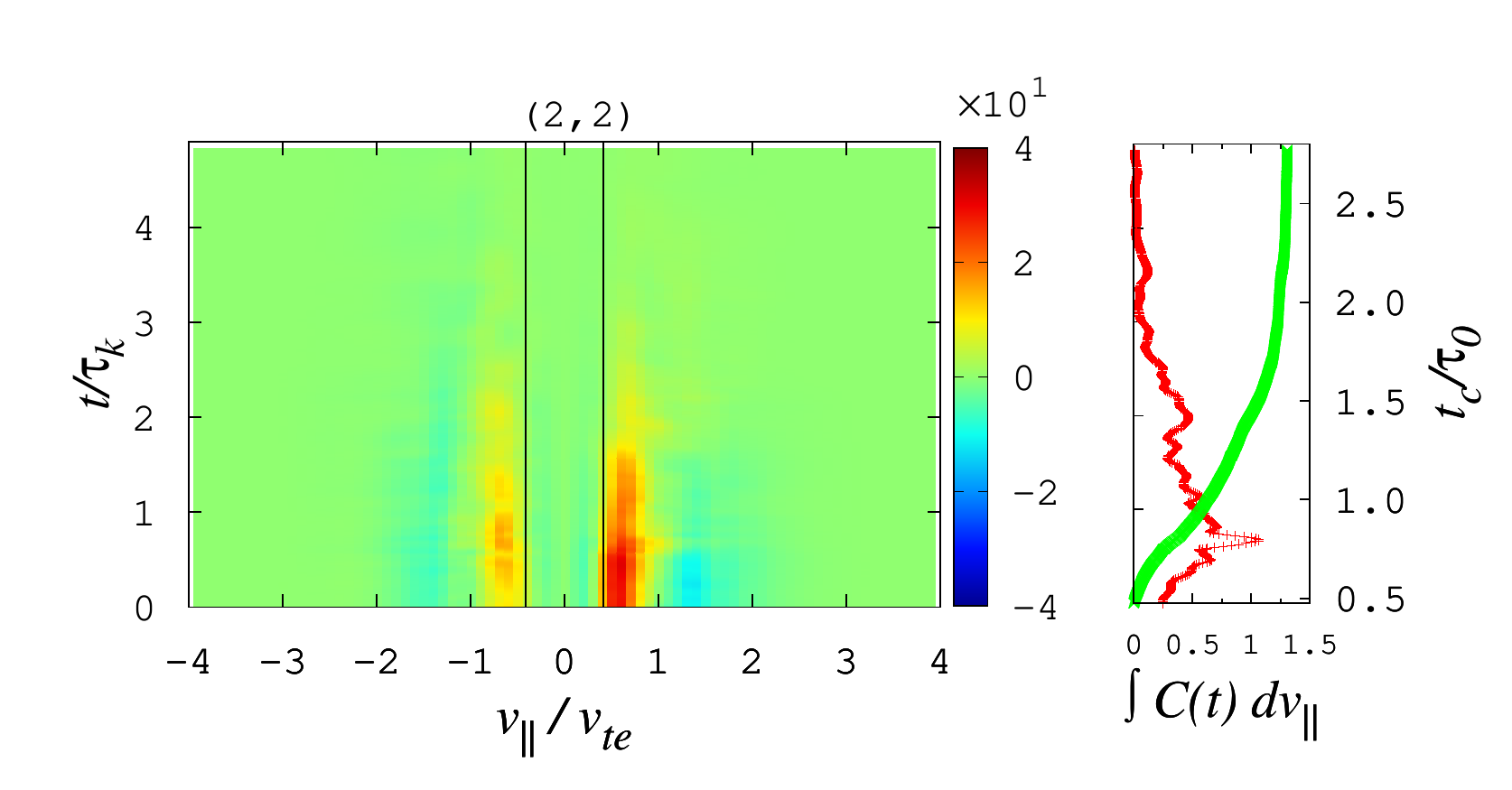}} \hfill  
  }  
 \caption{ \label{fig:ft3_1k22_363_F3dfn3} Electron parallel reduced
   correlations $C_{E_\parallel}(v_\parallel,t)$ for 7 $k_z$ modes and summed
   $C_{E_\parallel}(v_\parallel,t)$ for the $(2,2)$ Fourier mode. Correlation
   interval of $\tau$ = 2$\tau_k$ is chosen.
   Time $t$ is defined at the beginning 
   of the correlation interval. Vertical black lines indicate
   Landau resonant velocities: $v_{p \parallel}/v_{te}$ = $\pm$0.42. Arbitrary
   unit is used. Line plots on the right of each correlation are the
   $v_\parallel$-integrated correlation as a function of time (red
   curve) and accumulated over time (green curve). The $\times$10$^n$
   factor above each colorbar applies also to the x-axis of the line
   plots. Also plotted on the right of panel (h) is 
   an axis for the normalized centered time $t_c/\tau_0$. }
\end{figure}

Plotted in \figref{fig:ft3_1k22_363_F3dfn3} are timestack plots of the
parallel reduced correlations $C_{E_\parallel}(v_\parallel,t)$ for the
electrons for the $(2,2)$ Fourier mode, separately for each of
the 7 $k_z\in[-3,+3]$ modes in panels (a)--(g) and the sum of these
seven $k_z$ modes in panel (h), labelled by just $(2,2)$.  The
correlation interval chosen for all panels is $\tau$ = $2
\tau_k$. Plotted to the right of each timestack plot is the
$v_\parallel$-integrated correlation, giving the net transfer rate of
energy density for that Fourier mode $\int
C_{E_\parallel}(v_\parallel,t) dv_\parallel$ as a function of time
(red curve) and the accumulated energy transferred $\int_0^t dt \int
C_{E_\parallel}(v_\parallel,t) dv_\parallel$ over time (green
curve). The Landau resonant parallel phase velocity of each Fourier
mode is indicated by vertical black lines. We note that, for a given
value of $k_\perp$, the normalized phase velocity 
$\overline{\omega} = \omega/k_\parallel v_A$ is independent of 
$k_\parallel$, so the phase
velocity ($\omega/k_\parallel$) is \emph{constant} for all $k_z$ (note
that $k_\parallel$ and $k_z$ are the same in the linear dispersion
relation by which $\overline{\omega}$ is given). This leads to a
constant Landau resonant velocity, $v_{p \parallel}/v_{te}$ = $\pm$0.42, for
all $(2,2,k_z)$ Fourier modes in \figref{fig:ft3_1k22_363_F3dfn3}. 
Also plotted on the right of the $v_\parallel$-integrated correlation 
for (h) the sum of the 7 $k_z$ mode is an axis for the normalized 
centered time $t_c/\tau_0$.

A key characteristic of the parallel reduced correlations
for all $(2,2,k_z)$ Fourier modes is that the energy transfer is
largely dominated by particles connected to $v_{p \parallel}$, with an
increase in electron energy (red) just above $v_{p \parallel}$.
This reduced parallel velocity-space signature is consistent with
previous results showing Landau damping
\citep{Klein:2016,Howes:2017,KleinHT:2017,Howes:2018a,Chen:2019a}. This
localization of the energy transfer in parallel velocity indicates
that collisionless damping by Landau resonant electrons is the
dominant mechanism for transferring energy from the turbulent
electromagnetic fluctuations to the electrons, a primary result of
this study.

Furthermore, the timestack plots show clearly that this resonant
energization of the electrons by the parallel electric field sustains
over the course of the simulation, although the amplitude of this rate
of energization decreases in time as the initial fluctuations (the total
turbulence energy) damp in
time.  In fact, the decrease in the rate of electron energization by
the electric field over the course of the simulation is qualitatively
consistent with the decrease in the spatially integrated rate of electron
energization $ \dot{E}^{(fp)}_e$ (solid blue) in
\figref{fig:ft3_dedt}(b).
Additionally, the characteristic time scale of the energy transfer 
in the parallel reduced correlations is 
approximately $\tau_k$ of this $(2,2)$ Fourier mode. 
This time scale is consistent with the linear wave period of
this Fourier mode, as naturally expected.

The panels (a)--(g) for the 7 different $(2,2,k_z)$ Fourier modes in
\figref{fig:ft3_1k22_363_F3dfn3} show that most of the different
$k_z$ modes yield a net energy gain (green curve) at the end of the
run.  The (h) sum of these 7 $k_z$ components of the $(2,2)$ Fourier
mode indeed demonstrates a net heating of electrons over the course of
the simulation.
\begin{figure} \centering
\hbox{{(a)\hspace{-1.3em}} \hfill \resizebox{2.6in}{!}{\includegraphics*{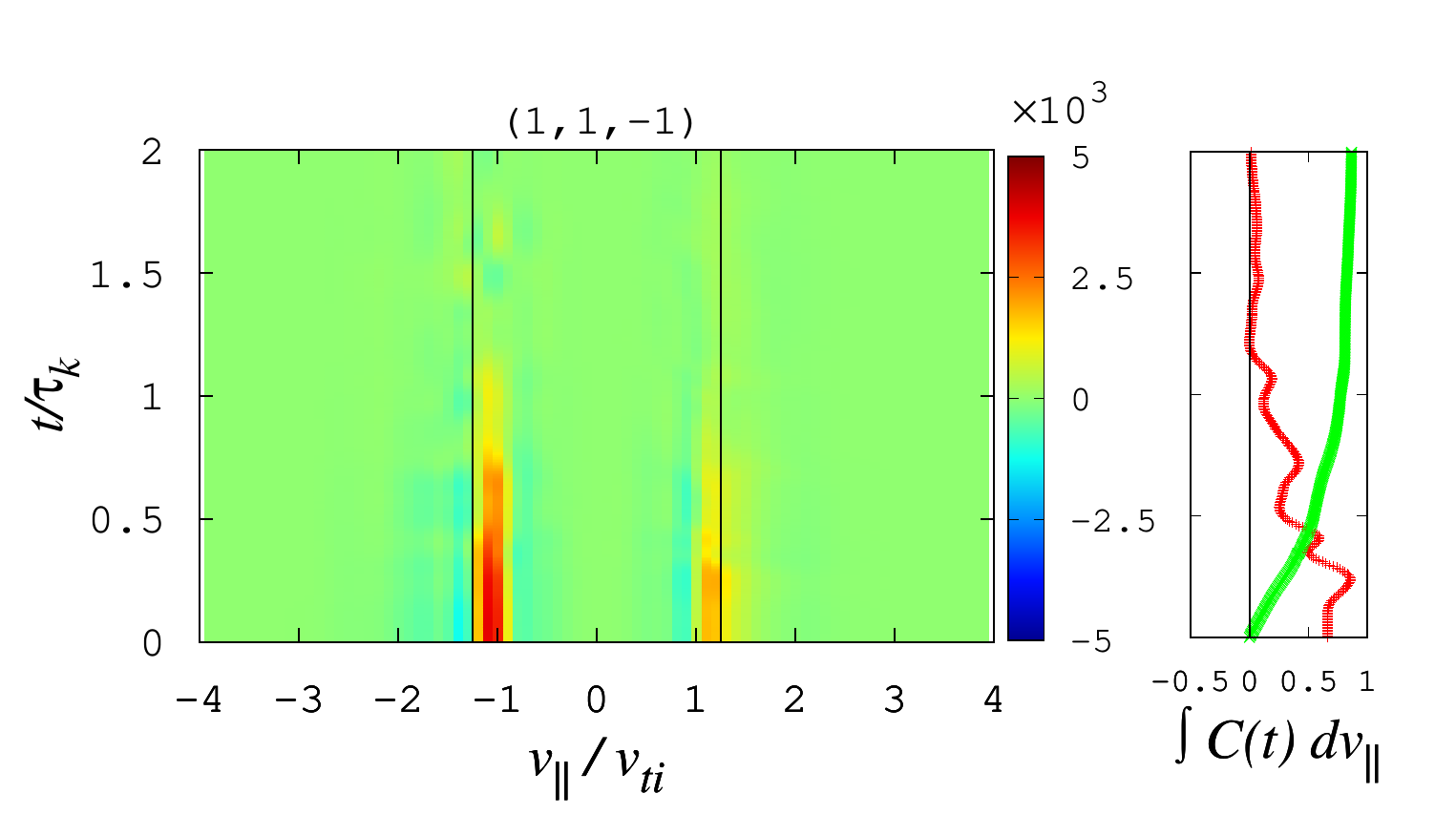}} \hfill  
{(b)\hspace{-1.3em}} \hfill \resizebox{2.6in}{!}{\includegraphics*{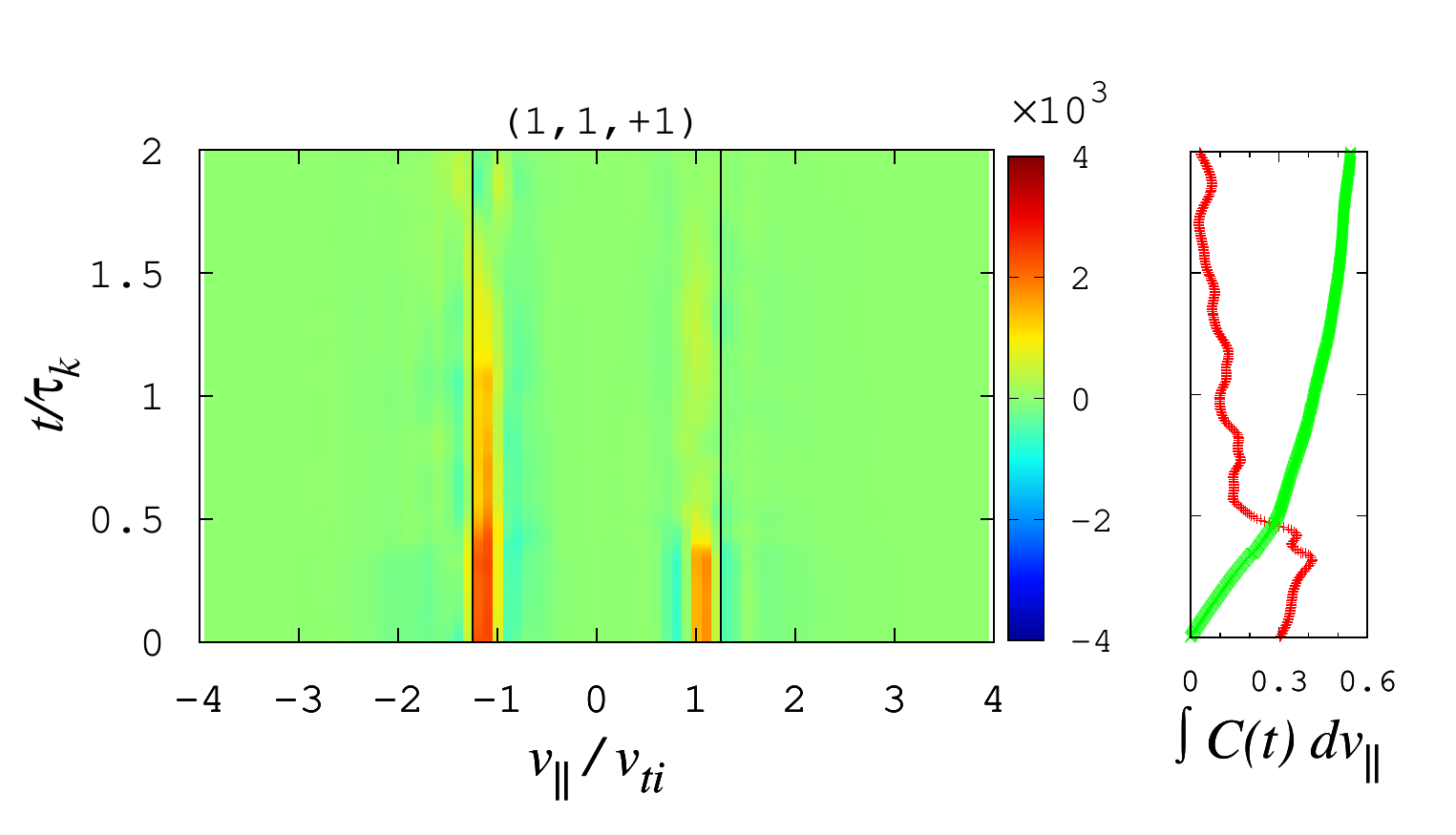}} \hfill  
  }  
\hbox{{(c)\hspace{-1.3em}} \hfill \resizebox{2.6in}{!}{\includegraphics*{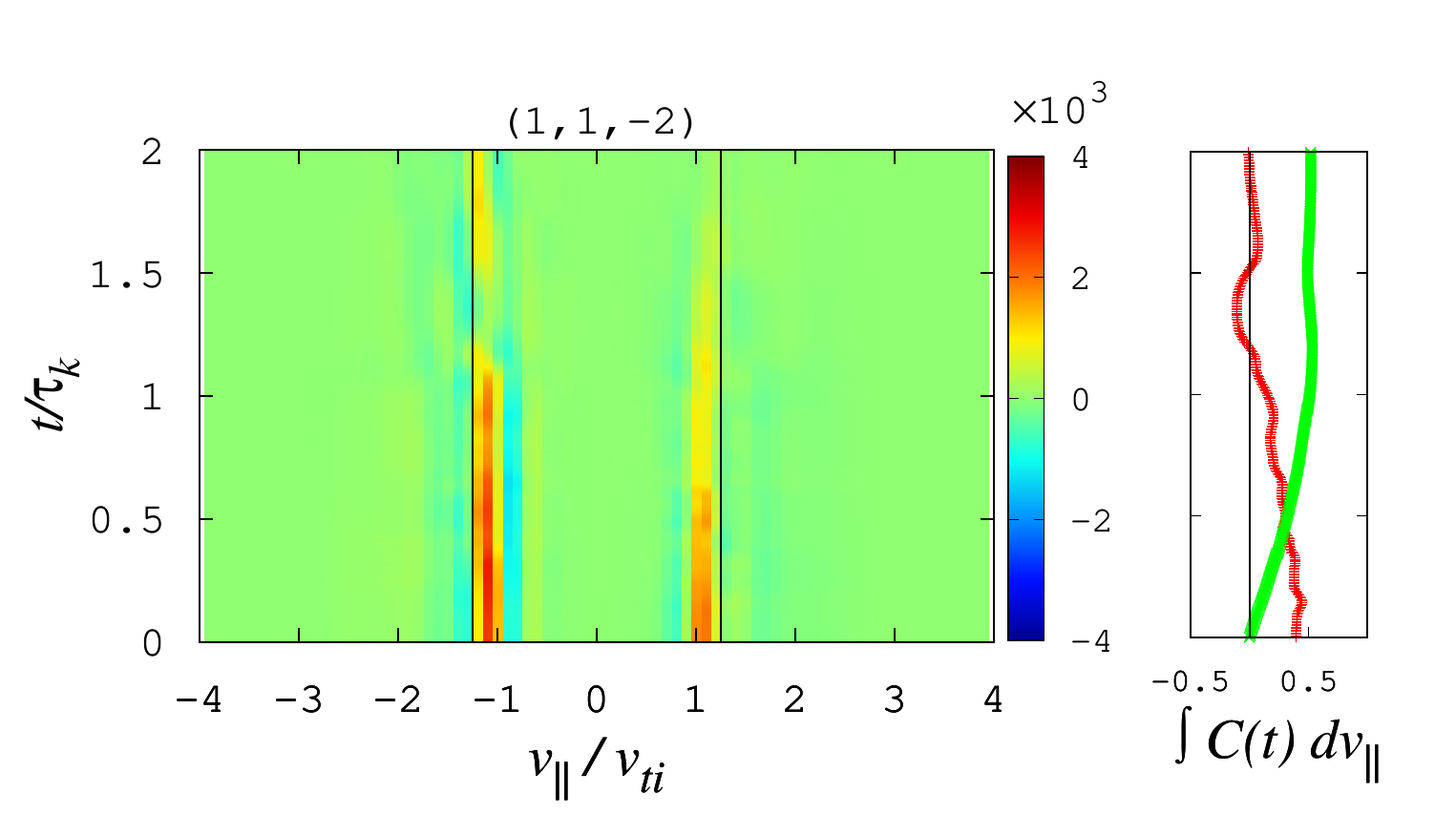}} \hfill  
{(d)\hspace{-1.3em}} \hfill \resizebox{2.6in}{!}{\includegraphics*{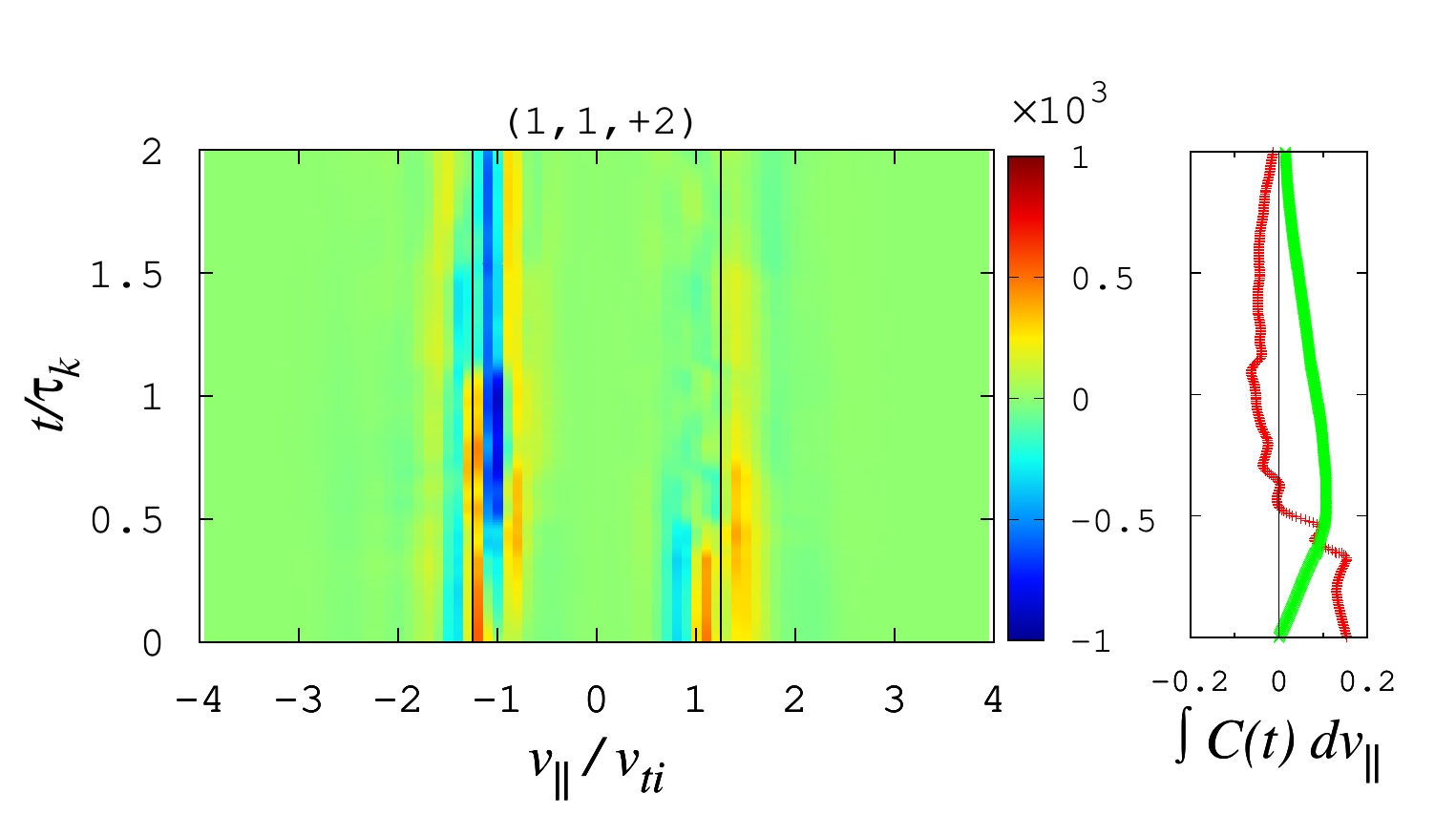}} \hfill  
  }  
\hbox{{(e)\hspace{-1.3em}} \hfill \resizebox{2.6in}{!}{\includegraphics*{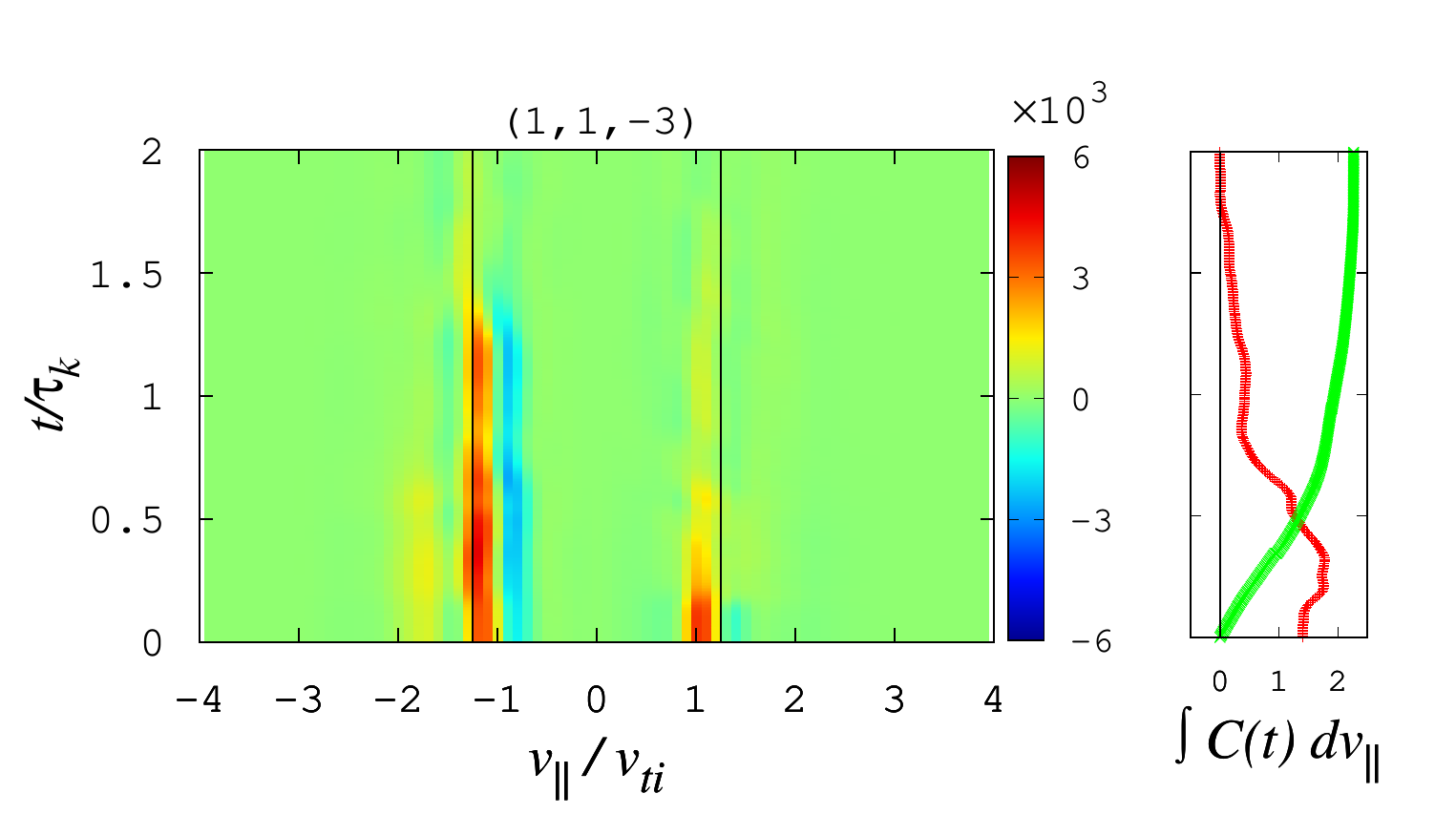}} \hfill  
{(f)\hspace{-1.3em}} \hfill \resizebox{2.6in}{!}{\includegraphics*{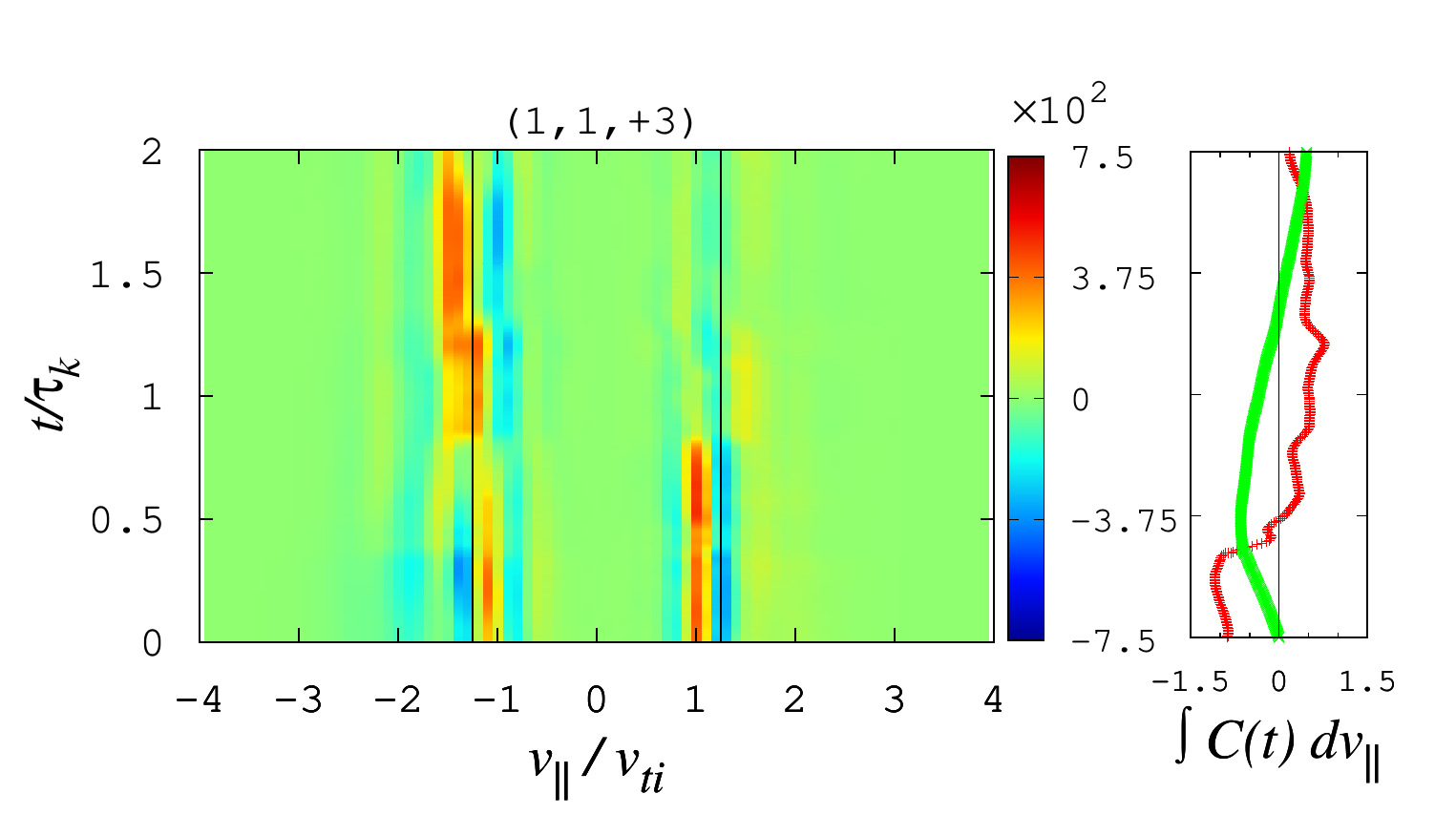}} \hfill  
  }  
\hbox{{(g)\hspace{-1.3em}} \hfill \resizebox{2.6in}{!}{\includegraphics*{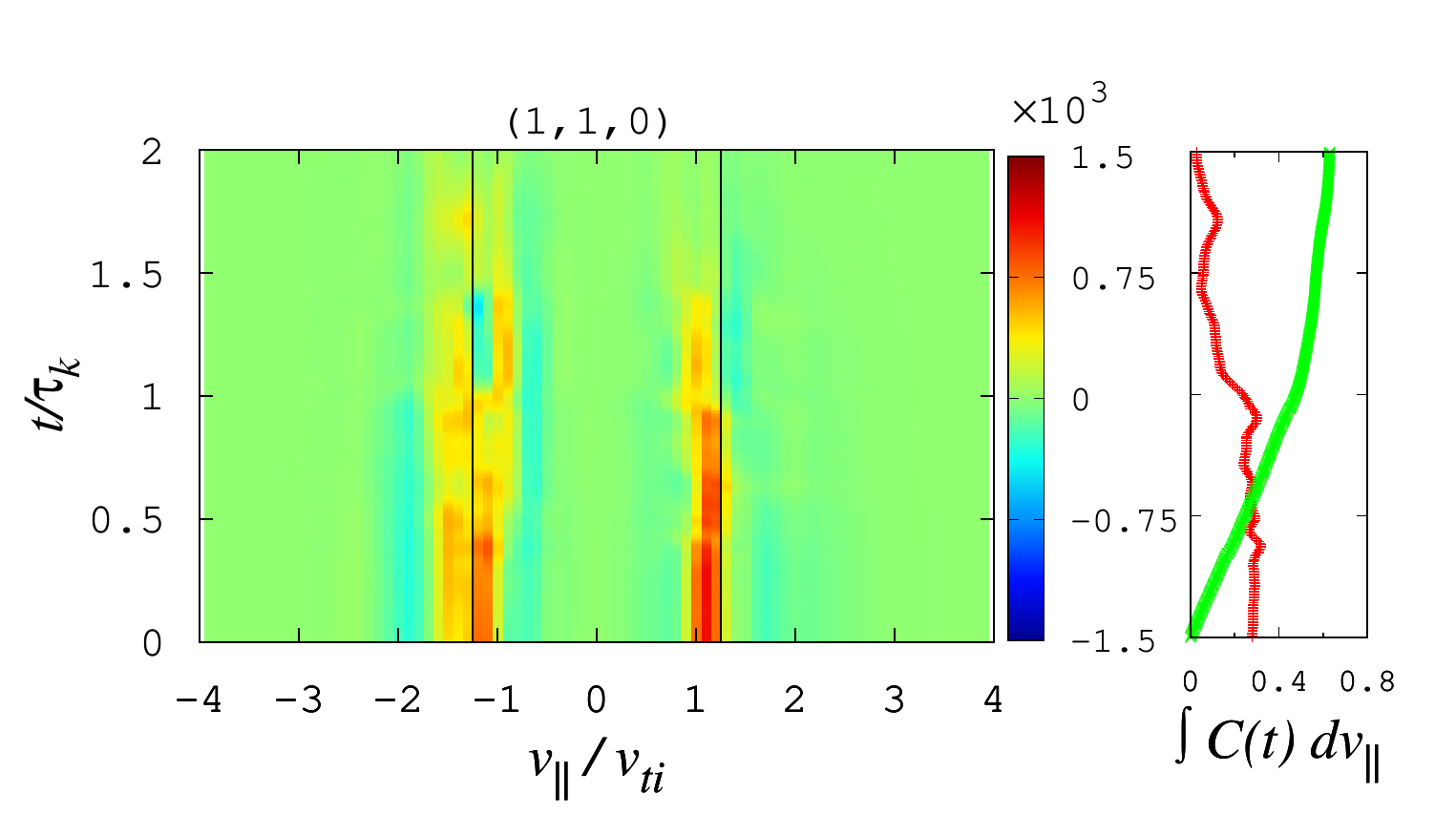}} \hfill  
{(h)\hspace{-1.3em}} \hfill \resizebox{2.8in}{!}{\includegraphics*[scale=1.2,trim=0cm 0cm 0cm 0cm, clip=true]{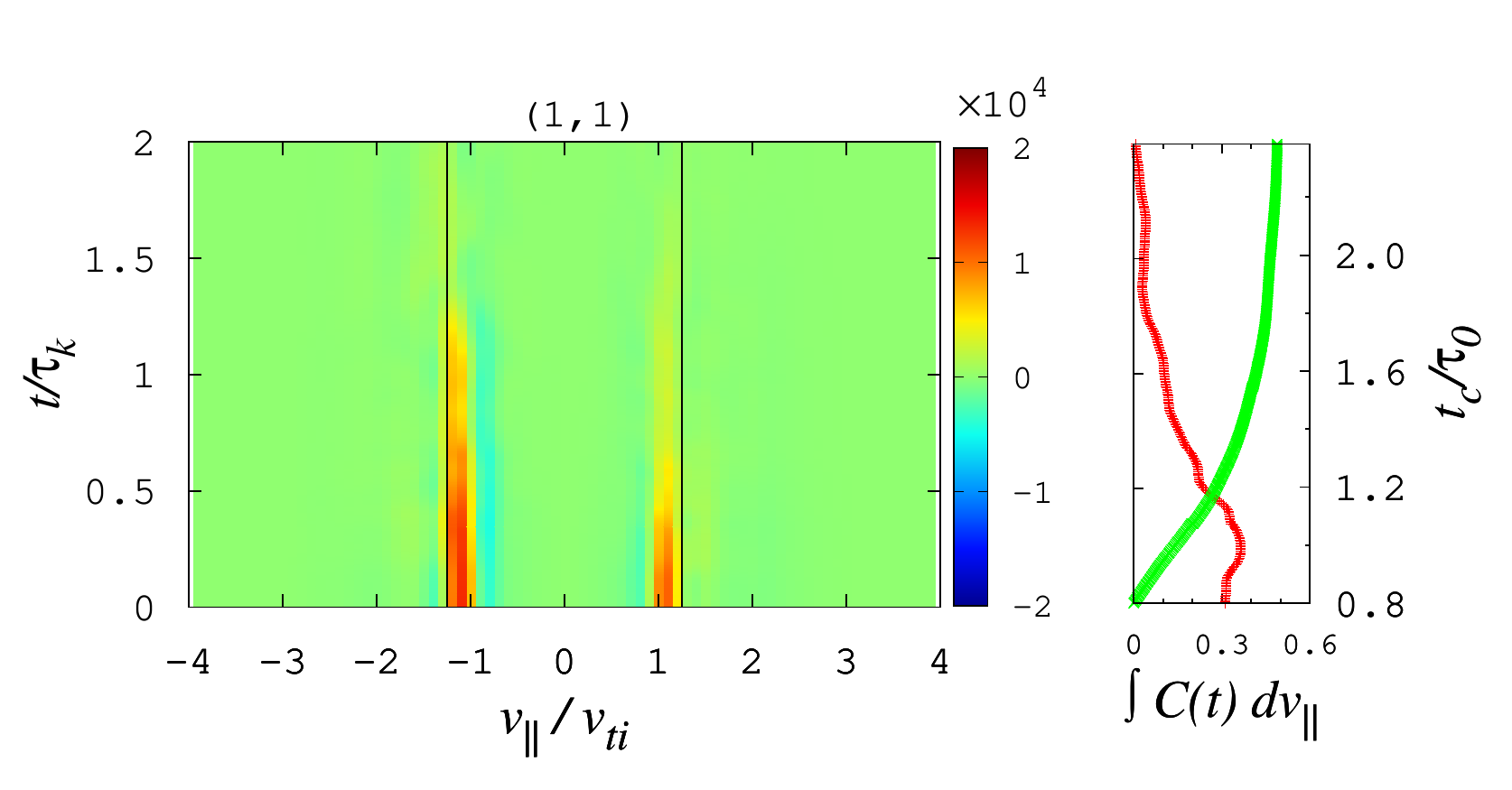}} \hfill  
  }  
 \caption{ \label{fig:ft3_1k11_336_F3dfn3_i} Ion parallel reduced correlations $C_{E_\parallel}(v_\parallel,t)$ for 7 $k_z$ modes and summed $C_{E_\parallel}(v_\parallel,t)$ for the $(1,1)$ Fourier mode. Correlation interval of $\tau$ = 2$\tau_k$ is used. $C_{E_\parallel}(v_\parallel,t)$ for each individual $k_z$ mode shows resonant signatures associated with Landau resonances, $v_{p \parallel}/v_{ti}$ = $\pm$1.3, indicated by vertical black lines. Highly localized energy transfer signals are observed for all seven Fourier modes. The same format is used as \figref{fig:ft3_1k22_363_F3dfn3}. Arbitrary unit
 is used.  }
\end{figure}



\subsubsection{Ion Parallel Reduced Correlations}\label{sec:1_ion_k11}


In \figref{fig:ft3_1k11_336_F3dfn3_i}, we plot timestack plots of the
parallel reduced correlations $C_{E_\parallel}(v_\parallel,t)$ for the
ions for the $(1,1)$ Fourier mode, in the same format as for the
electrons in \figref{fig:ft3_1k22_363_F3dfn3}. Again, a correlation
interval of $\tau$=$2 \tau_k$ is used for all plots. The Landau
resonant parallel phase velocity for these modes, normalized in terms
of the ion thermal velocity, is given by $v_{p \parallel}/v_{ti}=\pm 1.3$.

The ion reduced parallel correlations for the $(1,1,k_z)$ modes also
show that the energy transfer is largely dominated by particles near
the resonant velocity, indicating that collisionless damping by Landau
resonant ions is the dominant mechanism for transferring energy from
the parallel electric field to the ions. Although there
is significantly more variation of the accumulated energy transfer to
the ions among the different $k_z$ modes in panels (a)--(g), the
summed $k_z$ result in (h) is also qualitatively consistent with the
gradual decrease in time of the spatially integrated rate of ion
energization $ \dot{E}^{(fp)}_i$ (solid red) in
\figref{fig:ft3_dedt}(a). Furthermore, in the same way as the characteristic time scale of the electron 
energization rate being consistent with $\tau_k$ of the Fourier mode 
in \figref{fig:ft3_1k22_363_F3dfn3}, the characteristic time scale of the 
ion energization rate for the different $k_z$ modes is also 
approximately $\tau_k$ of this $(1,1)$ Fourier mode.

Consistent with the rapid decrease of the ion collisionless
damping rate $\gamma_i/k_\parallel v_A$ (cyan) in \figref
{fig:ft3_heat} at perpendicular wavenumbers $k_\perp \rho_i >2$, the
reduced parallel correlations (not shown) for all Fourier modes
other than $(1,1)$ do not exhibit clear resonant energy transfer
velocity-space signatures.

\subsubsection{Reduced Parallel Correlations: $k_\perp$ Dependence and $\beta_i$ Dependence }\label{sec:sum_fpc}
\begin{figure} \centering
\hbox{{(a)\hspace{-1.3em}} \hfill \resizebox{2.8in}{!}{\includegraphics*[scale=1.1,trim=0cm 0cm 0cm 0cm, clip=true]{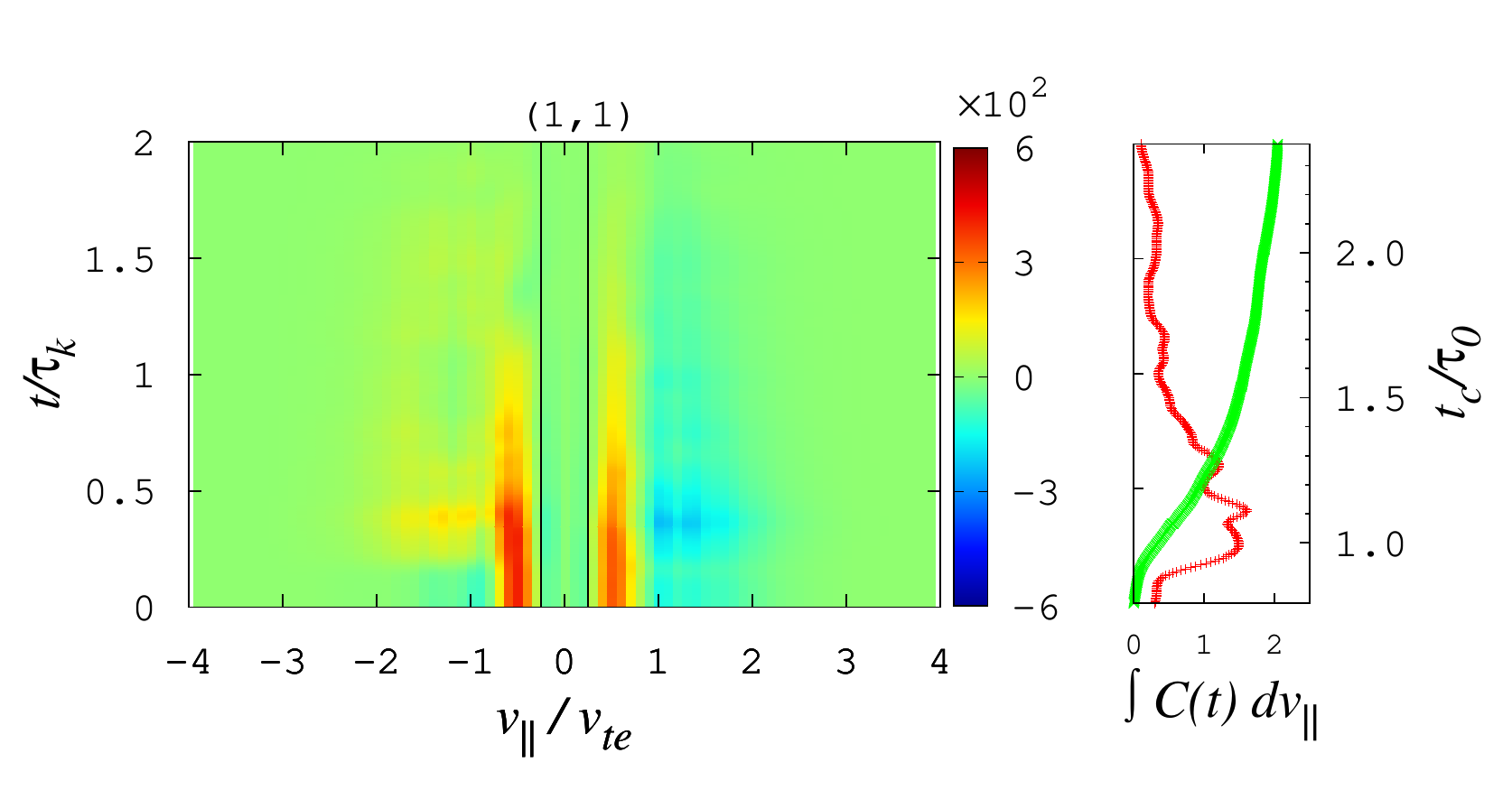}} \hfill  
{(b)\hspace{-1.3em}} \hfill \resizebox{2.8in}{!}{\includegraphics*[scale=1.1,trim=0cm 0cm 0cm 0cm, clip=true]{ft3_1k22_nc200_F3dfn3_Sum_dv_t0.pdf}} \hfill  
  }  \vspace{-1em}%
\hbox{{(c)\hspace{-1.3em}} \hfill \resizebox{2.8in}{!}{\includegraphics*[scale=1.1,trim=0cm 0cm 0cm 0cm, clip=true]{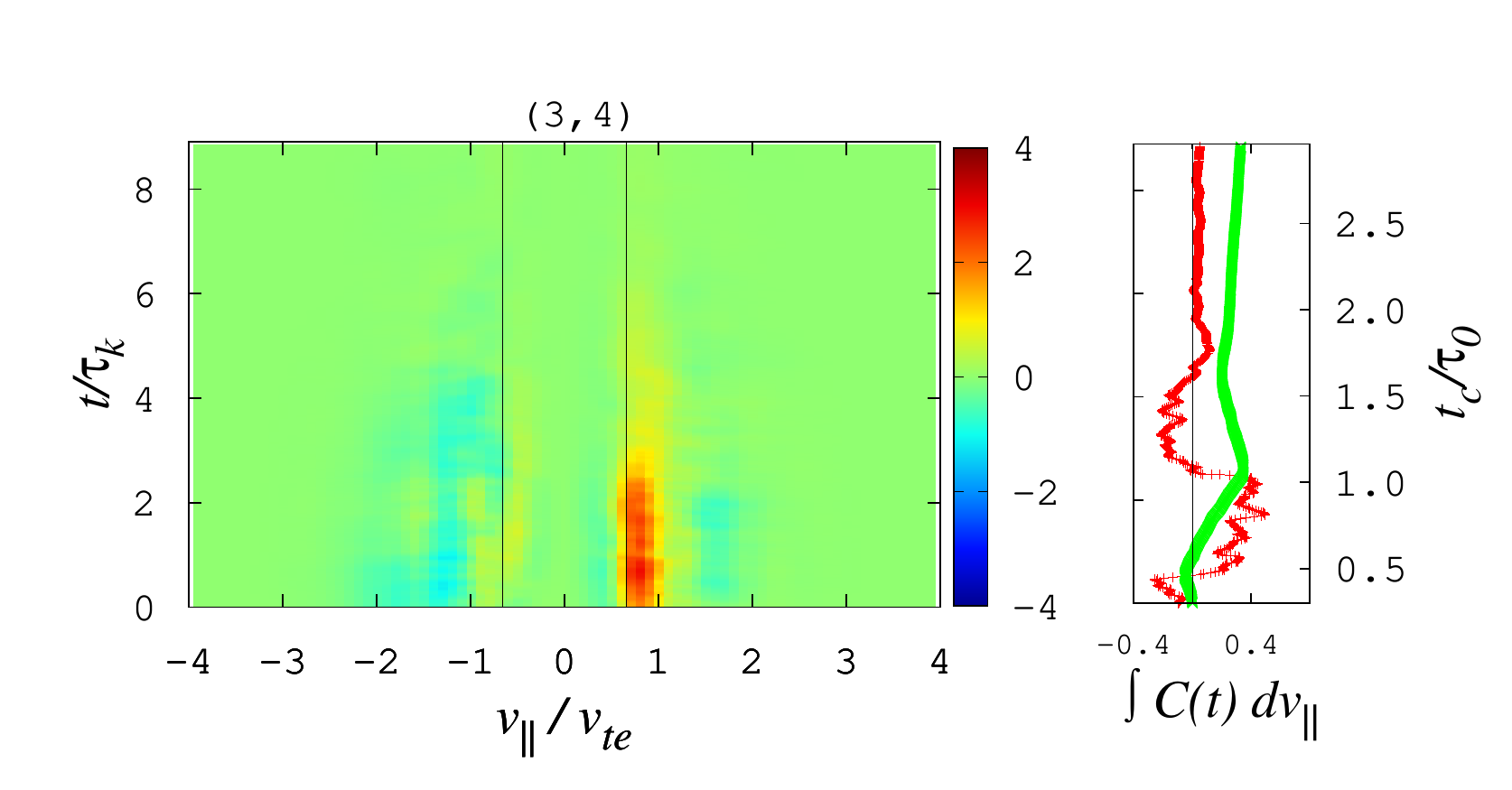}} \hfill  
{(d)\hspace{-1.3em}} \hfill \resizebox{2.8in}{!}{\includegraphics*[scale=1.1,trim=0cm 0cm 0cm 0cm, clip=true]{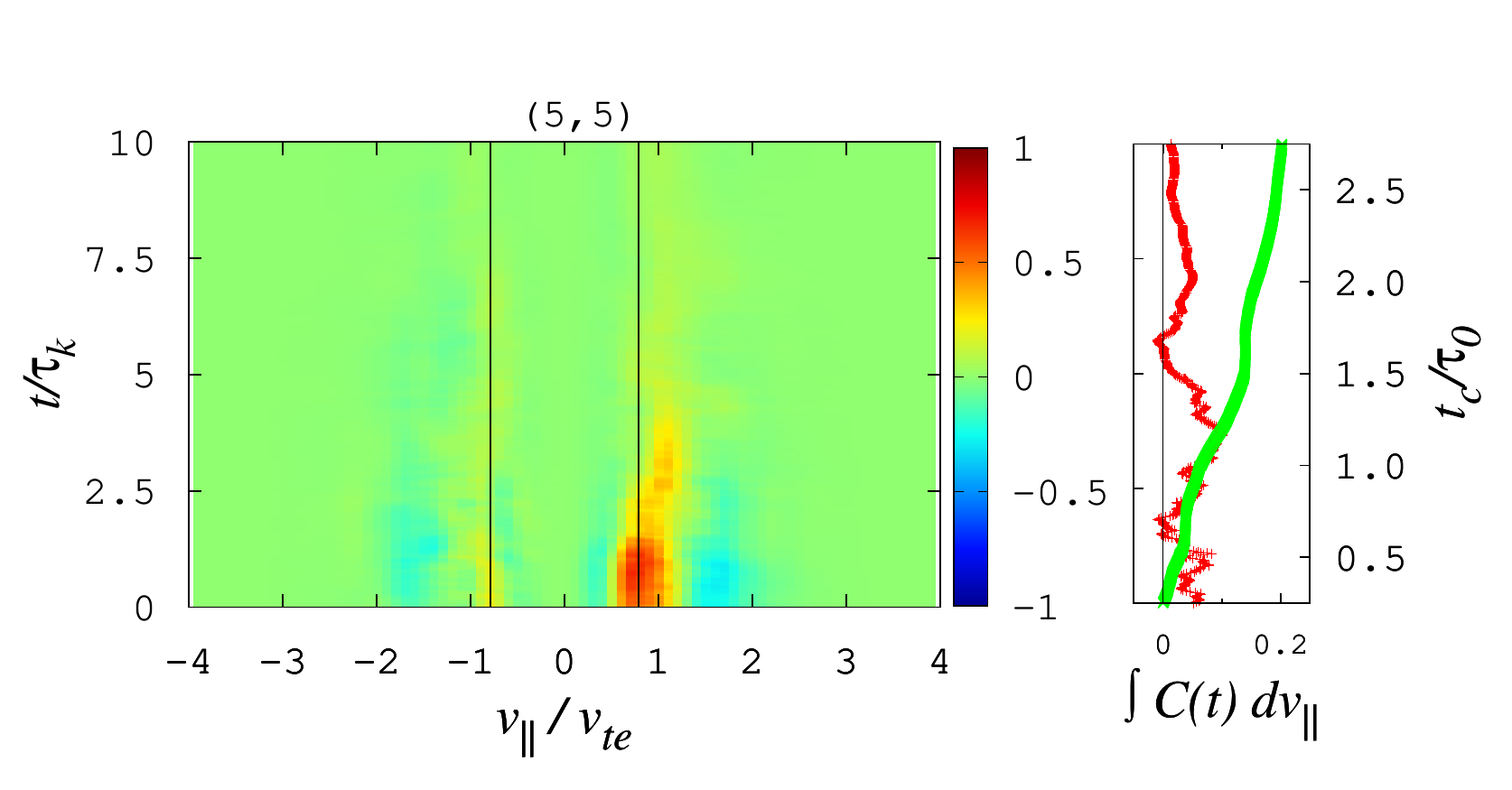}} \hfill  
  }  \vspace{-1em}%
\hbox{{(e)\hspace{-1.3em}} \hfill \resizebox{2.8in}{!}{\includegraphics*[scale=1.1,trim=0cm 0cm 0cm 0cm, clip=true]{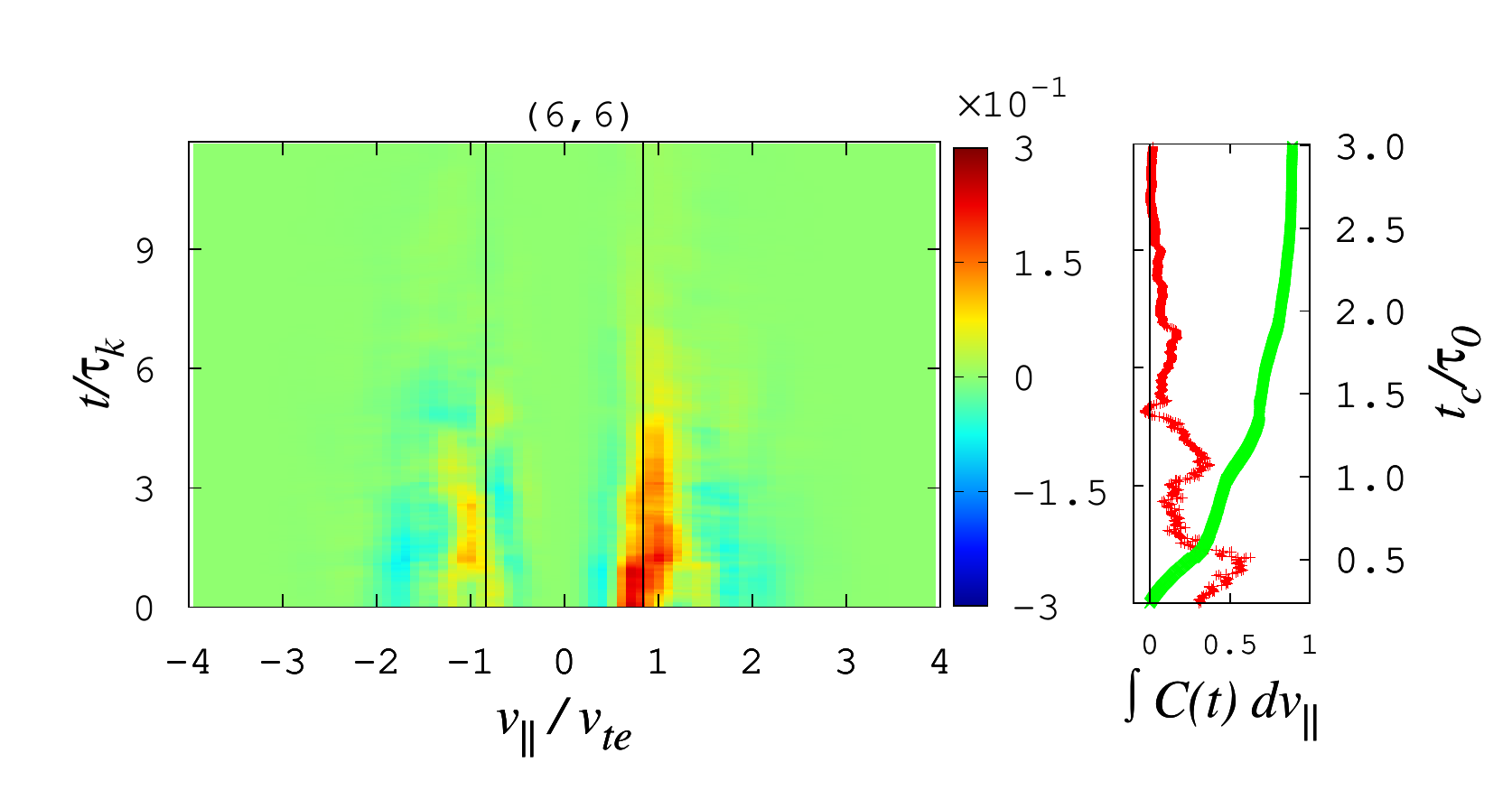}} \hfill  
{(f)\hspace{-1.3em}} \hfill \resizebox{2.8in}{!}{\includegraphics*[scale=1.1,trim=0cm 0cm 0cm 0cm, clip=true]{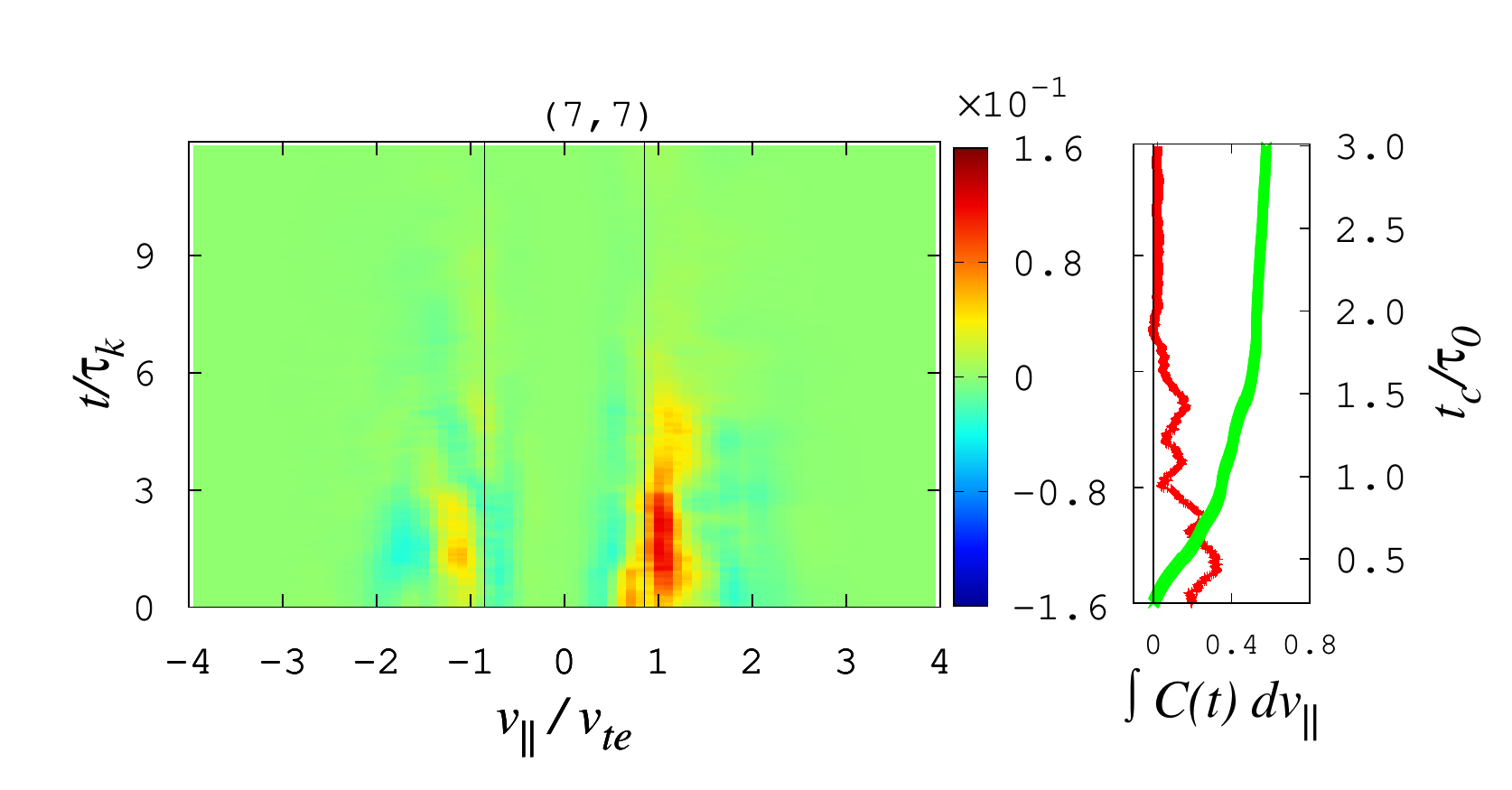}} \hfill  
  }
 \caption{ \label{fig:ft3_1_Sum} Parallel reduced correlations
   summed over 7 $k_z$ modes $\sum\limits_{k_z=-3}^{+3} C_{E_\parallel}(v_\parallel,t)$ 
   for 6 $(k_x,k_y)$ values: $(1,1)$, $(2,2)$, $(3,4)$, $(5,5)$, $(6,6)$
   and $(7,7)$, representing a total of 42 Fourier modes of electron
   parallel reduced correlations. Their Landau resonant velocities
   are: $v_{p \parallel}/v_{te}$ = $\pm$(0.25, 0.42, 0.66, 0.79, 0.84, 0.85)
   respectively. The same format is used as
   \figref{fig:ft3_1k22_363_F3dfn3} (h). }
 
\end{figure}


Previous applications of the field-particle correlation have analyzed
the energy transfer at a single-point in space
\citep{Klein:2016,Howes:2017,KleinHT:2017,Howes:2018a,Chen:2019a}. Although this
has the advantage of enabling the spatial localization of the energy
transfer between fields and particles to be studied, it cannot 
provide any information about how that energy transfer is mediated by
fluctutations at different scales. By performing the field-particle correlation analysis in Fourier
space, we can examine how the energy transfer as a function of the scale
of the Fourier modes plays a role in the energy transfer.
In the gyrokinetic limit $k_\parallel \ll k_\perp$, the resonant parallel
phase velocity $\omega/k_\parallel$ of the kinetic \Alfven wave modes
is strictly a function of $k_\perp \rho_i$ (as shown in \figref{fig:1om_g}(a)), 
so the reduced parallel
correlation for a single perpendicular wavevector $(k_x,k_y)$ summed
over $k_z$ has a single resonant velocity. We may obtain a clean velocity-space signature for each distinct $k_\perp\rho_i$ mode.





In \figref{fig:ft3_1_Sum}, we present timestack plots of the reduced
parallel correlations for the electrons summed over the 7 $k_z$ modes
for each diagnosed $k_\perp$ case spanning $1.4 \le k_\perp\rho_i \le 9.9$, 
or $ 0.3 \leq k_\perp\rho_e \le 2 $. The same format as the
summed-$k_z$ results in panel (h) of \figref{fig:ft3_1k22_363_F3dfn3}
is used.  The results clearly show the distinct advantages of the
Fourier implementation of the field-particle correlation technique.
First, the energy transfer signal for different $k_\perp \rho_i$ modes
follows closely the {\it increasing} Landau resonant velocity
(vertical black lines) as $k_\perp \rho_i$ increases from (a) through
(f). A net energy energization of the electrons 
is obtained in all of $k_\perp \rho_i$ modes, as shown in the accummulated
energy transferred (green curve). 
Clearly, Landau resonant interactions with the electrons must play a 
key role in the damping of fluctations across the sampled range of scales, 
consistent with expectations for a significant rate of collisionless 
damping by the electrons, $\gamma_e/\omega_{A0}\gtrsim\mathcal{O}(0.1)$, 
as shown in \figref{fig:1om_g}(b).  
Second, the magnitude of energy transfer
drops by around 3 orders of magnitude from $k_\perp \rho_i=1.4$ to
$k_\perp \rho_i=9.9$.  Of course, the number of perpendicular Fourier
modes increases linearly with $k_\perp$, so the decrease in amplitude
with $k_\perp \rho_i$ is partly compensated by the larger number of
Fourier modes at higher $k_\perp \rho_i$.  Nonetheless, 
the first one or two $k_\perp \rho_i$
modes that we diagnosed here (specifically $k_\perp \rho_i=1.4$ and
$k_\perp \rho_i=2.8$) would dominate the signals in a single-point
diagnostic, so one would not be
able to identify the energy transfer mediated by higher $k_\perp\rho_i$ modes
without Fourier decomposition. 
Finally, the characteristic time scale of the energy transfer is 
consistent with $\tau_k$ of each $(k_x,k_y)$ Fourier mode, even though
$\tau_k$ decreases with increasing $k_\perp \rho_i$. As seen in the 
the accummulated energy transferred (green curve), most of the electron 
energization occurs before a normalized centered time of $t_c/\tau_0\sim 1.5$.
This is consistent with 
with the spatially integrated rate of electron energization 
$ \dot{E}^{(fp)}_e$ (solid blue) in
\figref{fig:ft3_dedt}(b) being the most significant by $1.5 \tau_0$

\begin{figure}\centering
  \hbox{(a)\hspace{-1.3em} \hfill \resizebox{2.6in}{!}{\includegraphics*[scale=1,trim=0cm 0cm 0cm 0.5cm, clip=true]{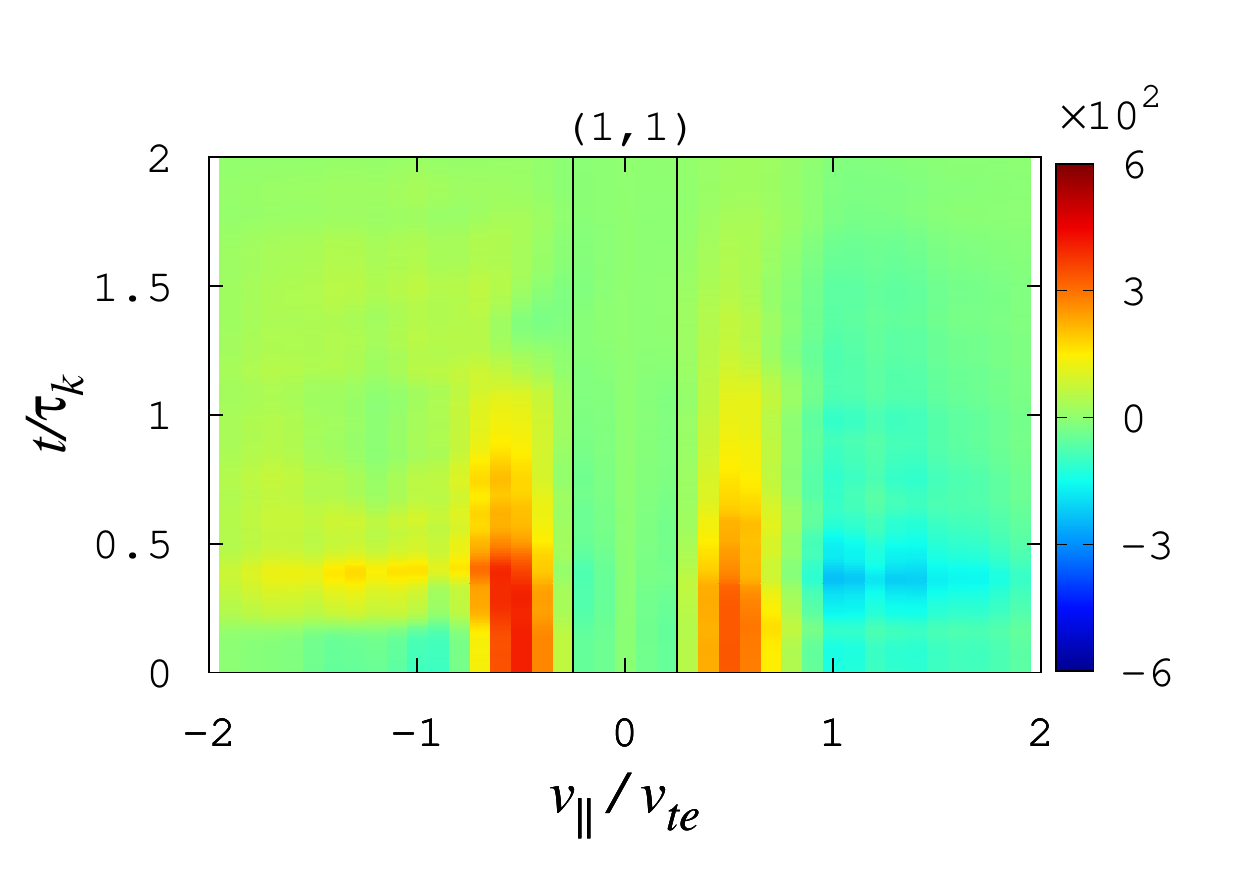}} \hfill
(d)\hspace{-1.3em} \hfill \resizebox{2.6in}{!}{\includegraphics*[scale=1,trim=0cm 0cm 0cm 0.5cm, clip=true]{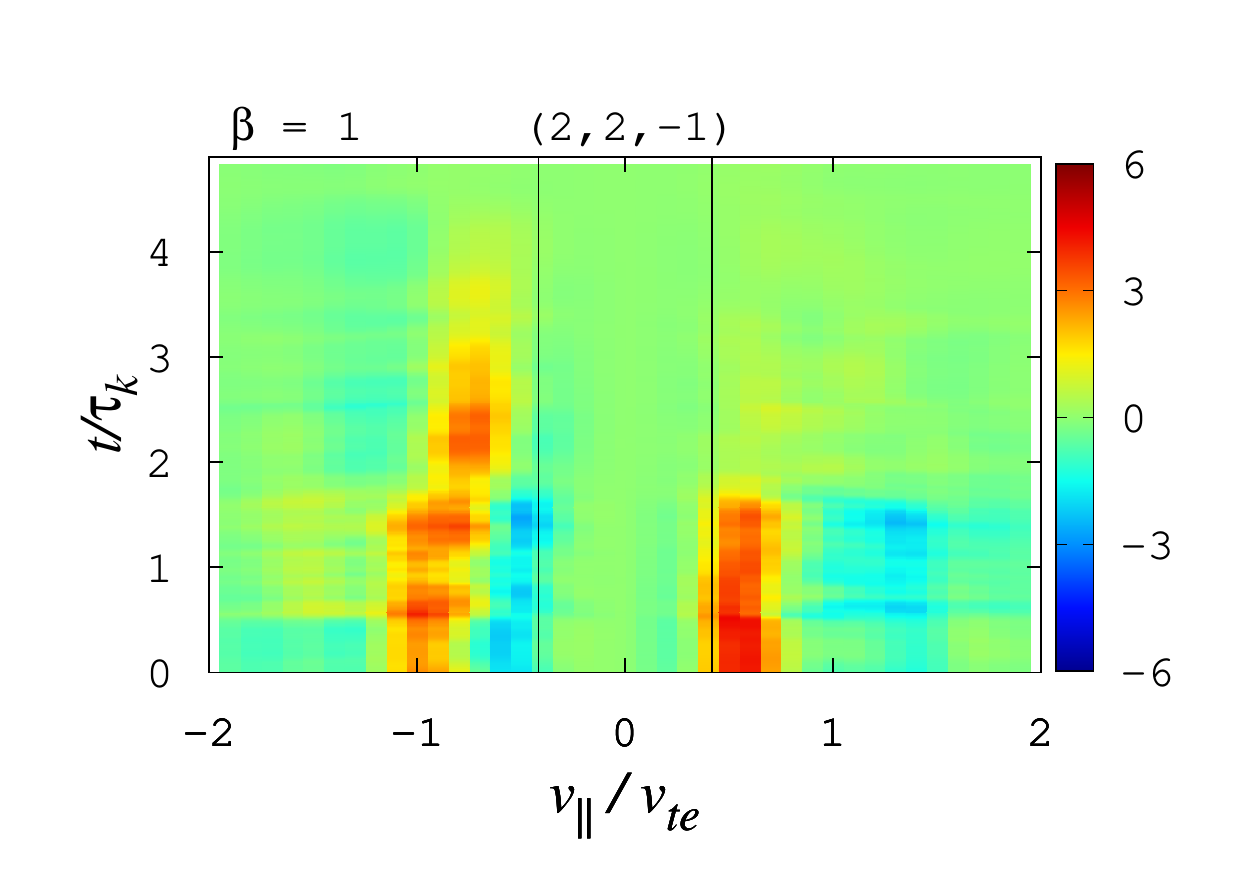}} \hfill}    
  \hbox{(b)\hspace{-1.3em} \hfill \resizebox{2.6in}{!}{\includegraphics*[scale=1,trim=0cm 0cm 0cm 0.5cm, clip=true]{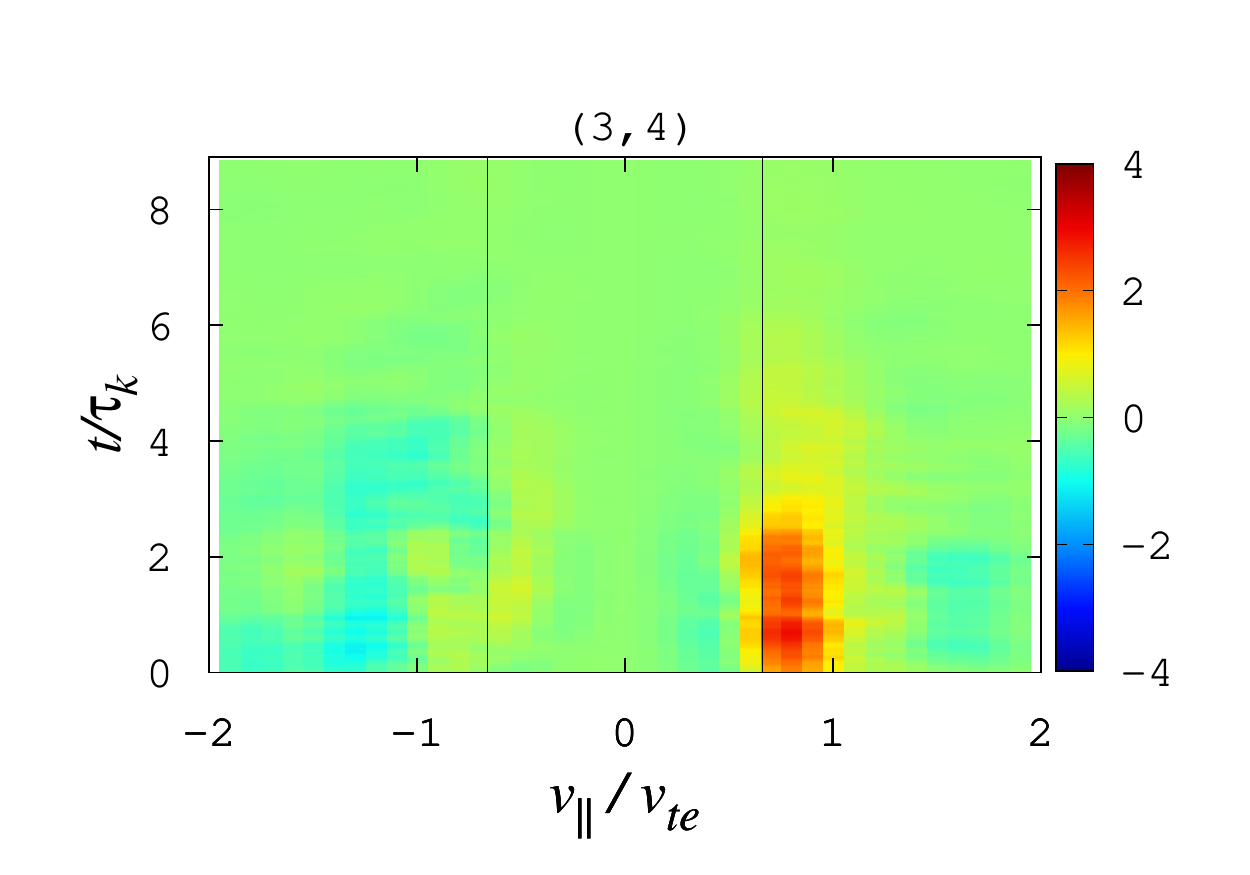}} \hfill 
    (e)\hspace{-1.3em} \hfill \resizebox{2.6in}{!}{\includegraphics*[scale=1,trim=0cm 0cm 0cm 0.5cm, clip=true]{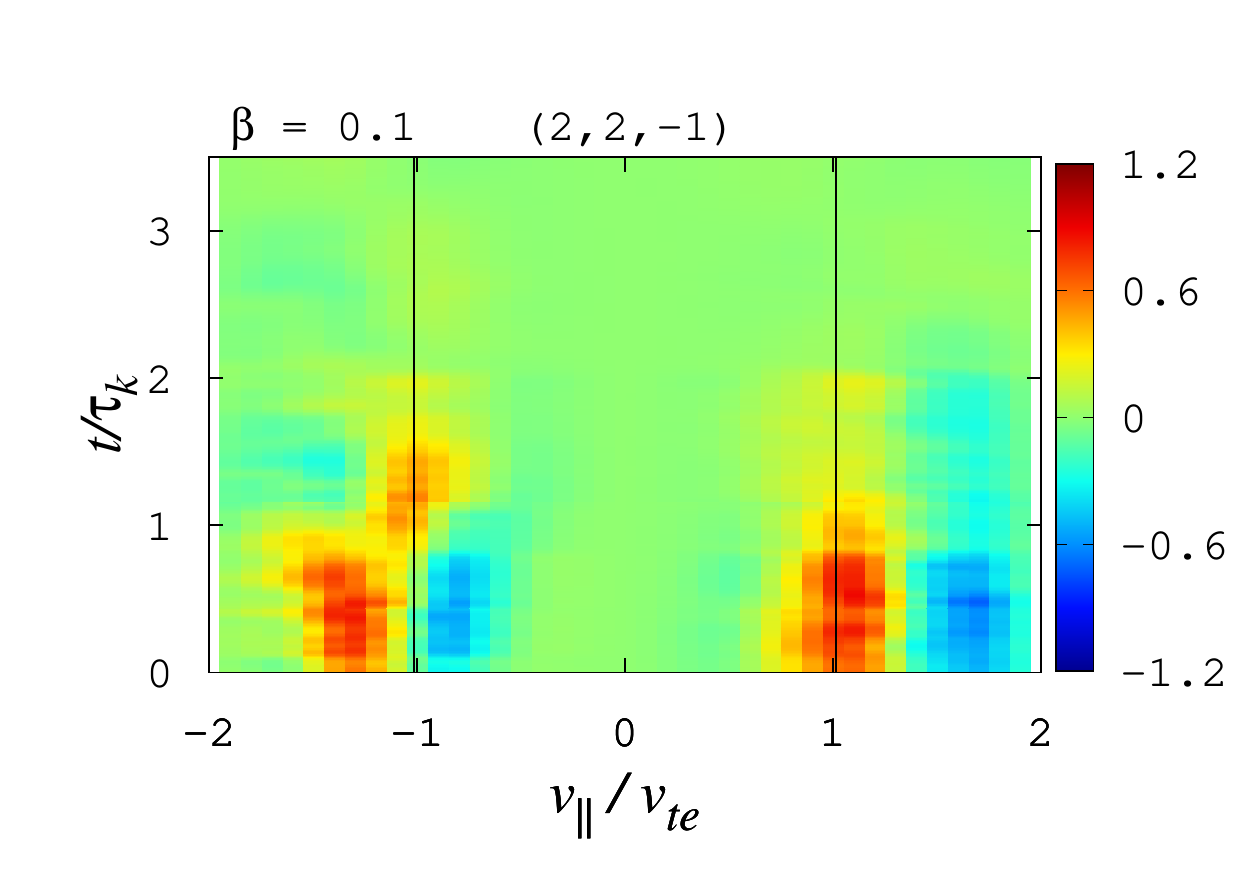}} \hfill    
}
  \hbox{(c)\hspace{-1.3em} \hfill \resizebox{2.6in}{!}{\includegraphics*[scale=1,trim=0cm 0cm 0cm 0.5cm, clip=true]{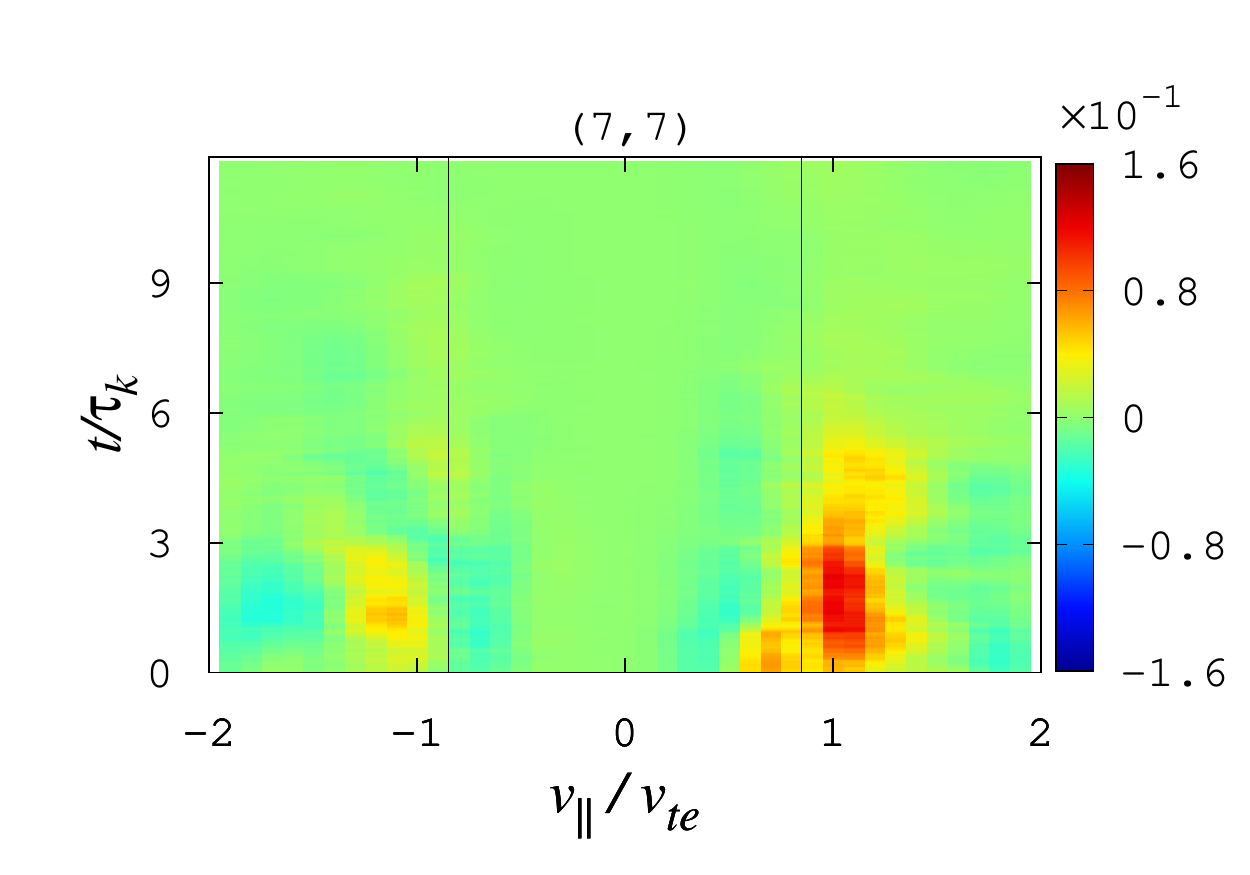}} \hfill
    (f)\hspace{-1.3em} \hfill \resizebox{2.6in}{!}{\includegraphics*[scale=1,trim=0cm 0cm 0cm 0.5cm, clip=true]{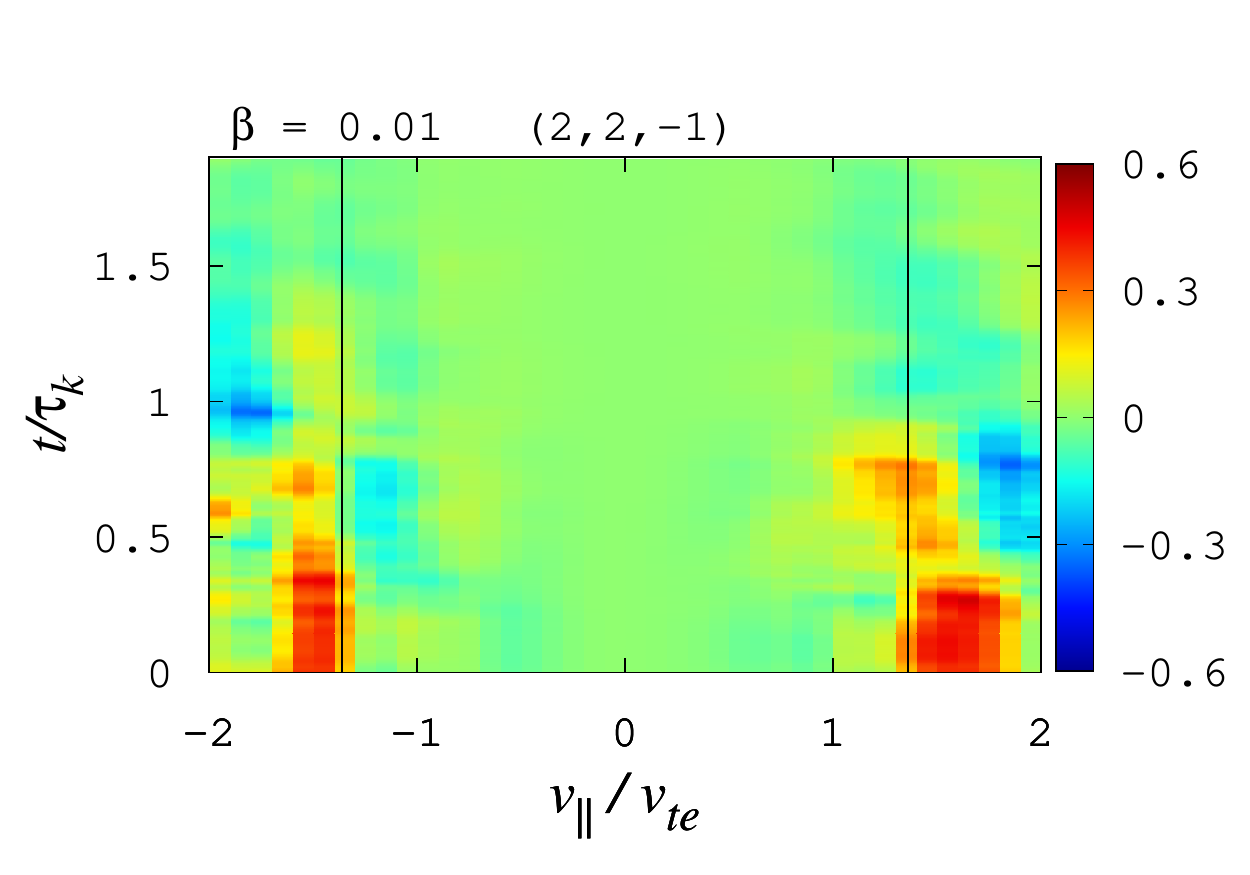}} \hfill }
\caption{ \label{fig:ft3_1_Sum_zoom} (a)--(c) Zoomed-in plot of
  summed-$k_z$ correlations from \figref{fig:ft3_1_Sum} for 3
  $(k_x,k_y)$ values--$(1,1)$, $(3,4)$ and $(7,7)$ with the
  corresponding Landau resonant velocities being $v_{p \parallel}/v_{te}$ =
  $\pm$(0.25, 0.66, 0.85), respectively--showing how the
  field-particle energy transfer rate closely tracks the 
  resonant parallel phase velocities as $(k_x,k_y)$ increases. 
  (d)--(f) Electron
  parallel reduced correlations $C_{E_\parallel}(v_\parallel,t)$ for the $(2,2,-1)$
  Fourier mode from simulations with $\beta_i$ = 1 (current run), 0.1
  and 0.01 in which the Landau resonant velocity, $v_{p \parallel}/v_{te}$ =
  $\pm$(0.42, 1.0, 1.4), respectively, increases with decreasing
  $\beta_i$. The energy transfer signals from all three simulations
  show remarkable agreement with the increasing parallel resonant velocity.  }
\end{figure}

In order to illustrate more clearly how the energy transfer to the electrons
closely tracks that Landau resonant velocity of kinetic \Alfven waves with a
given perpendicular wavenumber $k_\perp\rho_i$, we zoom into the
central part of the $v_\parallel$ range in
\figref{fig:ft3_1_Sum_zoom}(a)--(c). The localized region
dominating electron energization moves to
higher $v_\parallel$ with higher $k_\perp \rho_i$, consistent with the increasing
resonant parallel phase velocity of kinetic \Alfven waves as $k_\perp$ 
increases. This analysis
strongly suggests that collisionless damping via the Landau resonance
with electrons plays a key role in the removal of energy from the
turbulence and consequent energization of the electrons.

Another plasma parameter that has a strong impact on the resonant
parallel phase velocity of kinetic \Alfven waves is the plasma
$\beta_i$.  We performed simulations with an identical setup as the
$\beta_i=1$ simulation described in Sec.~\ref{sec:setup}, but with
values $\beta_i=0.1$ and $\beta_i=0.01$. To analyze both of these
additional simulations, we take a correlation interval $\tau=\tau_k$
for the $(2,2,-1)$ mode analyzed here.  In
\figref{fig:ft3_1_Sum_zoom}(d)--(f), we present timestack plots of the
reduced parallel correlation $C_{E_\parallel}(v_\parallel,t,\tau)$ for
electrons for Fourier mode $(2,2,-1)$.  The resonant parallel phase
velocity $v_{p\parallel}$ for the $(2,2)$ Fourier mode, when normalized to $v_{te}$,
increases with values $\omega/(k_\parallel v_{te})=\pm 0.42,\pm 1.0,\pm 1.4$ 
(vertical black lines) as the values of $\beta_i$
decreases for these three simulations $\beta_i=1, 0.1, 0.01$. We indeed
find a remarkably close association between $v_{p\parallel}$ and 
the region of velocity space where the electrons
participate in energy exchange with the parallel electric field
$E_\parallel$.  This close quantitative agreement with the $\beta_i$
dependence of phase velocities observed in the simulations again
suggests the dominant role that the Landau resonance plays in the
field-particle energy transfer for electrons in the turbulence
systems.



\subsubsection{Accumulated Particle Energization: $k_\perp$ Dependence }\label{sec:accum_E}

\begin{figure*} \centering
{\hfill\hbox{\hfill \resizebox{2.65in}{!}{\includegraphics[scale=1.]{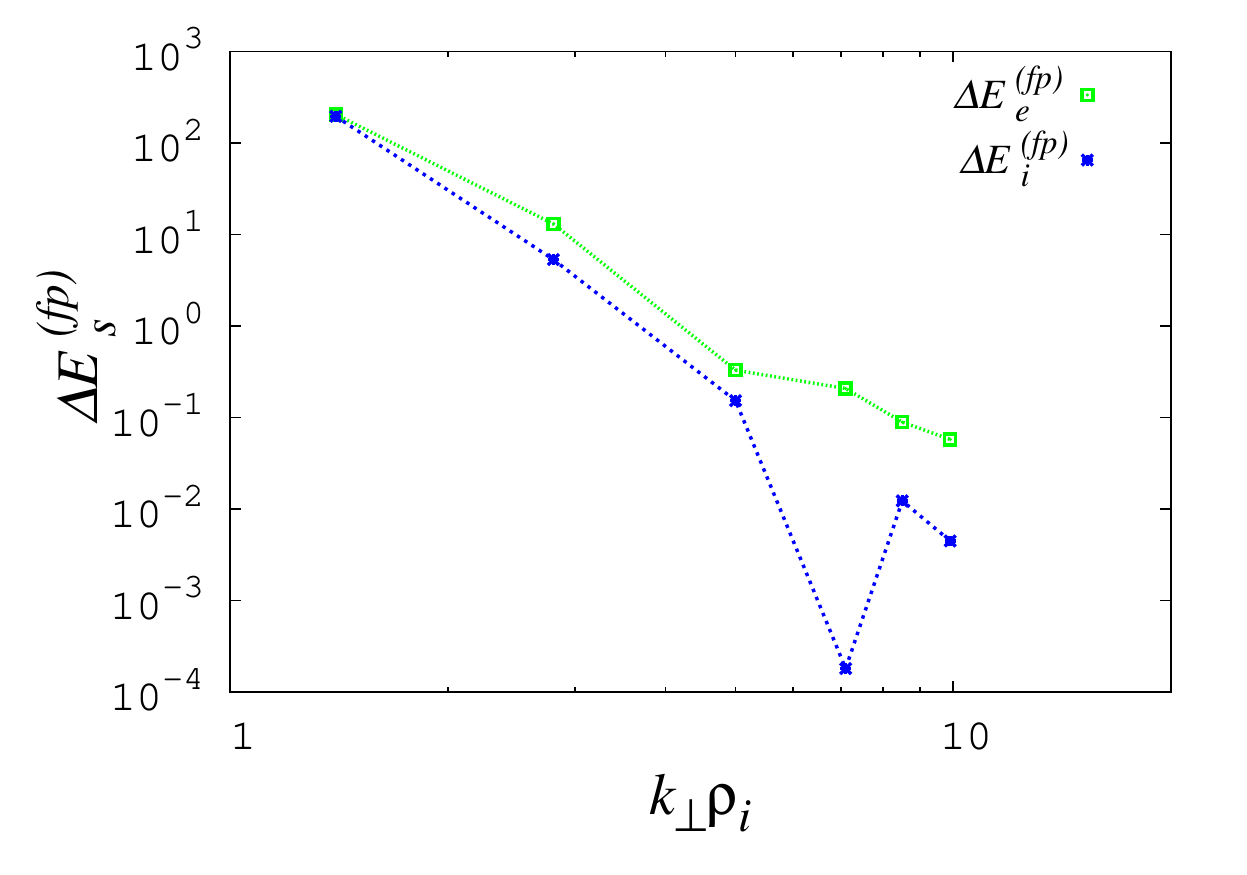}} \hfill  }\hfill}

\vspace{-.5em}
\caption{ \label{fig:tend} Accumulated particle energization $\Delta E^{(fp)}_s$
  at the end of the simulation for the sampled $(k_x,k_y)$ modes 
  (summed over $k_z$) as a function of $k_\perp\rho_i$. 
  Each $(k_x,k_y)$ mode is represented by a point and all points
  are connected by lines. The same normalization is used for both 
  ion $\Delta E^{(fp)}_i$ (blue) and electron $\Delta E^{(fp)}_e$
  (green) energization.  }
\end{figure*}

The collisionless field-particle energy transfer rate is directly measured by the field-particle correlations. With the sampled Fourier spectrum, we can see how the accumulated particle energization varies as a function of the length scales of the fluctuations. In \figref{fig:tend}, we plot this accumulated particle energization at the end of the simulation $\Delta E^{(fp)}_s$ as a function of $k_\perp\rho_i$. Each point represents the accumulated energization in each $(k_x,k_y)$ Fourier mode summed over $k_z$ for ions (blue) and electrons (green). The electron energization $\Delta E^{(fp)}_e$ is then the value of the green curve at the end of the simulation for each $(k_x,k_y)$ mode in \figref{fig:ft3_1_Sum}. The same normalization is used for both ion and electron energization, $\Delta E^{(fp)}_i$ and $\Delta E^{(fp)}_e$.


Several features of $\Delta E^{(fp)}_s$ are notable. First, the particle energization is dominated by the lowest $k_\perp$ mode that has the strongest amplitude, as expected. Second, the total ion and electron energization over the sampled spectrum are comparable given that both species are nearly equally energized at this dominant, lowest $k_\perp$ mode. This is consistent with the comparable species heating rates due to field-particle interactions $\dot{E}^{(fp)}_s$ (solid) in \figref{fig:ft3_dedt} in the turbulence system\footnote{Note 
that in order to use the field-particle correlations to quantitatively measure the total species heating (that is approximately twice larger for the 
electrons than for the ions), energization from the whole system 
and hence all possible Fourier modes needs to be taken into account.}. 
This is also consistent with the ion and electron collisionless damping rates, $\gamma_i$ and $\gamma_e$, being on the same order of magnitude at $k_\perp\rho_i\sim$ 1, represented by the lowest $k_\perp$ mode, in \figref{fig:1om_g}. Third, when plotted as as function of $k_\perp\rho_i$, $\Delta E^{(fp)}_s$ here can represent the energy transfer spectra; for electrons, $\Delta E^{(fp)}_e$ indicates a spectral rollover at the electron gyroradius scale of $k_\perp\rho_e$ = 1 ($k_\perp\rho_i$ = 5), becoming less steep at $k_\perp\rho_e >$ 1. Lastly, ion energization shows a dip at $k_\perp\rho_i$ = 7, rather than a monotonic drop for all $k_\perp\rho_i\geq$ 7 modes as would be expected from the ion collisionless damping rate, suggesting that collisionless damping cannot completely explain ion energization. These interesting features deserve further investigation in the future.  


\section{Conclusions}\label{sec:diss_conc}

The field-particle correlation technique is a new method both for
determining the rate of transfer of energy between electromagnetic
fields and plasma particles and for identifying the physical
mechanisms responsible for that energy transfer through their
velocity-space signatures.  Previous implementations of this novel
technique employ a time series of the particle velocity distribution
and electromagnetic field measurements performed at a single point in
space.  Here, we present an alternative Fourier implementation of the
field-particle correlation technique that determines the energy
transfer for a single wavevector in Fourier space instead of at a single
point in physical space.

The Fourier implementation has the capability to explore
 the energy transfer between fields and particles as a function of
the length scale of the fluctuations, information that cannot be
obtained through single-point measurements in physical space.
Furthermore, in a broadband
turbulent spectrum, the resonant
parallel phase velocity $\omega/k_\parallel$ is generally a function
of the wavevector, resonant collisionless interactions analyzed in
terms of Fourier wavevector modes are expected to yield a clean
velocity-space signature at the resonant velocity. This contrasts the
single-point implementation in which all wave modes will contribute to
the energy transfer at a given point, potentially smearing out 
the velocity-space signals over the range of resonant velocities for all
contributing wave modes.  Furthermore, the Fourier decomposition
separates the contributions to the energy transfer from each scale,
enabling the energy transfer for small amplitude modes at high
wavenumber to be observed even in the presence of much larger
amplitude modes at lower wavenumbers. This contrasts with 
the single-point implementation in which the much larger
amplitude modes at lower wavenumbers will dominate the energy transfer. 
Finally, in the Fourier
implementation, the correlation intervals need only extend over a few
periods of the mode under investigation, even in the presence of
larger amplitude, lower frequency modes.

The Fourier implementation employs information throughout
the spatial domain to decompose fluctuations as a function of scale,
so it cannot be applied to analyze
spacecraft observations that provide measurements at only one, or
possibly a few, points in space.
Nonetheless, it can be used in numerical simulations in which 
full spatial information is accessible. The key advantage of the Fourier implementation to investigate the particle energization as a function of the
length scale of the electromagnetic fluctuations provides insights 
that are complementary to 
the spatial information of particle energization provided by the
more standard, single-point implementation. 
In fact, dual implementation of the field-particle correlation technique
in both physical and Fourier space in 
simulations can take the advantages of both approaches in 
identifying collisionless energy transfer in simulations  
supporting spacecraft missions.

Here we apply the Fourier implementation of the field-particle
correlation technique to investigate the energization of the ions and
electrons in strong electromagnetic turbulence.
We simulate a 3D extension of the standard Orzsag-Tang
Vortex (3D OTV) problem, a setup previously used to explore the
differences between 2D and 3D turbulence \citep{Li:2016}. We follow
the flow of energy in the simulation from turbulent energy that is
first collisionlessly transferred to particles as non-thermal energy in the
velocity distribution functions of the ions and electrons, and later
is collisionally thermalized to ion and electron heat. For the
$\beta_i=1$ conditions of the simulation, 
we find that the electrons
are heated about twice as much as the ions.  We show that the particle
energization in the simulation, equal to $\V{J} \cdot \V{E}$
integrated over the simulation volume, occurs in a spatially
nonuniform manner, with the dominant heating confined to
narrow current sheet layers associated with strong parallel electric fields.

Applying the Fourier implementation of the field-particle correlation
technique to the simulation, we find that the velocity-space signature
of electron and ion energization for a particular Fourier mode is
consistent with Landau damping at the resonant parallel phase velocity
of the kinetic \Alfven wave for that mode.  Timestack plots of the
electron and ion energization by the parallel electric field
$E_\parallel$ also show a net energization of the plasma particles at
the expense of the turbulent energy.

The regions of velocity space in
which particles exchange energy with the electric field
closely follow the resonant parallel phase velocity for kinetic
\Alfven waves as the perpendicular wavenumber $k_\perp$ and plasma $\beta_i$ are
varied (\figref{fig:ft3_1_Sum_zoom}). This shows that collisionless damping via
the Landau resonance with ions and electrons is an important channel
of particle energization in strong electromagnetic turbulence,
relevant to space plasmas, such as the solar corona, solar wind, and
planetary magnetospheres. \\

\textbf{Acknowledgements}\\

TCL is grateful for Tulasi Parashar, Nuno Loureiro, 
Kevin Schoeffler and Alfred Mallet for valuable discussions. 
Supported by NSF CAREER Award AGS-1054061, 
NASA grants 80NSSC18K0754, 80NSSC18K0289, 80NSSC18K0643, 80NSSC18K1217, and
80NSSC18K1371, NASA HSR grant NNX16AM23G, and NSF SHINE award AGS-1622306. 
Computations used the Extreme Science and Engineering
Discovery Environment (XSEDE), which is supported by NSF grant
ACI-1053575.


\appendix
\section{Determination and Significance of Correlation Interval}\label{appA}


\begin{figure} \centering
\hbox{{(a)\hspace{-1.3em}} \hfill \resizebox{2.6in}{!}{\includegraphics*{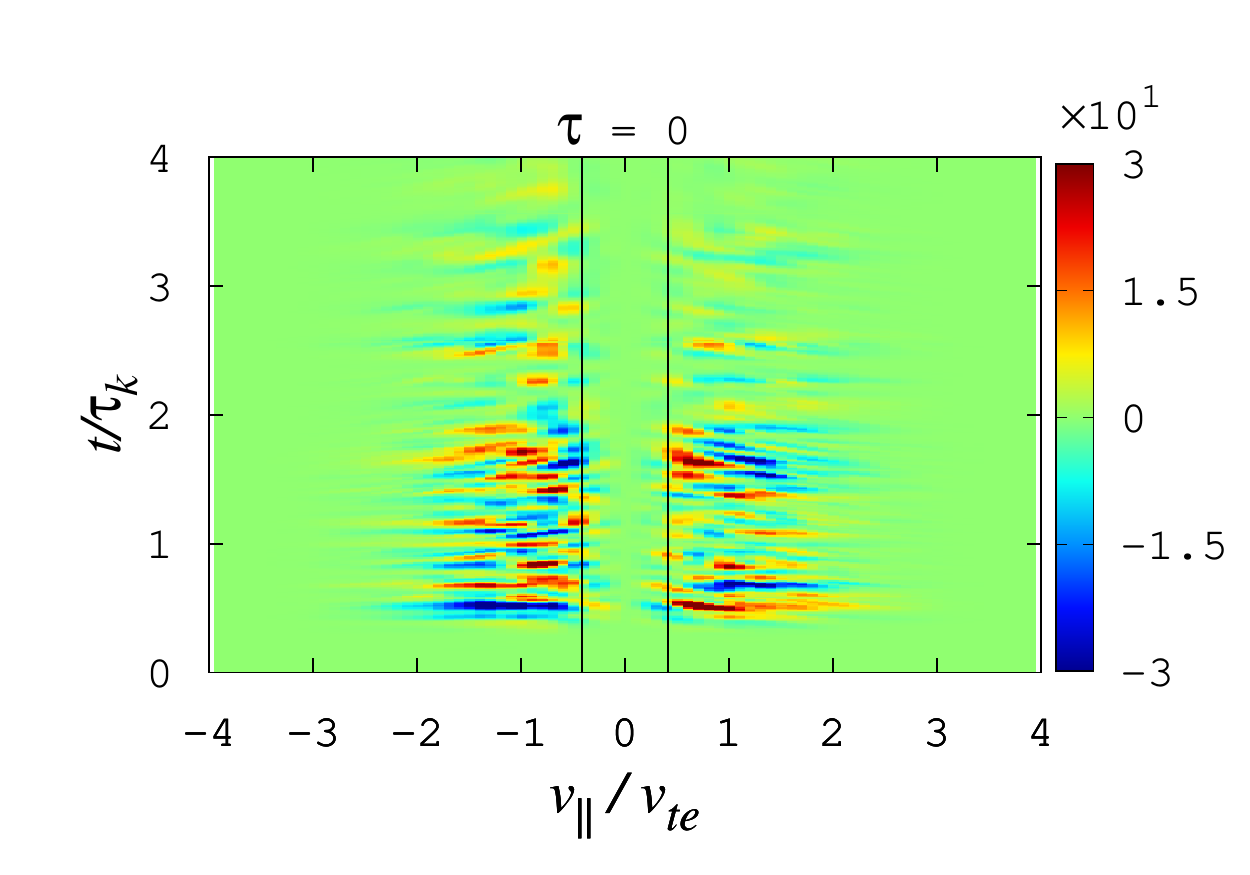}} \hfill  
{(b)\hspace{-1.3em}} \hfill \resizebox{2.6in}{!}{\includegraphics*{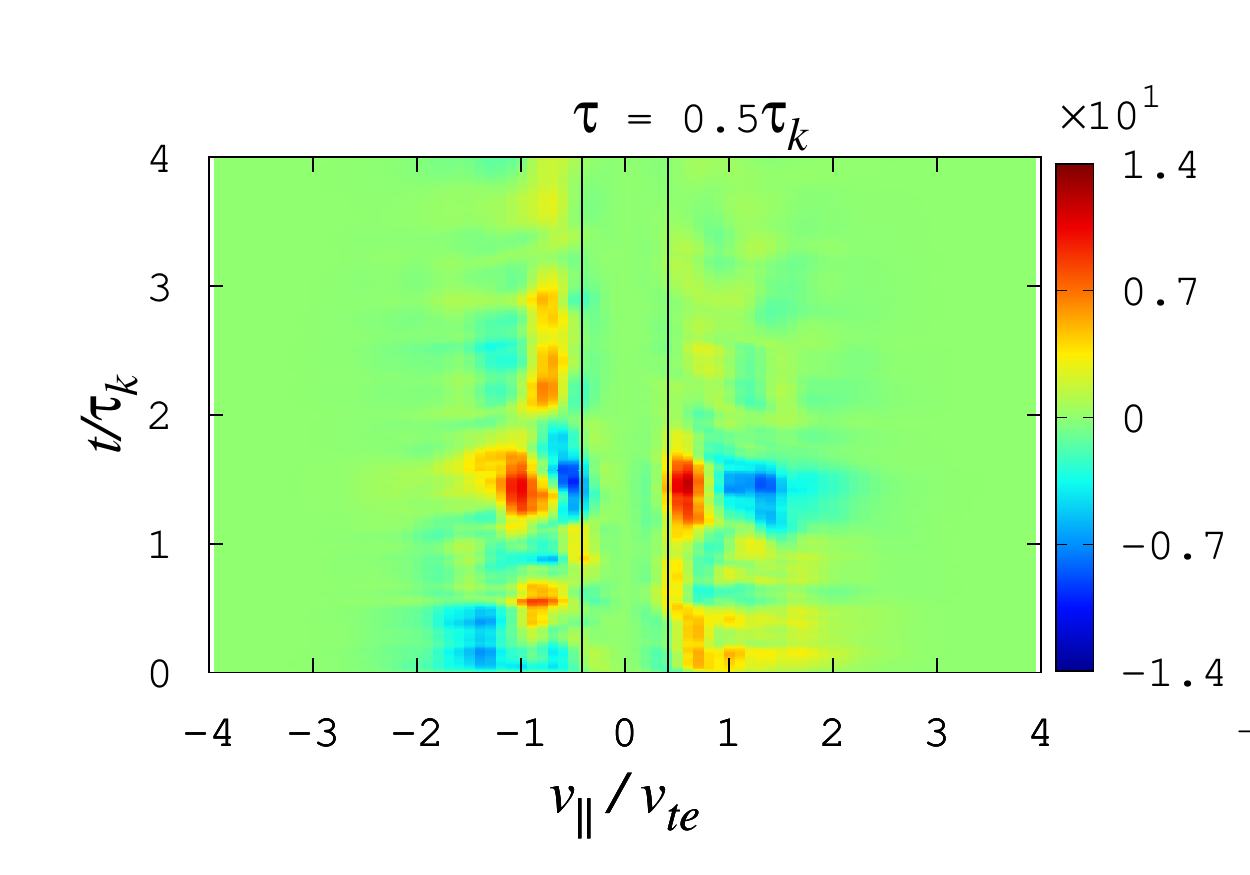}} \hfill  
  } 
\hbox{{(c)\hspace{-1.3em}} \hfill \resizebox{2.6in}{!}{\includegraphics*{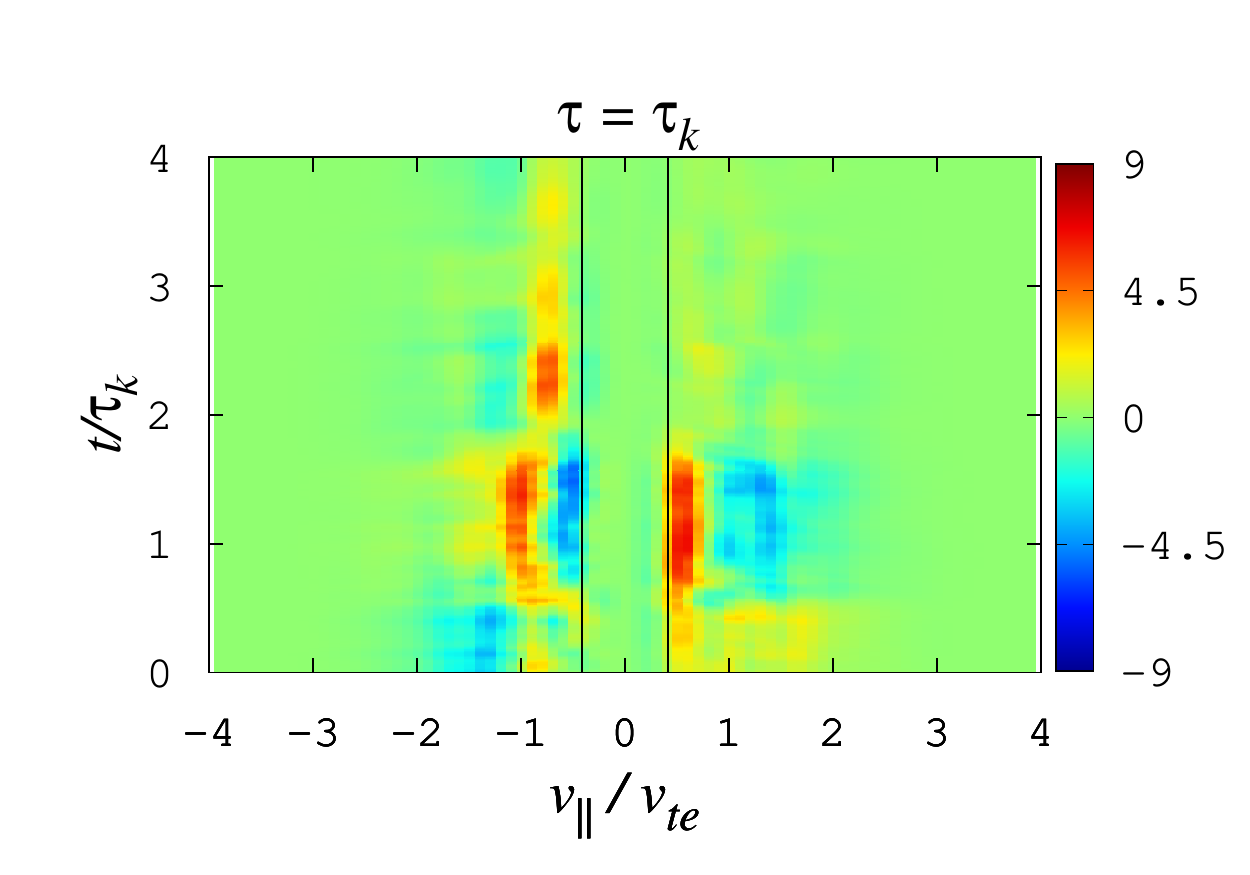}} \hfill  
{(d)\hspace{-1.3em}} \hfill \resizebox{2.6in}{!}{\includegraphics*{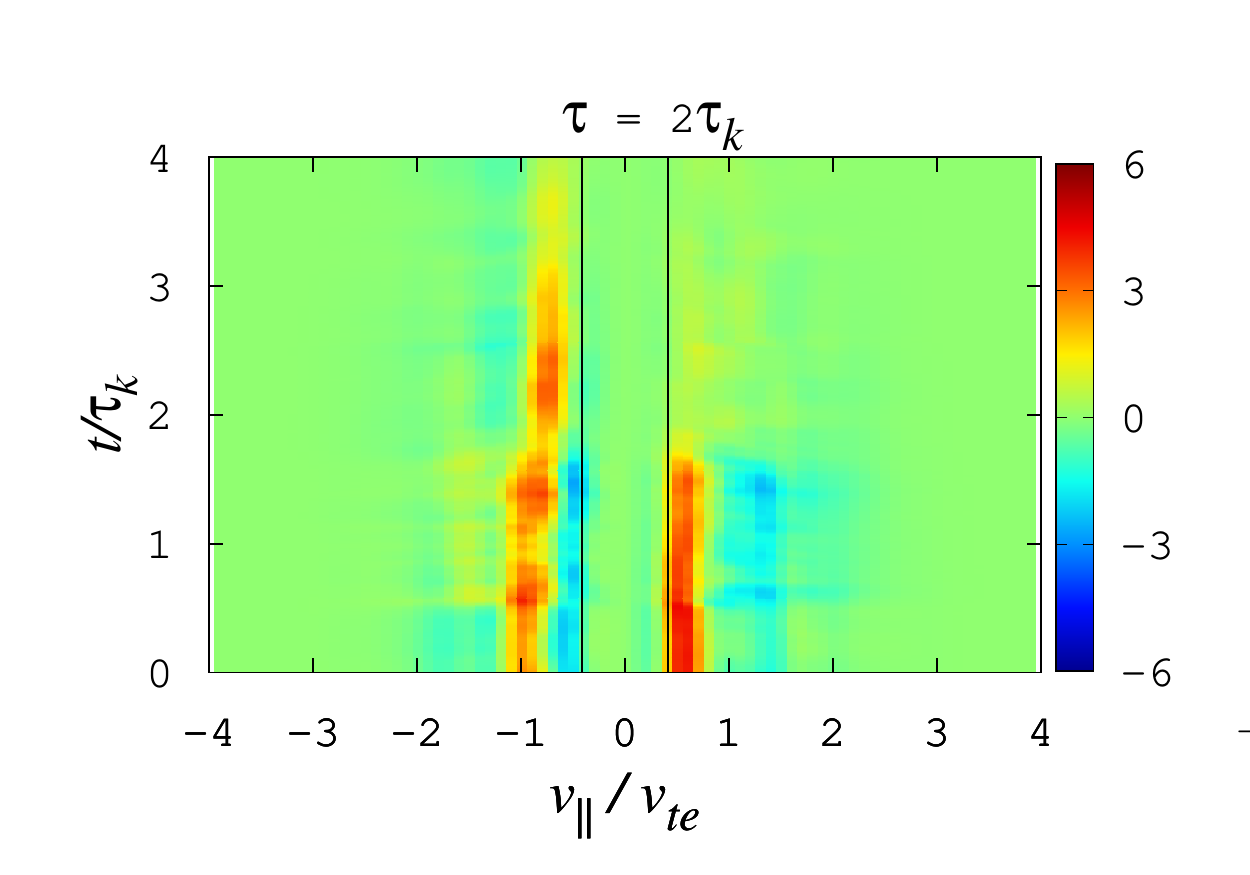}} \hfill  
  } 
\caption{ \label{fig:ft3_1k22m1_0_1p_F3dfn3} Electron parallel reduced
  correlations $C_{E_\parallel}(v_\parallel,t)$ using a correlation interval (a)
  $\tau$ = 0, (b) $\tau$ = 0.5$\tau_k$, (c) $\tau$ = $\tau_k$ and 
  (d) $\tau$ = 2$\tau_k$ for the $(2,2,-1)$ Fourier
  mode. The Landau resonant velocities:
  $v_{p \parallel}/v_{te}$ = $\pm$0.42, are indicated by vertical 
  black lines. Time $t$ is defined at the beginning of the correlation 
  interval.    }
\end{figure} 

Here we illustrate how a sufficiently long correlation interval is determined and its significance using electron parallel reduced correlations, which conveniently depict the energy transfer rate as a function of $v_\parallel$ and time $t$. Plotted in \figref{fig:ft3_1k22m1_0_1p_F3dfn3} are electron parallel reduced correlations for a correlation interval $\tau$ = 0, 0.5$\tau_k$, $\tau_k$ and 2$\tau_k$ for the $(2,2,-1)$ Fourier mode in (a)--(d). 

Using $\tau$ = 0, (a) the instantaneous energy transfer between $E_\parallel$ and electrons manifests as alternating red and blue signals over short time scales as they exchange energy back and forth many times throughout the course of the simulation. As $\tau$ increases in (b) through (d), the time evolution of the energy transfer becomes increasing concentrated to $v_{p \parallel}$. 
In (c) and (d), the time evolution of the energy transfer begins to be qualitatively consistent, with mostly red signals just above $v_{p \parallel}$. The maximum amplitude of energy transfer for $\tau=\tau_k$ and 2$\tau_k$ in (c) and (d) becomes comparable, whereas it is reduced by over twice from (a) $\tau$ = 0 to (b) $\tau= 0.5\tau_k$. This indicates that a correlation interval of $\tau\geq\tau_k$ leads to a qualitative convergence of the time evolution of the energy transfer rate as a function of $v_\parallel$. Hence, $\tau\geq\tau_k$ represents a sufficiently long correlation interval. The choice of $\tau= 2\tau_k$ used in this work is suitable for capturing the net secular energy transfer signals in velocity space.

An important feature again illustrated in \figref{fig:ft3_1k22m1_0_1p_F3dfn3} is that while electrons over a broad range of $v_\parallel$ participate in (a) the instantaneous energy transfer, the secular net energy transfer identified by correlating over a suitably long time period in (c) and (d) is dominated by "near-resonance" electrons that are close to the parallel resonant velocity $v_{p \parallel}$.




\begin{thebibliography}{88}
\expandafter\ifx\csname natexlab\endcsname\relax\def\natexlab#1{#1}\fi
\def\au#1{#1} \def\ed#1{#1} \def\yr#1{#1}\def\at#1{#1}\def\jt#1{\textit{#1}}
  \def\bt#1{#1}\def\bvol#1{\textbf{#1}} \def\vol#1{#1} \def\pg#1{#1}
  \def\publ#1{#1}\def\arxiv#1{#1}\def\org#1{#1}\def\st#1{\textit{#1}}

\bibitem[{Abel} {\em et~al.\/}(2008){Abel}, {Barnes}, {Cowley}, {Dorland} \&
  {Schekochihin}]{Abel:2008}
{\sc \au{{Abel}, I.~G.}, \au{{Barnes}, M.}, \au{{Cowley}, S.~C.},
  \au{{Dorland}, W.} \& \au{{Schekochihin}, A.~A.}} \yr{2008}  \at{{Linearized
  model Fokker-Planck collision operators for gyrokinetic simulations. I.
  Theory}}.  \jt{Phys.~Plasmas}  \bvol{15}~(12),  \pg{122509--+},
  \arxiv{arXiv: 0808.1300}.

\bibitem[Adkins \& Schekochihin(2018)]{Adkins:2018}
{\sc \au{Adkins, T.} \& \au{Schekochihin, A.~A.}} \yr{2018}  \at{A solvable
  model of vlasov-kinetic plasma turbulence in fourier–hermite phase space}.
  \jt{Journal of Plasma Physics}  \bvol{84}~(1),  \pg{905840107}.

\bibitem[{Alexandrova} {\em et~al.\/}(2013){Alexandrova}, {Chen},
  {Sorriso-Valvo}, {Horbury} \& {Bale}]{Alexandrova:2013b}
{\sc \au{{Alexandrova}, O.}, \au{{Chen}, C.~H.~K.}, \au{{Sorriso-Valvo}, L.},
  \au{{Horbury}, T.~S.} \& \au{{Bale}, S.~D.}} \yr{2013}  \at{{Solar Wind
  Turbulence and the Role of Ion Instabilities}}.  \jt{Space Sci.~Rev.}
  \bvol{178},  \pg{101--139},  \arxiv{arXiv: 1306.5336}.

\bibitem[{Arzamasskiy} {\em et~al.\/}(2019){Arzamasskiy}, {Kunz}, {Chandran} \&
  {Quataert}]{Arzamasskiy:2019}
{\sc \au{{Arzamasskiy}, L.}, \au{{Kunz}, M.~W.}, \au{{Chandran}, B.~D.~G.} \&
  \au{{Quataert}, E.}} \yr{2019}  \at{{Hybrid-kinetic Simulations of Ion
  Heating in Alfv{\'e}nic Turbulence}}.  \jt{Astrophys. J.}  \bvol{879},
  \pg{53},  \arxiv{arXiv: 1901.11028}.

\bibitem[{Barnes}(1966)]{Barnes:1966}
{\sc \au{{Barnes}, A.}} \yr{1966}  \at{{Collisionless Damping of Hydromagnetic
  Waves}}.  \jt{Phys.~Fluids}  \bvol{9},  \pg{1483--1495}.

\bibitem[{Barnes} {\em et~al.\/}(2009){Barnes}, {Abel}, {Dorland}, {Ernst},
  {Hammett}, {Ricci}, {Rogers}, {Schekochihin} \& {Tatsuno}]{Barnes:2009}
{\sc \au{{Barnes}, M.}, \au{{Abel}, I.~G.}, \au{{Dorland}, W.}, \au{{Ernst},
  D.~R.}, \au{{Hammett}, G.~W.}, \au{{Ricci}, P.}, \au{{Rogers}, B.~N.},
  \au{{Schekochihin}, A.~A.} \& \au{{Tatsuno}, T.}} \yr{2009}  \at{{Linearized
  model Fokker-Planck collision operators for gyrokinetic simulations. II.
  Numerical implementation and tests}}.  \jt{Phys.~Plasmas}  \bvol{16}~(7),
  \pg{072107--+}.

\bibitem[{Boldyrev} {\em et~al.\/}(2013){Boldyrev}, {Horaites}, {Xia} \&
  {Perez}]{Boldyrev:2013}
{\sc \au{{Boldyrev}, S.}, \au{{Horaites}, K.}, \au{{Xia}, Q.} \& \au{{Perez},
  J.~C.}} \yr{2013}  \at{{Toward a Theory of Astrophysical Plasma Turbulence at
  Subproton Scales}}.  \jt{Astrophys.~J.}  \bvol{777},  \pg{41}.

\bibitem[{Brizard} \& {Hahm}(2007)]{Brizard:2007}
{\sc \au{{Brizard}, A.~J.} \& \au{{Hahm}, T.~S.}} \yr{2007}  \at{{Foundations
  of nonlinear gyrokinetic theory}}.  \jt{Rev. Mod. Phys.}  \bvol{79},
  \pg{421--468}.

\bibitem[Burch {\em et~al.\/}(2016)Burch, Moore, Torbert \& Giles]{Burch:2016}
{\sc \au{Burch, J.~L.}, \au{Moore, T.~E.}, \au{Torbert, R.~B.} \& \au{Giles,
  B.~L.}} \yr{2016}  \at{Magnetospheric multiscale overview and science
  objectives}.  \jt{Space Science Reviews}  \bvol{199}~(1),  \pg{5--21}.

\bibitem[Cerri {\em et~al.\/}(2018)Cerri, Kunz \& Califano]{Cerri:2018}
{\sc \au{Cerri, S.~S.}, \au{Kunz, M.~W.} \& \au{Califano, F.}} \yr{2018}
  \at{Dual phase-space cascades in 3d hybrid-vlasov–maxwell turbulence}.
  \jt{The Astrophysical Journal Letters}  \bvol{856}~(1),  \pg{L13}.

\bibitem[{Chandran} {\em et~al.\/}(2010){Chandran}, {Li}, {Rogers}, {Quataert}
  \& {Germaschewski}]{Chandran:2010a}
{\sc \au{{Chandran}, B.~D.~G.}, \au{{Li}, B.}, \au{{Rogers}, B.~N.},
  \au{{Quataert}, E.} \& \au{{Germaschewski}, K.}} \yr{2010}
  \at{{Perpendicular Ion Heating by Low-frequency Alfv{\'e}n-wave Turbulence in
  the Solar Wind}}.  \jt{Astrophys.~J.}  \bvol{720},  \pg{503--515}.

\bibitem[{Chang} {\em et~al.\/}(2014){Chang}, {Peter Gary} \&
  {Wang}]{Chang:2014}
{\sc \au{{Chang}, O.}, \au{{Peter Gary}, S.} \& \au{{Wang}, J.}} \yr{2014}
  \at{{Energy dissipation by whistler turbulence: Three-dimensional
  particle-in-cell simulations}}.  \jt{Phys.~Plasmas}  \bvol{21}~(5),
  \pg{052305}.

\bibitem[{Chen} {\em et~al.\/}(2013){Chen}, {Boldyrev}, {Xia} \&
  {Perez}]{Chen:2013b}
{\sc \au{{Chen}, C.~H.~K.}, \au{{Boldyrev}, S.}, \au{{Xia}, Q.} \& \au{{Perez},
  J.~C.}} \yr{2013}  \at{{Nature of Subproton Scale Turbulence in the Solar
  Wind}}.  \jt{Phys.~Rev.~Lett.}  \bvol{110}~(22),  \pg{225002},  \arxiv{arXiv:
  1305.2950}.

\bibitem[{Chen} {\em et~al.\/}(2019){Chen}, {Klein} \& {Howes}]{Chen:2019a}
{\sc \au{{Chen}, C.~H.~K.}, \au{{Klein}, K.~G.} \& \au{{Howes}, G.~G.}}
  \yr{2019}  \at{{Evidence for electron Landau damping in space plasma
  turbulence}}.  \jt{Nature Communications}  \bvol{10},  \pg{740},
  \arxiv{arXiv: 1902.05785}.

\bibitem[{Chen} {\em et~al.\/}(2001){Chen}, {Lin} \& {White}]{Chen:2001}
{\sc \au{{Chen}, L.}, \au{{Lin}, Z.} \& \au{{White}, R.}} \yr{2001}  \at{{On
  resonant heating below the cyclotron frequency}}.  \jt{Phys.~Plasmas}
  \bvol{8},  \pg{4713--4716}.

\bibitem[{Coleman}(1968)]{Coleman:1968}
{\sc \au{{Coleman}, Jr., P.~J.}} \yr{1968}  \at{{Turbulence, Viscosity, and
  Dissipation in the Solar-Wind Plasma}}.  \jt{Astrophys.~J.}  \bvol{153},
  \pg{371--388}.

\bibitem[{Dahlburg} \& {Picone}(1989)]{Dahlburg:1989}
{\sc \au{{Dahlburg}, R.~B.} \& \au{{Picone}, J.~M.}} \yr{1989}  \at{{Evolution
  of the Orszag-Tang vortex system in a compressible medium. I - Initial
  average subsonic flow}}.  \jt{Phys.~Fluids B}  \bvol{1},  \pg{2153--2171}.

\bibitem[{Dahlin} {\em et~al.\/}(2016){Dahlin}, {Drake} \&
  {Swisdak}]{Dahlin:16}
{\sc \au{{Dahlin}, J.~T.}, \au{{Drake}, J.~F.} \& \au{{Swisdak}, M.}} \yr{2016}
   \at{{Parallel electric fields are inefficient drivers of energetic electrons
  in magnetic reconnection}}.  \jt{Physics of Plasmas}  \bvol{23}~(12),
  \pg{120704}.

\bibitem[Franci {\em et~al.\/}(2015)Franci, Landi, Matteini, Verdini \&
  Hellinger]{Franci:2015b}
{\sc \au{Franci, L.}, \au{Landi, S.}, \au{Matteini, L.}, \au{Verdini, A.} \&
  \au{Hellinger, P.}} \yr{2015}  \at{High-resolution hybrid simulations of
  kinetic plasma turbulence at proton scales}.  \jt{The Astrophysical Journal}
  \bvol{812}~(1),  \pg{21}.

\bibitem[{Frieman} \& {Chen}(1982)]{Frieman:1982}
{\sc \au{{Frieman}, E.~A.} \& \au{{Chen}, L.}} \yr{1982}  \at{{Nonlinear
  gyrokinetic equations for low-frequency electromagnetic waves in general
  plasma equilibria}}.  \jt{Phys.~Fluids}  \bvol{25},  \pg{502--508}.

\bibitem[{Gershman} {\em et~al.\/}(2017){Gershman}, {F-Vi{\~n}as}, {Dorelli},
  {Boardsen}, {Avanov}, {Bellan}, {Schwartz}, {Lavraud}, {Coffey}, {Chandler},
  {Saito}, {Paterson}, {Fuselier}, {Ergun}, {Strangeway}, {Russell}, {Giles},
  {Pollock}, {Torbert} \& {Burch}]{Gershman:2017}
{\sc \au{{Gershman}, D.~J.}, \au{{F-Vi{\~n}as}, A.}, \au{{Dorelli}, J.~C.},
  \au{{Boardsen}, S.~A.}, \au{{Avanov}, L.~A.}, \au{{Bellan}, P.~M.},
  \au{{Schwartz}, S.~J.}, \au{{Lavraud}, B.}, \au{{Coffey}, V.~N.},
  \au{{Chandler}, M.~O.}, \au{{Saito}, Y.}, \au{{Paterson}, W.~R.},
  \au{{Fuselier}, S.~A.}, \au{{Ergun}, R.~E.}, \au{{Strangeway}, R.~J.},
  \au{{Russell}, C.~T.}, \au{{Giles}, B.~L.}, \au{{Pollock}, C.~J.},
  \au{{Torbert}, R.~B.} \& \au{{Burch}, J.~L.}} \yr{2017}  \at{{Wave-particle
  energy exchange directly observed in a kinetic Alfv{\'e}n-branch wave}}.
  \jt{Nature Communications}  \bvol{8},  \pg{14719}.

\bibitem[Goldreich \& Sridhar(1995)]{Goldreich:1995}
{\sc \au{Goldreich, P.} \& \au{Sridhar, S.}} \yr{1995}  \at{{Toward a Theery of
  Interstellar Turbulence II. Strong Alfv\'enic Turbulence}}.
  \jt{Astrophys.~J.}  \bvol{438},  \pg{763--775}.

\bibitem[{Grauer} \& {Marliani}(2000)]{Grauer:2000}
{\sc \au{{Grauer}, R.} \& \au{{Marliani}, C.}} \yr{2000}  \at{{Current-Sheet
  Formation in 3D Ideal Incompressible Magnetohydrodynamics}}.
  \jt{Phys.~Rev.~Lett.}  \bvol{84},  \pg{4850}.

\bibitem[{Gro\'selj} {\em et~al.\/}(2017){Gro\'selj}, Cerri, Navarro, Willmott,
  Told, Loureiro, Califano \& Jenko]{Grošelj:2017}
{\sc \au{{Gro\'selj}, D.}, \au{Cerri, S.~S.}, \au{Navarro, A.~B.},
  \au{Willmott, C.}, \au{Told, D.}, \au{Loureiro, N.~F.}, \au{Califano, F.} \&
  \au{Jenko, F.}} \yr{2017}  \at{Fully kinetic versus reduced-kinetic modeling
  of collisionless plasma turbulence}.  \jt{The Astrophysical Journal}
  \bvol{847}~(1),  \pg{28}.

\bibitem[{Gro\'selj} {\em et~al.\/}(2018){Gro\'selj}, {Chen}, {Mallet},
  {Samtaney}, {Schneider} \& {Jenko}]{Grošelj:2018}
{\sc \au{{Gro\'selj}, D.}, \au{{Chen}, C.~H.~K.}, \au{{Mallet}, A.},
  \au{{Samtaney}, R.}, \au{{Schneider}, K.} \& \au{{Jenko}, F.}} \yr{2018}
  \at{{Kinetic Turbulence in Astrophysical Plasmas: Waves and/or Structures?}}
  \jt{ArXiv e-prints} ,  \arxiv{arXiv: 1806.05741}.

\bibitem[{He} {\em et~al.\/}(2012){He}, {Tu}, {Marsch} \& {Yao}]{He:2012}
{\sc \au{{He}, J.}, \au{{Tu}, C.}, \au{{Marsch}, E.} \& \au{{Yao}, S.}}
  \yr{2012}  \at{{Do Oblique Alfv{\'e}n/Ion-cyclotron or Fast-mode/Whistler
  Waves Dominate the Dissipation of Solar Wind Turbulence near the Proton
  Inertial Length?}}  \jt{Astrophys.~J.~Lett.}  \bvol{745},  \pg{L8}.

\bibitem[{Howes}(2008)]{Howes:2008c}
{\sc \au{{Howes}, G.~G.}} \yr{2008}  \at{{Inertial range turbulence in kinetic
  plasmas}}.  \jt{Phys.~Plasmas}  \bvol{15}~(5),  \pg{055904}.

\bibitem[Howes(2015)]{Howes:2015b}
{\sc \au{Howes, G.~G.}} \yr{2015}  \at{A dynamical model of plasma turbulence
  in the solar wind}.  \jt{Philosophical Transactions of the Royal Society of
  London A: Mathematical, Physical and Engineering Sciences}
  \bvol{373}~(2041),  \pg{20140145}.

\bibitem[{Howes}(2016)]{Howes:2016b}
{\sc \au{{Howes}, G.~G.}} \yr{2016}  \at{{The Dynamical Generation of Current
  Sheets in Astrophysical Plasma Turbulence}}.  \jt{Astrophys.~J.~Lett.}
  \bvol{82},  \pg{L28},  \arxiv{arXiv: 1607.07465}.

\bibitem[Howes(2017)]{Howes:2017b}
{\sc \au{Howes, G.~G.}} \yr{2017}  \at{A prospectus on kinetic heliophysics}.
  \jt{Physics of Plasmas}  \bvol{24}~(5),  \pg{055907},  \arxiv{arXiv:
  https://doi.org/10.1063/1.4983993}.

\bibitem[{Howes}(2017)]{Howes:2017c}
{\sc \au{{Howes}, G.~G.}} \yr{2017}  \at{A prospectus on kinetic heliophysics}.
   \jt{Phys.~Plasmas}  \bvol{24}~(5),  \pg{055907}.

\bibitem[{Howes} {\em et~al.\/}(2006){Howes}, {Cowley}, {Dorland}, {Hammett},
  {Quataert} \& {Schekochihin}]{Howes:2006}
{\sc \au{{Howes}, G.~G.}, \au{{Cowley}, S.~C.}, \au{{Dorland}, W.},
  \au{{Hammett}, G.~W.}, \au{{Quataert}, E.} \& \au{{Schekochihin}, A.~A.}}
  \yr{2006}  \at{{Astrophysical Gyrokinetics: Basic Equations and Linear
  Theory}}.  \jt{Astrophys.~J.}  \bvol{651},  \pg{590--614},  \arxiv{arXiv:
  astro-ph/0511812}.

\bibitem[{Howes} {\em et~al.\/}(2008{\natexlab{{\em a\/}}}){Howes}, {Cowley},
  {Dorland}, {Hammett}, {Quataert} \& {Schekochihin}]{Howes:2008b}
{\sc \au{{Howes}, G.~G.}, \au{{Cowley}, S.~C.}, \au{{Dorland}, W.},
  \au{{Hammett}, G.~W.}, \au{{Quataert}, E.} \& \au{{Schekochihin}, A.~A.}}
  \yr{2008{\natexlab{{\em a\/}}}}  \at{{A model of turbulence in magnetized
  plasmas: Implications for the dissipation range in the solar wind}}.
  \jt{J.~Geophys.~Res.}  \bvol{113}~(A12),  \pg{A05103},  \arxiv{arXiv:
  arXiv:0707.3147}.

\bibitem[{Howes} {\em et~al.\/}(2008{\natexlab{{\em b\/}}}){Howes}, {Dorland},
  {Cowley}, {Hammett}, {Quataert}, {Schekochihin} \& {Tatsuno}]{Howes:2008a}
{\sc \au{{Howes}, G.~G.}, \au{{Dorland}, W.}, \au{{Cowley}, S.~C.},
  \au{{Hammett}, G.~W.}, \au{{Quataert}, E.}, \au{{Schekochihin}, A.~A.} \&
  \au{{Tatsuno}, T.}} \yr{2008{\natexlab{{\em b\/}}}}  \at{{Kinetic Simulations
  of Magnetized Turbulence in Astrophysical Plasmas}}.  \jt{Phys.~Rev.~Lett.}
  \bvol{100}~(6),  \pg{065004}.

\bibitem[Howes {\em et~al.\/}(2017)Howes, Klein \& Li]{Howes:2017}
{\sc \au{Howes, G.~G.}, \au{Klein, K.~G.} \& \au{Li, T.~C.}} \yr{2017}
  \at{Diagnosing collisionless energy transfer using field–particle
  correlations: Vlasov–poisson plasmas}.  \jt{Journal of Plasma Physics}
  \bvol{83}~(1),  \pg{705830102}.

\bibitem[Howes {\em et~al.\/}(2018)Howes, McCubbin \& Klein]{Howes:2018a}
{\sc \au{Howes, G.~G.}, \au{McCubbin, A.~J.} \& \au{Klein, K.~G.}} \yr{2018}
  \at{Spatially localized particle energization by landau damping in current
  sheets produced by strong alfvén wave collisions}.  \jt{Journal of Plasma
  Physics}  \bvol{84}~(1),  \pg{905840105}.

\bibitem[Howes {\em et~al.\/}(2011)Howes, TenBarge, Dorland, Quataert,
  Schekochihin, Numata \& Tatsuno]{Howes:2011a}
{\sc \au{Howes, G.~G.}, \au{TenBarge, J.~M.}, \au{Dorland, W.}, \au{Quataert,
  E.}, \au{Schekochihin, A.~A.}, \au{Numata, R.} \& \au{Tatsuno, T.}} \yr{2011}
   \at{Gyrokinetic simulations of solar wind turbulence from ion to electron
  scales}.  \jt{Phys.~Rev.~Lett.}  \bvol{107},  \pg{035004}.

\bibitem[Hughes {\em et~al.\/}(2017)Hughes, Gary, Wang \&
  Parashar]{Hughes:2017}
{\sc \au{Hughes, R.~S.}, \au{Gary, S.~P.}, \au{Wang, J.} \& \au{Parashar,
  T.~N.}} \yr{2017}  \at{Kinetic alfvén turbulence: Electron and ion heating
  by particle-in-cell simulations}.  \jt{The Astrophysical Journal Letters}
  \bvol{847}~(2),  \pg{L14}.

\bibitem[{Isenberg} \& {Hollweg}(1983)]{Isenberg:1983}
{\sc \au{{Isenberg}, P.~A.} \& \au{{Hollweg}, J.~V.}} \yr{1983}  \at{{On the
  preferential acceleration and heating of solar wind heavy ions}}.   \textit{J. Geophys. Res.}
  \bvol{88},  \pg{3923--3935}.

\bibitem[{Karimabadi} {\em et~al.\/}(2013){Karimabadi}, {Roytershteyn}, {Wan},
  {Matthaeus}, {Daughton}, {Wu}, {Shay}, {Loring}, {Borovsky}, {Leonardis},
  {Chapman} \& {Nakamura}]{Karimabadi:2013}
{\sc \au{{Karimabadi}, H.}, \au{{Roytershteyn}, V.}, \au{{Wan}, M.},
  \au{{Matthaeus}, W.~H.}, \au{{Daughton}, W.}, \au{{Wu}, P.}, \au{{Shay}, M.},
  \au{{Loring}, B.}, \au{{Borovsky}, J.}, \au{{Leonardis}, E.}, \au{{Chapman},
  S.~C.} \& \au{{Nakamura}, T.~K.~M.}} \yr{2013}  \at{{Coherent structures,
  intermittent turbulence, and dissipation in high-temperature plasmas}}.
  \jt{Phys.~Plasmas}  \bvol{20}~(1),  \pg{012303}.

\bibitem[{Kawazura} {\em et~al.\/}(2019){Kawazura}, {Barnes} \&
  {Schekochihin}]{Kawazura:2018b}
{\sc \au{{Kawazura}, Y.}, \au{{Barnes}, M.} \& \au{{Schekochihin}, A.~A.}}
  \yr{2019}  \at{{Thermal disequilibration of ions and electrons by
  collisionless plasma turbulence}}.  \jt{Proceedings of the National Academy
  of Science}  \bvol{116},  \pg{771--776},  \arxiv{arXiv: 1807.07702}.

\bibitem[Klein(2017)]{Klein:2017}
{\sc \au{Klein, K.~G.}} \yr{2017}  \at{Characterizing fluid and kinetic
  instabilities using field-particle correlations on single-point time series}.
   \jt{Physics of Plasmas}  \bvol{24}~(5),  \pg{055901},  \arxiv{arXiv:
  https://doi.org/10.1063/1.4977465}.

\bibitem[Klein \& Howes(2016)]{Klein:2016}
{\sc \au{Klein, K.~G.} \& \au{Howes, G.~G.}} \yr{2016}  \at{Measuring
  collisionless damping in heliospheric plasmas using field–particle
  correlations}.  \jt{The Astrophysical Journal Letters}  \bvol{826}~(2),
  \pg{L30}.

\bibitem[Klein {\em et~al.\/}(2017)Klein, Howes \& TenBarge]{KleinHT:2017}
{\sc \au{Klein, K.~G.}, \au{Howes, G.~G.} \& \au{TenBarge, J.~M.}} \yr{2017}
  \at{Diagnosing collisionless energy transfer using field–particle
  correlations: gyrokinetic turbulence}.  \jt{Journal of Plasma Physics}
  \bvol{83}~(4),  \pg{535830401}.

\bibitem[Kobayashi {\em et~al.\/}(2014)Kobayashi, Rogers \&
  Numata]{Kobayashi:2014}
{\sc \au{Kobayashi, S.}, \au{Rogers, B.~N.} \& \au{Numata, R.}} \yr{2014}
  \at{Gyrokinetic simulations of collisionless reconnection in turbulent
  non-uniform plasmas}.  \jt{Physics of Plasmas}  \bvol{21}~(4),  \pg{040704},
  \arxiv{arXiv: https://doi.org/10.1063/1.4873703}.

\bibitem[{Kruskal} \& {Oberman}(1958)]{Kruskal:1958}
{\sc \au{{Kruskal}, M.~D.} \& \au{{Oberman}, C.~R.}} \yr{1958}  \at{{On the
  Stability of Plasma in Static Equilibrium}}.  \jt{Phys.~Fluids}  \bvol{1},
  \pg{275--280}.

\bibitem[Landau(1946)]{Landau:1946}
{\sc \au{Landau, L.~D.}} \yr{1946}  \at{On the vibrations of the electronic
  plasma}.  \jt{Zh. Eksp. Teor. Fiz.}  \bvol{16},  \pg{574}.

\bibitem[Li {\em et~al.\/}(2016)Li, Howes, Klein \& TenBarge]{Li:2016}
{\sc \au{Li, T.~C.}, \au{Howes, G.~G.}, \au{Klein, K.~G.} \& \au{TenBarge,
  J.~M.}} \yr{2016}  \at{Energy dissipation and landau damping in two- and
  three-dimensional plasma turbulence}.  \jt{The Astrophysical Journal Letters}
   \bvol{832}~(2),  \pg{L24}.

\bibitem[{Mininni} {\em et~al.\/}(2006){Mininni}, {Pouquet} \&
  {Montgomery}]{Mininni:2006}
{\sc \au{{Mininni}, P.~D.}, \au{{Pouquet}, A.~G.} \& \au{{Montgomery}, D.~C.}}
  \yr{2006}  \at{{Small-Scale Structures in Three-Dimensional
  Magnetohydrodynamic Turbulence}}.  \jt{Phys.~Rev.~Lett.}  \bvol{97}~(24),
  \pg{244503},  \arxiv{arXiv: physics/0607269}.

\bibitem[{Morrison}(1994)]{Morrison:1994}
{\sc \au{{Morrison}, P.~J.}} \yr{1994}  \at{{The energy of perturbations for
  Vlasov plasmas}}.  \jt{Phys.~Plasmas}  \bvol{1},  \pg{1447--1451}.

\bibitem[{Narita} {\em et~al.\/}(2011){Narita}, {Gary}, {Saito}, {Glassmeier}
  \& {Motschmann}]{Narita:2011}
{\sc \au{{Narita}, Y.}, \au{{Gary}, S.~P.}, \au{{Saito}, S.}, \au{{Glassmeier},
  K.-H.} \& \au{{Motschmann}, U.}} \yr{2011}  \at{{Dispersion relation analysis
  of solar wind turbulence}}.  \jt{Geophys.~Res.~Lett.}  \bvol{38},
  \pg{L05101}.

\bibitem[Narita {\em et~al.\/}(2016)Narita, Nakamura, Baumjohann, Glassmeier,
  Motschmann, Giles, Magnes, Fischer, Torbert, Russell, Strangeway, Burch,
  Nariyuki, Saito \& Gary]{Narita:2016}
{\sc \au{Narita, Y.}, \au{Nakamura, R.}, \au{Baumjohann, W.}, \au{Glassmeier,
  K.-H.}, \au{Motschmann, U.}, \au{Giles, B.}, \au{Magnes, W.}, \au{Fischer,
  D.}, \au{Torbert, R.~B.}, \au{Russell, C.~T.}, \au{Strangeway, R.~J.},
  \au{Burch, J.~L.}, \au{Nariyuki, Y.}, \au{Saito, S.} \& \au{Gary, S.~P.}}
  \yr{2016}  \at{On electron-scale whistler turbulence in the solar wind}.
  \jt{The Astrophysical Journal Letters}  \bvol{827}~(1),  \pg{L8}.

\bibitem[Navarro {\em et~al.\/}(2016)Navarro, Teaca, Told, Groselj, Crandall \&
  Jenko]{Navarro:2016}
{\sc \au{Navarro, A. B.~n.}, \au{Teaca, B.}, \au{Told, D.}, \au{Groselj, D.},
  \au{Crandall, P.} \& \au{Jenko, F.}} \yr{2016}  \at{Structure of plasma
  heating in gyrokinetic alfv\'enic turbulence}.  \jt{Phys. Rev. Lett.}
  \bvol{117},  \pg{245101}.

\bibitem[{Nielson} {\em et~al.\/}(2013){Nielson}, {Howes} \&
  {Dorland}]{Nielson:2013a}
{\sc \au{{Nielson}, K.~D.}, \au{{Howes}, G.~G.} \& \au{{Dorland}, W.}}
  \yr{2013}  \at{{Alfv{\'e}n wave collisions, the fundamental building block of
  plasma turbulence. II. Numerical solution}}.  \jt{Physics of Plasmas}
  \bvol{20}~(7),  \pg{072303},  \arxiv{arXiv: 1306.1456}.

\bibitem[{Numata} {\em et~al.\/}(2011){Numata}, {Dorland}, {Howes}, {Loureiro},
  {Rogers} \& {Tatsuno}]{Numata:2011}
{\sc \au{{Numata}, R.}, \au{{Dorland}, W.}, \au{{Howes}, G.~G.},
  \au{{Loureiro}, N.~F.}, \au{{Rogers}, B.~N.} \& \au{{Tatsuno}, T.}} \yr{2011}
   \at{{Gyrokinetic simulations of the tearing instability}}.  \jt{Physics of
  Plasmas}  \bvol{18}~(11),  \pg{112106},  \arxiv{arXiv: 1107.5842}.

\bibitem[{Numata} {\em et~al.\/}(2010){Numata}, {Howes}, {Tatsuno}, {Barnes} \&
  {Dorland}]{Numata:2010}
{\sc \au{{Numata}, R.}, \au{{Howes}, G.~G.}, \au{{Tatsuno}, T.}, \au{{Barnes},
  M.} \& \au{{Dorland}, W.}} \yr{2010}  \at{{AstroGK: Astrophysical
  gyrokinetics code}}.  \jt{J.~Comp.~Phys.}  \bvol{229},  \pg{9347},
  \arxiv{arXiv: 1004.0279}.

\bibitem[{Numata} \& {Loureiro}(2015)]{Numata:2015}
{\sc \au{{Numata}, R.} \& \au{{Loureiro}, N.~F.}} \yr{2015}  \at{{Ion and
  electron heating during magnetic reconnection in weakly collisional
  plasmas}}.  \jt{Journal of Plasma Physics}  \bvol{81},  \pg{3001},
  \arxiv{arXiv: 1406.6456}.

\bibitem[{Orszag} \& {Tang}(1979)]{Orszag:1979}
{\sc \au{{Orszag}, S.~A.} \& \au{{Tang}, C.-M.}} \yr{1979}  \at{{Small-scale
  structure of two-dimensional magnetohydrodynamic turbulence}}.  \jt{J.~Fluid
  Mech.}  \bvol{90},  \pg{129--143}.

\bibitem[Parashar \& Matthaeus(2016)]{Parashar:2016}
{\sc \au{Parashar, T.~N.} \& \au{Matthaeus, W.~H.}} \yr{2016}  \at{Propinquity
  of current and vortex structures: Effects on collisionless plasma heating}.
  \jt{The Astrophysical Journal}  \bvol{832}~(1),  \pg{57}.

\bibitem[{Parashar} {\em et~al.\/}(2009){Parashar}, {Shay}, {Cassak} \&
  {Matthaeus}]{Parashar:2009}
{\sc \au{{Parashar}, T.~N.}, \au{{Shay}, M.~A.}, \au{{Cassak}, P.~A.} \&
  \au{{Matthaeus}, W.~H.}} \yr{2009}  \at{{Kinetic dissipation and anisotropic
  heating in a turbulent collisionless plasma}}.  \jt{Phys.~Plasmas}
  \bvol{16}~(3),  \pg{032310--+}.

\bibitem[{Parashar} {\em et~al.\/}(2014){Parashar}, {Vasquez} \&
  {Markovskii}]{Parashar:2014b}
{\sc \au{{Parashar}, T.~N.}, \au{{Vasquez}, B.~J.} \& \au{{Markovskii}, S.~A.}}
  \yr{2014}  \at{{The role of electron equation of state in heating partition
  of protons in a collisionless plasma}}.  \jt{Physics of Plasmas}
  \bvol{21}~(2),  \pg{022301}.

\bibitem[Passot \& Sulem(2015)]{Passot:2015}
{\sc \au{Passot, T.} \& \au{Sulem, P.~L.}} \yr{2015}  \at{A model for the
  non-universal power law of the solar wind sub-ion-scale magnetic spectrum}.
  \jt{The Astrophysical Journal Letters}  \bvol{812}~(2),  \pg{L37}.

\bibitem[Perrone {\em et~al.\/}(2016)Perrone, Alexandrova, Mangeney,
  Maksimovic, Lacombe, Rakoto, Kasper \& Jovanovic]{Perrone:2016}
{\sc \au{Perrone, D.}, \au{Alexandrova, O.}, \au{Mangeney, A.}, \au{Maksimovic,
  M.}, \au{Lacombe, C.}, \au{Rakoto, V.}, \au{Kasper, J.~C.} \& \au{Jovanovic,
  D.}} \yr{2016}  \at{Compressive coherent structures at ion scales in the slow
  solar wind}.  \jt{The Astrophysical Journal}  \bvol{826}~(2),  \pg{196}.

\bibitem[{Picone} \& {Dahlburg}(1991)]{Picone:1991}
{\sc \au{{Picone}, J.~M.} \& \au{{Dahlburg}, R.~B.}} \yr{1991}  \at{{Evolution
  of the Orszag-Tang vortex system in a compressible medium. II - Supersonic
  flow}}.  \jt{Phys.~Fluids B}  \bvol{3},  \pg{29--44}.

\bibitem[{Politano} {\em et~al.\/}(1989){Politano}, {Pouquet} \&
  {Sulem}]{Politano:1989}
{\sc \au{{Politano}, H.}, \au{{Pouquet}, A.} \& \au{{Sulem}, P.~L.}} \yr{1989}
  \at{{Inertial ranges and resistive instabilities in two-dimensional
  magnetohydrodynamic turbulence}}.  \jt{Phys.~Fluids B}  \bvol{1},
  \pg{2330--2339}.

\bibitem[{Politano} {\em et~al.\/}(1995){Politano}, {Pouquet} \&
  {Sulem}]{Politano:1995b}
{\sc \au{{Politano}, H.}, \au{{Pouquet}, A.} \& \au{{Sulem}, P.~L.}} \yr{1995}
  \at{{Current and vorticity dynamics in three-dimensional magnetohydrodynamic
  turbulence}}.  \jt{Physics of Plasmas}  \bvol{2},  \pg{2931--2939}.

\bibitem[{Quataert}(1998)]{Quataert:1998}
{\sc \au{{Quataert}, E.}} \yr{1998}  \at{{Particle Heating by Alfv\'enic
  Turbulence in Hot Accretion Flows}}.  \jt{Astrophys.~J.}  \bvol{500},
  \pg{978--991},  \arxiv{arXiv: astro-ph/9710127}.

\bibitem[Roberts {\em et~al.\/}(2017)Roberts, Alexandrova, Kajdič, Turc,
  Perrone, Escoubet \& Walsh]{Roberts:2017Nov}
{\sc \au{Roberts, O.~W.}, \au{Alexandrova, O.}, \au{Kajdič, P.}, \au{Turc,
  L.}, \au{Perrone, D.}, \au{Escoubet, C.~P.} \& \au{Walsh, A.}} \yr{2017}
  \at{Variability of the magnetic field power spectrum in the solar wind at
  electron scales}.  \jt{The Astrophysical Journal}  \bvol{850}~(2),  \pg{120}.

\bibitem[{Roberts} {\em et~al.\/}(2015){Roberts}, {Li} \&
  {Jeska}]{Roberts:2015b}
{\sc \au{{Roberts}, O.~W.}, \au{{Li}, X.} \& \au{{Jeska}, L.}} \yr{2015}
  \at{{A Statistical Study of the Solar Wind Turbulence at Ion Kinetic Scales
  Using the k-filtering Technique and Cluster Data}}.  \jt{The Astrophysical Journal}  \bvol{802},
  \pg{2}.

\bibitem[{Roberts} {\em et~al.\/}(2013){Roberts}, {Li} \& {Li}]{Roberts:2013}
{\sc \au{{Roberts}, O.~W.}, \au{{Li}, X.} \& \au{{Li}, B.}} \yr{2013}
  \at{{Kinetic Plasma Turbulence in the Fast Solar Wind Measured by Cluster}}.
  \jt{Astrophys.~J.}  \bvol{769},  \pg{58},  \arxiv{arXiv: 1303.5129}.

\bibitem[{Sahraoui} {\em et~al.\/}(2010){Sahraoui}, {Goldstein}, {Belmont},
  {Canu} \& {Rezeau}]{Sahraoui:2010b}
{\sc \au{{Sahraoui}, F.}, \au{{Goldstein}, M.~L.}, \au{{Belmont}, G.},
  \au{{Canu}, P.} \& \au{{Rezeau}, L.}} \yr{2010}  \at{{Three Dimensional
  Anisotropic k Spectra of Turbulence at Subproton Scales in the Solar Wind}}.
  \jt{Phys.~Rev.~Lett.}  \bvol{105}~(13),  \pg{131101--+}.

\bibitem[{Schekochihin} {\em et~al.\/}(2008){Schekochihin}, {Cowley},
  {Dorland}, {Hammett}, {Howes}, {Plunk}, {Quataert} \&
  {Tatsuno}]{Schekochihin:2008}
{\sc \au{{Schekochihin}, A.~A.}, \au{{Cowley}, S.~C.}, \au{{Dorland}, W.},
  \au{{Hammett}, G.~W.}, \au{{Howes}, G.~G.}, \au{{Plunk}, G.~G.},
  \au{{Quataert}, E.} \& \au{{Tatsuno}, T.}} \yr{2008}  \at{{Gyrokinetic
  turbulence: a nonlinear route to dissipation through phase space}}.
  \jt{Plasma Physics and Controlled Fusion}  \bvol{50}~(12),  \pg{124024},
  \arxiv{arXiv: 0806.1069}.

\bibitem[{Schekochihin} {\em et~al.\/}(2009){Schekochihin}, {Cowley},
  {Dorland}, {Hammett}, {Howes}, {Quataert} \& {Tatsuno}]{Schekochihin:2009}
{\sc \au{{Schekochihin}, A.~A.}, \au{{Cowley}, S.~C.}, \au{{Dorland}, W.},
  \au{{Hammett}, G.~W.}, \au{{Howes}, G.~G.}, \au{{Quataert}, E.} \&
  \au{{Tatsuno}, T.}} \yr{2009}  \at{{Astrophysical Gyrokinetics: Kinetic and
  Fluid Turbulent Cascades in Magnetized Weakly Collisional Plasmas}}.
  \jt{Astrophys.~J.~Supp.}  \bvol{182},  \pg{310--377}.

\bibitem[{Schekochihin} {\em et~al.\/}(2019){Schekochihin}, {Kawazura} \&
  {Barnes}]{Schekochihin:2019}
{\sc \au{{Schekochihin}, A.~A.}, \au{{Kawazura}, Y.} \& \au{{Barnes}, M.~A.}}
  \yr{2019}  \at{{Constraints on ion versus electron heating by plasma
  turbulence at low beta}}.  \jt{Journal of Plasma Physics}  \bvol{85}~(3),
  \pg{905850303},  \arxiv{arXiv: 1812.09792}.

\bibitem[Schekochihin {\em et~al.\/}(2016)Schekochihin, Parker, Highcock,
  Dellar, Dorland \& Hammett]{Schekochihin:2016}
{\sc \au{Schekochihin, A.~A.}, \au{Parker, J.~T.}, \au{Highcock, E.~G.},
  \au{Dellar, P.~J.}, \au{Dorland, W.} \& \au{Hammett, G.~W.}} \yr{2016}
  \at{Phase mixing versus nonlinear advection in drift-kinetic plasma
  turbulence}.  \jt{Journal of Plasma Physics}  \bvol{82}~(2),  \pg{905820212}.

\bibitem[{Schoeffler} {\em et~al.\/}(2014){Schoeffler}, {Loureiro}, {Fonseca}
  \& {Silva}]{Schoeffler:2014}
{\sc \au{{Schoeffler}, K.~M.}, \au{{Loureiro}, N.~F.}, \au{{Fonseca}, R.~A.} \&
  \au{{Silva}, L.~O.}} \yr{2014}  \at{{Magnetic-Field Generation and
  Amplification in an Expanding Plasma}}.  \jt{Physical Review Letters}
  \bvol{112}~(17),  \pg{175001},  \arxiv{arXiv: 1308.3421}.

\bibitem[Servidio {\em et~al.\/}(2017)Servidio, Chasapis, Matthaeus, Perrone,
  Valentini, Parashar, Veltri, Gershman, Russell, Giles, Fuselier, Phan \&
  Burch]{Servidio:2017}
{\sc \au{Servidio, S.}, \au{Chasapis, A.}, \au{Matthaeus, W.~H.}, \au{Perrone,
  D.}, \au{Valentini, F.}, \au{Parashar, T.~N.}, \au{Veltri, P.}, \au{Gershman,
  D.}, \au{Russell, C.~T.}, \au{Giles, B.}, \au{Fuselier, S.~A.}, \au{Phan,
  T.~D.} \& \au{Burch, J.}} \yr{2017}  \at{Magnetospheric multiscale
  observation of plasma velocity-space cascade: Hermite representation and
  theory}.  \jt{Phys. Rev. Lett.}  \bvol{119},  \pg{205101}.

\bibitem[{Tatsuno} {\em et~al.\/}(2009){Tatsuno}, {Schekochihin}, {Dorland},
  {Plunk}, {Barnes}, {Cowley} \& {Howes}]{Tatsuno:2009}
{\sc \au{{Tatsuno}, T.}, \au{{Schekochihin}, A.~A.}, \au{{Dorland}, W.},
  \au{{Plunk}, G.}, \au{{Barnes}, M.~A.}, \au{{Cowley}, S.~C.} \& \au{{Howes},
  G.~G.}} \yr{2009}  \at{{Nonlinear phase mixing and phase-space cascade of
  entropy in gyrokinetic plasma turbulence}}.  \jt{Phys.~Rev.~Lett.}
  \bvol{103}~(1),  \pg{015003}.

\bibitem[TenBarge {\em et~al.\/}(2014)TenBarge, Daughton, Karimabadi, Howes \&
  Dorland]{TenBarge:2014b}
{\sc \au{TenBarge, J.~M.}, \au{Daughton, W.}, \au{Karimabadi, H.}, \au{Howes,
  G.~G.} \& \au{Dorland, W.}} \yr{2014}  \at{Collisionless reconnection in the
  large guide field regime: Gyrokinetic versus particle-in-cell simulations}.
  \jt{Phys.~Plasmas}  \bvol{21}~(2),  \pg{020708}.

\bibitem[{TenBarge} \& {Howes}(2012)]{TenBarge:2012a}
{\sc \au{{TenBarge}, J.~M.} \& \au{{Howes}, G.~G.}} \yr{2012}  \at{{Evidence of
  critical balance in kinetic Alfv{\'e}n wave turbulence simulations}}.
  \jt{Phys.~Plasmas}  \bvol{19}~(5),  \pg{055901}.

\bibitem[{TenBarge} \& {Howes}(2013)]{TenBarge:2013a}
{\sc \au{{TenBarge}, J.~M.} \& \au{{Howes}, G.~G.}} \yr{2013}  \at{{Current
  Sheets and Collisionless Damping in Kinetic Plasma Turbulence}}.
  \jt{Astrophys.~J.~Lett.}  \bvol{771},  \pg{L27},  \arxiv{arXiv: 1304.2958}.

\bibitem[{TenBarge} {\em et~al.\/}(2013){TenBarge}, {Howes} \&
  {Dorland}]{TenBarge:2013b}
{\sc \au{{TenBarge}, J.~M.}, \au{{Howes}, G.~G.} \& \au{{Dorland}, W.}}
  \yr{2013}  \at{{Collisionless Damping at Electron Scales in Solar Wind
  Turbulence}}.  \jt{Astrophys.~J.}  \bvol{774},  \pg{139}.

\bibitem[{Told} {\em et~al.\/}(2015){Told}, {Jenko}, {TenBarge}, {Howes} \&
  {Hammett}]{Told:2015}
{\sc \au{{Told}, D.}, \au{{Jenko}, F.}, \au{{TenBarge}, J.~M.}, \au{{Howes},
  G.~G.} \& \au{{Hammett}, G.~W.}} \yr{2015}  \at{{Multiscale Nature of the
  Dissipation Range in Gyrokinetic Simulations of Alfv{\'e}nic Turbulence}}.
  \jt{Phys.~Rev.~Lett.}  \bvol{115}~(2),  \pg{025003},  \arxiv{arXiv:
  1505.02204}.

\bibitem[{V\'asconez} {\em et~al.\/}(2014){V\'asconez}, {Valentini},
  {Camporeale} \& {Veltri}]{Vásconez:2014}
{\sc \au{{V\'asconez}, C.~L.}, \au{{Valentini}, F.}, \au{{Camporeale}, E.} \&
  \au{{Veltri}, P.}} \yr{2014}  \at{Vlasov simulations of kinetic alfvén waves
  at proton kinetic scales}.  \jt{Physics of Plasmas}  \bvol{21}~(11),
  \pg{112107},  \arxiv{arXiv: https://doi.org/10.1063/1.4901583}.

\bibitem[Vech {\em et~al.\/}(2017)Vech, Klein \& Kasper]{Vech:2017}
{\sc \au{Vech, D.}, \au{Klein, K.~G.} \& \au{Kasper, J.~C.}} \yr{2017}
  \at{Nature of stochastic ion heating in the solar wind: Testing the
  dependence on plasma beta and turbulence amplitude}.  \jt{The Astrophysical
  Journal Letters}  \bvol{850}~(1),  \pg{L11}.

\bibitem[Wan {\em et~al.\/}(2016)Wan, Matthaeus, Roytershteyn, Parashar, Wu \&
  Karimabadi]{Wan:2016}
{\sc \au{Wan, M.}, \au{Matthaeus, W.~H.}, \au{Roytershteyn, V.}, \au{Parashar,
  T.~N.}, \au{Wu, P.} \& \au{Karimabadi, H.}} \yr{2016}  \at{Intermittency,
  coherent structures and dissipation in plasma turbulence}.  \jt{Physics of
  Plasmas}  \bvol{23}~(4),  \pg{042307}.

\bibitem[Wang {\em et~al.\/}(2018)Wang, Tu, He \& Wang]{Wang:2018}
{\sc \au{Wang, X.}, \au{Tu, C.-Y.}, \au{He, J.-S.} \& \au{Wang, L.-H.}}
  \yr{2018}  \at{Ion-scale spectral break in the normal plasma beta range in
  the solar wind turbulence}.  \jt{Journal of Geophysical Research: Space
  Physics}  \bvol{123}~(1),  \pg{68--75},  \arxiv{arXiv:
  https://agupubs.onlinelibrary.wiley.com/doi/pdf/10.1002/2017JA024813}.

\bibitem[Yang {\em et~al.\/}(2017)Yang, Matthaeus, Parashar, Haggerty,
  Roytershteyn, Daughton, Wan, Shi \& Chen]{Yang:2017}
{\sc \au{Yang, Y.}, \au{Matthaeus, W.~H.}, \au{Parashar, T.~N.}, \au{Haggerty,
  C.~C.}, \au{Roytershteyn, V.}, \au{Daughton, W.}, \au{Wan, M.}, \au{Shi, Y.}
  \& \au{Chen, S.}} \yr{2017}  \at{Energy transfer, pressure tensor, and
  heating of kinetic plasma}.  \jt{Physics of Plasmas}  \bvol{24}~(7),
  \pg{072306}.

\end{thebibliography}

\end{document}